\documentclass[useAMS,usenatbib]{mn2e}
\usepackage{natbib}
\usepackage{epsfig}
\usepackage{longtable}
\usepackage{times}
\def \apj {ApJ}
\def \apjl {ApJL}

\def \aap {A\&A}

\def \aj {Astronomical Journal}
\def \mnras {MNRAS}
\def \araa {Annu. Rev. Astro. Astrophys.}
\title[The structure of blue supergiant winds and the accretion in supergiant High Mass X--ray Binaries]{The structure of blue supergiant winds and the accretion in supergiant High Mass X--ray Binaries}
\author[L. Ducci et al.]{L. Ducci,$^{1,2}$ L. Sidoli,$^{2}$ S. Mereghetti,$^{2}$ A. Paizis,$^{2}$ P. Romano,$^{3}$\\
$^{1}$ Dipartimento di Fisica e Matematica, Universit\`a degli Studi dell'Insubria, Via Valleggio 11, I-22100 Como, Italy \\
$^{2}$ INAF, Istituto di Astrofisica Spaziale e Fisica Cosmica, Via E. Bassini 15, I-20133 Milano, Italy \\
$^{3}$ INAF, Istituto di Astrofisica Spaziale e Fisica Cosmica, Via U. La Malfa 153, I-90146 Palermo, Italy}
\begin{document}

\pagerange{\pageref{firstpage}--\pageref{lastpage}} \pubyear{2009}

\maketitle

\label{firstpage}

\begin{abstract}
We have developed a stellar wind model for OB supergiants
to investigate the effects of  accretion from a clumpy wind
on the luminosity and variability properties of High Mass X--ray
Binaries. Assuming that the clumps are confined by ram pressure of
the ambient gas and exploring different distributions for their
mass and radii, we computed the expected X--ray light curves in
the framework of the Bondi-Hoyle accretion theory, modified to
take into account the presence of clumps. The resulting
variability properties are found to depend not only on the assumed
orbital parameters but also on the wind characteristics. We have
then applied this model to reproduce the X-ray light curves of
three representative High Mass X-ray Binaries: two persistent
supergiant systems (Vela X-1 and 4U 1700--377) and the Supergiant
Fast X-ray Transient IGR~J11215$-$5952. The model can reproduce
well the observed light curves, but requiring in all cases an
overall mass loss from the supergiant about a factor 3$-$10 
smaller than the values inferred from UV lines studies that assume
a homogeneous wind.
\end{abstract}

\begin{keywords}
X--ray: individuals: Vela~X$-$1/4U~1900$-$40, 4U 1700$-$377, IGR~J11215$-$5952. stars: supergiants.
\end{keywords}

%--------------------------
\section{Introduction}
\label{Introduction}
%--------------------------

A new class of massive X-ray binaries has been recognized in the
last few years, mainly  thanks to observations carried out with
the \emph{INTEGRAL} satellite. They are transient X-ray sources
associated to O or B supergiant stars and characterized by short
outbursts. These \emph{Supergiant Fast X$-$ray Transients} 
(SFXTs) [\citet{Sguera-et-al.-2005}; \citet{Negueruela-et-al.-2006}] 
are remarkably different from the classical 
High Mass X-ray Binaries (HMXBs) with supergiant
companions, that are bright persistent sources, and also differ
from the Be transients for their optical companions and shorter
outbursts.

The outbursts of SFXTs involve a high dynamic range, spanning 3 to
5 orders of magnitudes, from a quiescent luminosity of
$\sim10^{32}$~erg~s$^{-1}$ up to the peak luminosity of $10^{36} -
10^{37}$~erg~s$^{-1}$. The outbursts typically last a few days and
are composed of many short flares with duration of a few hours. 
Besides these bright outbursts, the SFXTs can
display a fainter flaring activity with  luminosity $L_{\rm
x}=10^{33}-10^{34}$~erg~s$^{-1}$ \citep{Sidoli-et-al.-2008}.
Different mechanisms have been proposed to explain  the SFXTs
properties (see \citet{Sidoli-2009} and references therein for a recent review).

\citet{Sidoli-et-al.-2007},
proposed that SFXTs outbursts are due to the presence of an
equatorial wind component, denser and slower than the spherically
symmetric wind from the supergiant, and possibly inclined with respect to
the orbital plane of the system.  The enhanced accretion rate
occurring when the neutron star crosses this wind component can
explain SFXTs showing periodic outbursts,  such as
IGR~J11215$-$5952 and also other SFXTs, assuming 
different geometries for the outflowing equatorial wind.

Another possibility  involves the gated mechanisms due to
transitions across the centrifugal barrier
\citep{Grebenev-and-Sunyaev-2007}.\
\citet{Bozzo-et-al.-2008} showed that a centrifugal
or a magnetic barrier can explain the
SFXTs properties only if the supergiant wind is inhomogeneous, and
the accreting neutron star has a strong magnetic field ($\ga 10^{14}$~G) and a
long spin period ($P_{\rm spin} \geq 10^3$~s).

\citet{in't-Zand-2005} proposed that the SFXTs flares are produced
by accretion of clumps of matter from the companion wind. 
In the framework of the clumpy wind model proposed by
\citet{Oskinova-et-al.-2007},
\citet{Walter-and-Zurita-Heras-2007} and
\citet{Negueruela-et-al.-2008} proposed that what distinguishes
the SFXTs from the persistent HMXBs with supergiant companions is
their different orbital separation:
in persistently bright sources the compact object orbits the
companion at a small distance ($<$2 stellar radii) where there is
a high number density of clumps, while the transient emission in SFXTs is
produced by accretion of much rarer clumps present at larger
distances.
On the other hand, the monitoring with \emph{Swift} of a sample
of four SFXTs \citep{Sidoli-et-al.-2008} has demonstrated that these
sources accrete matter also outside the bright outbursts, so any
model should also account for this important observational
property.

The winds of O and B type stars are driven by the momentum
transfer of the radiation field and \citet{Lucy-and-Solomon-1970}
showed that the dominant mechanism is the line scattering. The
analytical formulation for line-driven winds has been  developed
by \citet{Castor-et-al.-1975} (CAK). In the CAK theory the mass
lost by the star is smoothly accelerated by the momentum
transferred from the stellar continuum radiation, and forms a
stationary and homogeneous wind. However, both observational
evidence and theoretical considerations indicate that the stellar
winds are variable and non homogeneous. Changes in the UV line
profiles, revealing wind variability, 
have been observed on time scales shorter than a day \citep{Kudritzki-and-Puls-2000}. 
The X-ray variability observed in 4U~1700$-$377,
Vela~X$-$1 and other HMXBs can be explained in terms of wind
inhomogeneity [\citet{White-et-al.-1983}; \citet{Kreykenbohm-et-al.-2008}] 
and further indications for the presence of
clumps come from  X$-$ray spectroscopy. For example, the X$-$ray
spectrum of Vela~X$-$1 during the eclipse phase shows
recombination lines produced by a hot ionized gas and fluorescent
K-shell lines produced by cool and dense gas of near-neutral ions
\citep{Sako-et-al.-1999}. These authors proposed that the
coexistence of highly ionized and near neutral ions can be
explained with an inhomogeneous wind, where cool, dense clumps are
embedded in a lower density, highly ionized medium.

\citet{Lucy-and-White-1980} suggested that the wind acceleration
is subject to a strong instability since small perturbations in
the velocity or density distribution grow with time producing a
variable and strongly structured wind. The first time-dependent
hydrodynamical simulations of unstable line-driven winds were
performed by \citet{Owocki-et-al.-1988} and, more recently, by
\citet{Runacres-and-Owocki-2002} and
\citet{Runacres-and-Owocki-2005}. All these simulations show that
the line-driven instability produces a highly structured wind,
with reverse and forward shocks that compress the gas into clumps.
Moreover, the shock heating can generate a hot inter-clump medium
into which the  colder clumps are immersed
\citep{Carlberg-et-al.-1980}.

Based on these considerations, we study in this work the expected
variability and X$-$ray luminosity properties of neutron stars accreting
from a clumpy wind. In the next two Sections we describe our
clumpy wind model and the assumptions for the mass accretion. 
In Section \ref{Application of the clumpy wind model} 
we study the dependence of the parameters introduced in the previous sections.
In Sections \ref{Study of the HMXB Vela X-1}, \ref{Study of the SGXB 4U 1700-377 } 
and \ref{Study of the SFXT IGR J11215-5952} we compare the X-ray light curves predicted
with our model to the observations of three high mass X-ray
binaries, showing that it is possible to reproduce well the
observed flaring behaviors.

%----------------------------------------
\section{Clumpy stellar wind properties}
\label{Section Clumpy stellar winds properties}
%----------------------------------------

In our model, where the dynamical problem has not been treated, 
we assume that a fraction  of the stellar wind is in
form of clumps with a power law mass distribution

\begin{equation} \label{Npunto}
p(M_{\rm cl})=k \left ( \frac{M_{\rm cl}}{M_{\rm a}} \right)^{-\zeta}
\end{equation}
in the mass range $M_{\rm a}$ - $M_{\rm b}$.
The rate of clumps produced by the supergiant is related to the
total mass loss rate $\dot{M}_{\rm tot}$ by:
\begin{equation} \label{Ndot emitted}
\dot{N}_{\rm cl} = \frac{f \dot{M}_{\rm tot}}{<M>} \mbox{ \ \
clumps s}^{-1},
\end{equation}
where $f=\dot{M}_{\rm cl} / \dot{M}_{\rm tot}$ is the fraction of
mass lost in  clumps  and $<M>$ is the average clump mass, which can
be computed from Equation (\ref{Npunto}).
Clumps are driven radially outward by absorption of UV spectral
lines \citep{Castor-et-al.-1975}. From spectroscopic observations
of O stars, \citet{Lepine-and-Moffat-2008} suggest that clumps
follow on average the same velocity law of a smooth stellar wind.
We can then assume the following clump velocity profile
without solving the dynamical problem:
\begin{equation} \label{legge_velocita}
v_{\rm cl}(r) = v_{\infty}\left (1 - 0.9983\frac{R_{\rm OB}}{r}
\right )^{\beta}
\end{equation}
where $v_{\infty}$ is the terminal wind velocity, $R_{\rm OB}$ is the
radius of the supergiant, $0.9983$ is a parameter which ensures
that $v(R_{\rm OB}) \approx 10$~km~s$^{-1}$, and $\beta$ is a
constant in the range $\sim$$0.5$--$1.5$
[\citet{Lamers-and-Cassinelli-1999};
\citet{Kudritzki-et-al.-1989}].

Assuming that the clumps are confined by the ram pressure of the
ambient gas, their size can be derived by the balance pressure
equation. Following \citet{Lucy-and-White-1980} and
\citet{Howk-et-al.-2000}, the average density of a clump is:
\begin{equation} \label{lucy_white}
\bar{\rho}_{\rm cl} = \rho_{\rm w}(r) \left ( \frac{a_{\rm w}^2 +
C_{\rho}\omega^2}{a_c^2} \right )
\end{equation}
where $\rho_{\rm w}(r)$ is the density profile of the homogeneous
(inter-clump) wind, $a_{\rm w}$ and $a_c$ are the inter-clump wind
and the clump thermal velocity, respectively: $a_{\rm
w}^2=\frac{kT_{\rm w}}{\mu m_H}$ and $a_c^2=\frac{kT_c}{\mu m_H}$.
$k$ is the Boltzmann constant, $T_{\rm w}$ and $T_c$ are the
temperatures of the inter-clump wind and of the clumps,
respectively, and $\mu$ is the mean atomic weight. The constant
$C_{\rho}=0.29$ accounts for the confining effect  of the bow shock
produced by the ram pressure around the clump
\citep{Lucy-and-White-1980}. $\omega$ is the relative velocity
between the wind and the clump ($\omega = v_{\rm w} - v_{\rm
cl}$). Adopting $\bar{\omega} \sim 5 \times 10^7$~cm~s$^{-1}$,
$\bar{T}_c \sim 10^5$~K,  $\bar{T}_{\rm w} \sim 10^7$~K and
$\mu=1.3$, we obtain\footnote{This value is just an estimate of a typical value
of ratio of clump to ambient density, and will not adopted throughout.}:
\begin{equation} \label{lucy_white_200}
\left ( \frac{a_{\rm w}^2 + C_{\rho}\omega^2}{a_c^2} \right )
\approx 200
\end{equation}
Since the density radial profile $\rho_{\rm w}(r)$ of the
homogeneous inter-clump wind (obtained from the continuity
equation $\dot{M} = 4 \pi r^2 \rho_{\rm w}(r) v(r) = \ constant$)
is:
\begin{equation} \label{legge_densita_vento_omogeneo}
\rho_{\rm w}(r) = \rho_{\rm w}(r_0) \frac{r_0^2 v(r_0)}{r^2 v(r)}
\end{equation}
where $r_0$ is a generic distance from the supergiant, from
Equations (\ref{lucy_white}), (\ref{lucy_white_200}),
(\ref{legge_densita_vento_omogeneo}), we obtain:
\begin{equation} \label{legge_densita_clump}
\bar{\rho}_{\rm cl}(r) = \bar{\rho}_{\rm cl}(r_0) \frac{r_0^2
v(r_0)}{r^2 v(r)}
\end{equation}
where $\bar{\rho}_{\rm cl}(r_0) = \rho_{\rm w}(r_0) \times ( a_{\rm w}^2 + C_{\rho}\omega^2)/a_c^2 $.
\citet{Bouret-et-al.-2005} analyzed the far-ultraviolet spectrum
of O-type supergiants and found that clumping starts deep in the
wind, just above the sonic point $R_s$, at velocity $v(R_s)
\approx 30$~km~s$^{-1}$. In the CAK model the sonic point is
defined as the point where the wind velocity is equal to the sound
speed ($v(R_s)=c_s$):
\begin{equation} \label{SonicPoint}
R_s = \frac{0.9983 R_{\rm OB}}{1 - (c_s/v_{\infty})^{1/\beta}}
\end{equation}
Adopting typical parameters for O supergiants, from Equation
(\ref{SonicPoint}) we obtain that the clumping phenomenon starts
close to the photosphere ($R_s \approx R_{\rm OB}$).
Assuming spherical geometry for the clumps and that the mass of
each clump is conserved, it is possible to obtain the expansion
law of the clumps from Equation (\ref{legge_densita_clump}), with
$r_0=R_s$:
\begin{equation} \label{legge_Rcl_r}
R_{\rm cl}(r) = R_{\rm cl}(R_s) \left [ \frac{r^2 v_{\rm
cl}(r)}{R_s^2 v(R_s)} \right ]^{1/3}
\end{equation}
From Equation (\ref{legge_Rcl_r}) we find that the clump size
increases with the distance from the supergiant star.

For the initial clump dimensions we tried two different
distributions, a power law
\begin{equation} \label{distrib R_cl}
\dot{N}_{M_{cl}} \propto R_{\rm cl}^{\gamma} \mbox{ \ \ clumps s}^{-1}
\end{equation}
and a truncated gaussian function:
\begin{equation} \label{distrib R_cl gauss}
\dot{N}_{M_{cl}} \propto \frac{1}{\sigma \sqrt{2 \pi}} e^{-\frac{1}{2}
\left ( \frac{R_{\rm cl} - \bar{R}_{\rm cl}}{\sigma} \right )^2}
\mbox{ \ \ clumps s}^{-1}
\end{equation}
where $\sigma=(R_{\rm cl,max}-R_{\rm cl,min})/(2N_{\sigma})$, and
$N_{\sigma}$ is a free parameter.

For any given  mass clump we derived the minimum and maximum
values for the initial radii as follows. The minimum radius is
that below which the clump is optically thick for the UV resonance
lines. In this case  gravity dominates over the radiative force
causing the clump to fall back onto the supergiant.
The momentum equation of a radiatively driven clump is:
\begin{equation} \label{CAK_equation}
v_{\rm cl} \frac{dv_{\rm cl}}{dr} = - \frac{GM_{\rm OB}}{r^2} +
g_{\rm e} + g_{\rm L}
\end{equation}
where
$M_{\rm OB}$ is the mass of the supergiant,
$g_{\rm e}$ is the radiative acceleration due to the continuum
opacity by electron scattering, $g_{\rm L}$ is the radiative
acceleration due to line scattering. While Equation
(\ref{CAK_equation}) is an approximation that ignores the pressure
gradient, the solution of the momentum equation differs only
slightly from the accurate solution derived by
\citet{Kudritzki-et-al.-1989}; another assumption is that the
photosphere has been treated as a point source
\citep{Lamers-and-Cassinelli-1999}.
The radiative acceleration due to electron scattering is:
\begin{equation} \label{radiative acceleration due to electron scattering}
g_{\rm e}(r) = \frac{\sigma_{\rm e}(r) L_{\rm OB}}{4 \pi r^2 c}
\end{equation}
where $\sigma_{\rm e}$ is the opacity for electron scattering, and
is given by $\sigma_{\rm e} = \sigma_T \frac{n_e}{\rho_{\rm cl}}$,
where $\sigma_T$ 
is the Thomson
cross section, $n_e$ is the number density of electrons,
$\rho_{\rm cl}$ is the density of the clump and $L_{\rm OB}$ is
the luminosity of the OB supergiant.
\citet{Lamers-and-Cassinelli-1999} found $0.28 < \sigma_{\rm e} <
0.35 \ cm^2 \ g^{-1}$. Assuming a constant degree of ionization in
the wind both $\sigma_{\rm e}$ and $g_{\rm e}$ are constant.

The radiative acceleration due to the line scattering is:
\begin{equation} \label{g_L}
g_{\rm L} = \frac{\sigma_{\rm e}^{\rm ref} L_{\rm OB}}{4 \pi c
r^2} k t^{-\alpha} \left ( 10^{-11} \frac{n_e}{W} \right
)^{\delta}
\end{equation}
where $\sigma_{\rm e}^{\rm ref} = 0.325$~cm$^2$~g$^{-1}$, $k$,
$\alpha$, $\delta$ are the force multiplier parameters, which are
obtained with the calculation of the line radiative force
\citep{Abbott-1982}. $t$ is the dimensionless optical depth
parameter \citep{Castor-et-al.-1975}. According to the model of
\citet{Howk-et-al.-2000}, we assume no velocity gradient inside
the clump (of size $l \approx 2 R_{\rm cl}$), then we utilize the
dimensionless optical depth parameter for a static atmosphere $t =
\sigma_{\rm e} \int_l \rho dr$. $W(r)$ is the geometrical dilution
factor \citep{Lamers-and-Cassinelli-1999}, given by:
\begin{equation} \label{geometrical dilution factor}
W(r) = \frac{1}{2} \left [ 1 - \sqrt{1 - \left ( \frac{R_{\rm
OB}}{r}\right )^2} \right ]
\end{equation}
According to the Equation (\ref{legge_densita_clump}), we find
that the number density of electrons in each clump is given by:
\begin{equation} \label{number density of electrons in each clump}
n_e(r) = n_0 \frac{R_{\rm s}^2 v(R_{\rm s})}{r^2 v(r)}
\end{equation}
where $n_0=n_e(R_{\rm s})$ such that:
\begin{equation} \label{n_e}
n_0 = \frac{\rho_{\rm cl}(R_{\rm s})}{\mu_e m_H} \mbox{ .}
\end{equation}

From Equation (\ref{CAK_equation}) we obtain that the minimum radius
of the clump is given by:
\begin{equation} \label{condizione Rclmin}
g_{\rm L} + g_{\rm e} -g_{\rm g} = 0
\end{equation}
where $g_{\rm g}$ is the acceleration due to gravity.
Approximating $t$ as
\begin{equation} \label{t approssimata}
t \approx \sigma_{\rm e} \frac{M_{\rm cl}}{V_{\rm cl}} 2R_{\rm cl}
= \frac{3}{2} \sigma_{\rm e} \frac{M_{\rm cl}}{\pi R_{\rm cl}^2}
\end{equation}
and assuming the force multiplier parameters calculated by
\citet{Shimada-et-al.-1994}, with $v(R_{\rm s})=30$~km~s$^{-1}$ and
$r_0=R_{\rm s}$, we finally obtain
the lower-limit for the clump radius:
\begin{equation} \label{R_cl_min}
R_{\rm cl,min}(M_{\rm cl}) = \left ( \frac{A \cdot B}{C} \right
)^{1/(3\delta - 2\alpha)}
\end{equation}
where:
\begin{eqnarray}
A & = & \left ( \frac{3 \sigma_{\rm e} M_{\rm cl}}{2 \pi} \right )^{- \alpha}\\
B & = & \left ( \frac{3 M_{\rm cl} 10^{-11}}{4 \pi \mu_e m_H W(R_s)} \right )^{\delta} \\
C & = & (g_{\rm g} - g_{\rm e})\frac{4 \pi c R_{\rm
S}^2}{\sigma_{\rm e}^{\rm ref} L_{\rm OB} k}
\end{eqnarray}
We found that for the interesting range of the clump masses, 
the drag force \citep{Lucy-and-White-1980} values
are less than 3\% of the forces resulting from Equations 
(\ref{radiative acceleration due to electron scattering}) and 
(\ref{g_L}). Thus, in the determination of the minimum clump radius,
we can neglect this contribution.

The upper-limit to the clump radius is obtained from the
definition of a clump as an over-density with respect to the
inter-clump smooth wind:

\begin{equation} \label{rho_cl_max}
\rho_{\rm cl} \geq \frac{\dot{M}_{\rm w}}{4 \pi R_{\rm s}^2 v_{\rm
w}(R_{\rm s})}
\end{equation}
where $\dot{M}_{\rm w}$ is the mass loss rate of the homogeneous
wind component (inter-clump). From Equation (\ref{rho_cl_max}) we
obtain the upper-limit for the clump radius
\begin{equation} \label{R_cl_max}
R_{\rm cl,max}(M_{\rm cl}) = \left ( \frac{3 M_{\rm cl} R_{\rm
S}^2 v_{\rm w}(R_{\rm s})}{\dot{M}_{\rm w}}   \right )^{1/3}
\end{equation}
The lower and the upper limits for the clump radius (Equations
\ref{R_cl_min} and \ref{rho_cl_max}) depend on  the supergiant
parameter $T_{\rm eff}$, $L_{\rm OB}$, $R_{\rm OB}$, $v_{\infty}$,
$\beta$, $\dot{M}_{\rm w}$. Thus supergiants of different spectral
type have, for any given mass of the clump, different minimum and
maximum values for the initial radii distribution. However, these
differences are smaller than a factor 10 for OB supergiants, and
the intersection of the upper-limit and lower-limit functions
ranges from $M_{\rm cl}\approx 10^{20}$~g to $\approx 10^{23}$~g.

%----------------------------------------
\section{Luminosity computation}
\label{Luminosity computation}
%----------------------------------------

%
\begin{figure*}
\begin{center}
\includegraphics[bb= 415 2 687 218,clip]{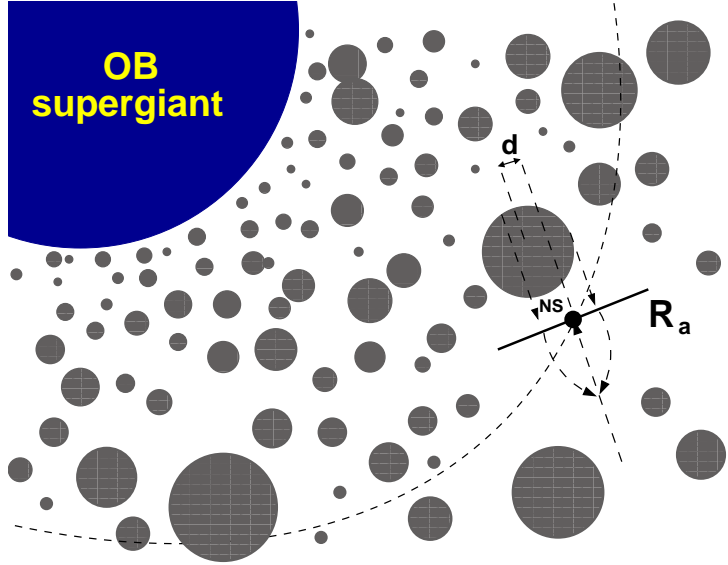}
\end{center}
\caption{Schematic representation of our clumpy wind model.  $d$
is the distance between the centre of the clump and the centre of
accreting compact object. $R_{\rm a}$ is the accretion radius.}
\label{figura accrescimento clump}
\end{figure*}
The Bondi-Hoyle-Lyttleton accretion theory
[\citet{Hoyle-and-Lyttleton-1939}; \citet{Bondi-and-Hoyle-1944}]
is usually applied to the HMXBs where
a OB supergiant loses mass in the form of a fast stellar wind,
(terminal velocity, $v_{\infty} \approx 1000 - 2000$~km~s$^{-1}$),
that is assumed to be homogeneous and spherically symmetric.
Only matter within a distance
smaller than the \emph{accretion radius} ($R_{\rm a}$) is
accreted:
\begin{equation} \label{accretion radius}
R_{\rm a}(r) = \frac{2 G M_{\rm NS}}{v_{\rm rel}^2(r) + c_s^2}
\end{equation}
where $M_{\rm NS}$ is the neutron star mass
and $v_{\rm rel}(r)$ is the relative velocity
between the neutron star and the wind.
The fraction of stellar wind gravitationally captured by the
neutron star is given by:
\begin{equation} \label{M_accr}
\dot{M}_{\rm accr}= \rho_{\rm wind}(r)v_{\rm rel}(r) \pi R_{\rm a}^2
\end{equation}
where $\rho_{\rm wind}(r)$ is the density of the wind
\citep{Davidson-and-Ostriker-1973}.

\begin{figure*}
\begin{center}
\begin{tabular}{cccc}
\includegraphics[height=4.5cm]{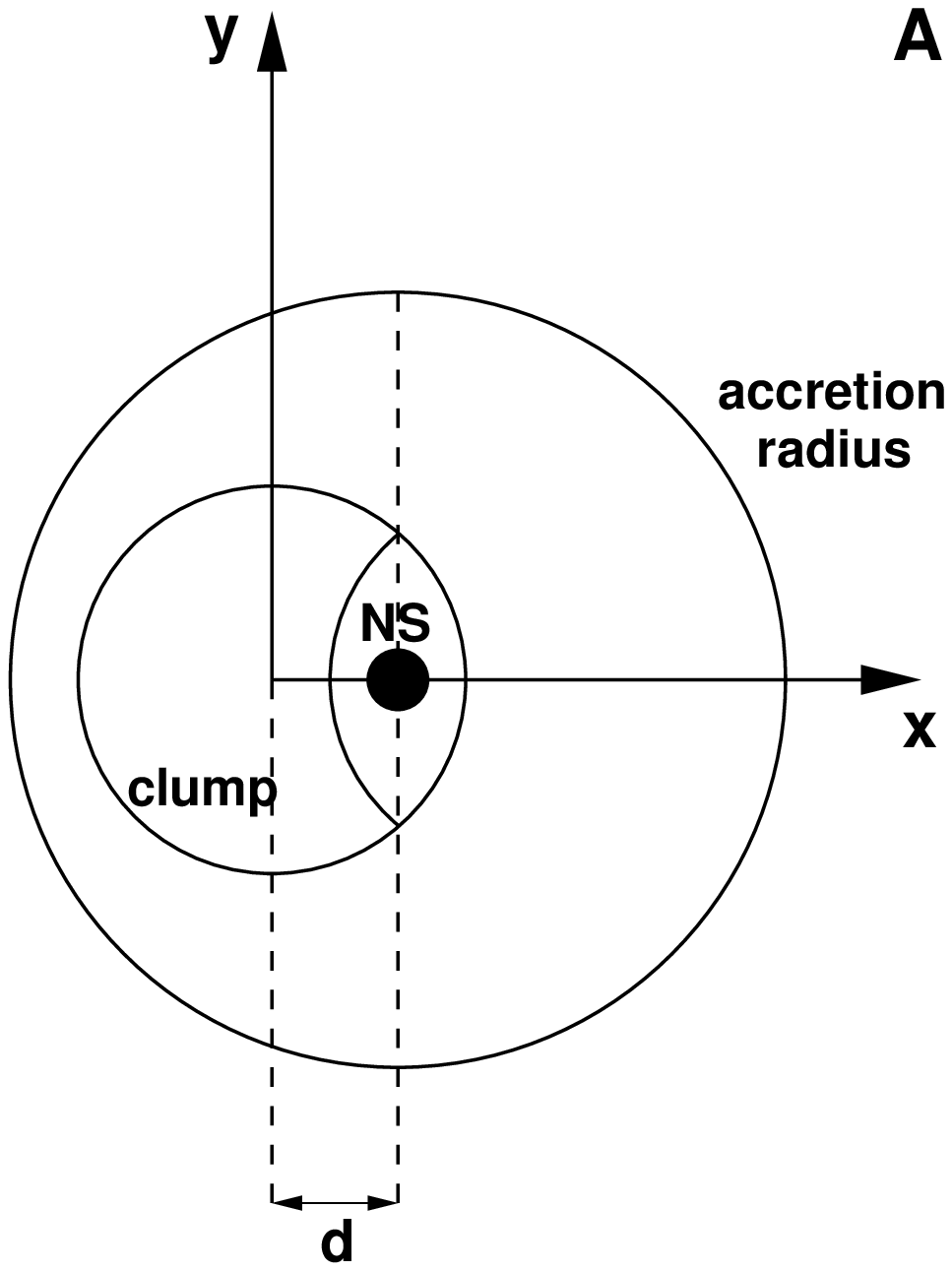} &
\includegraphics[height=4.5cm]{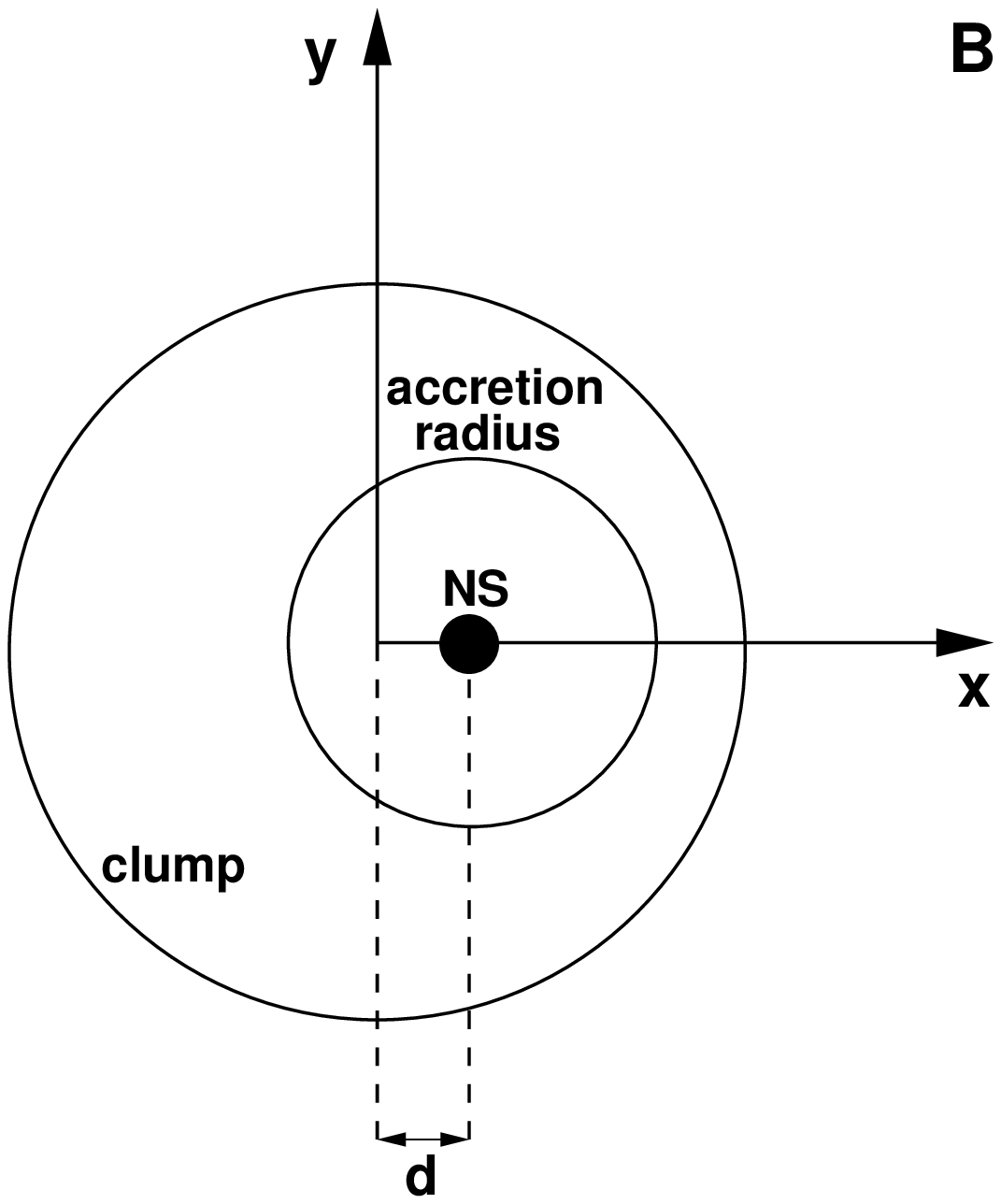} &
\includegraphics[height=4.5cm]{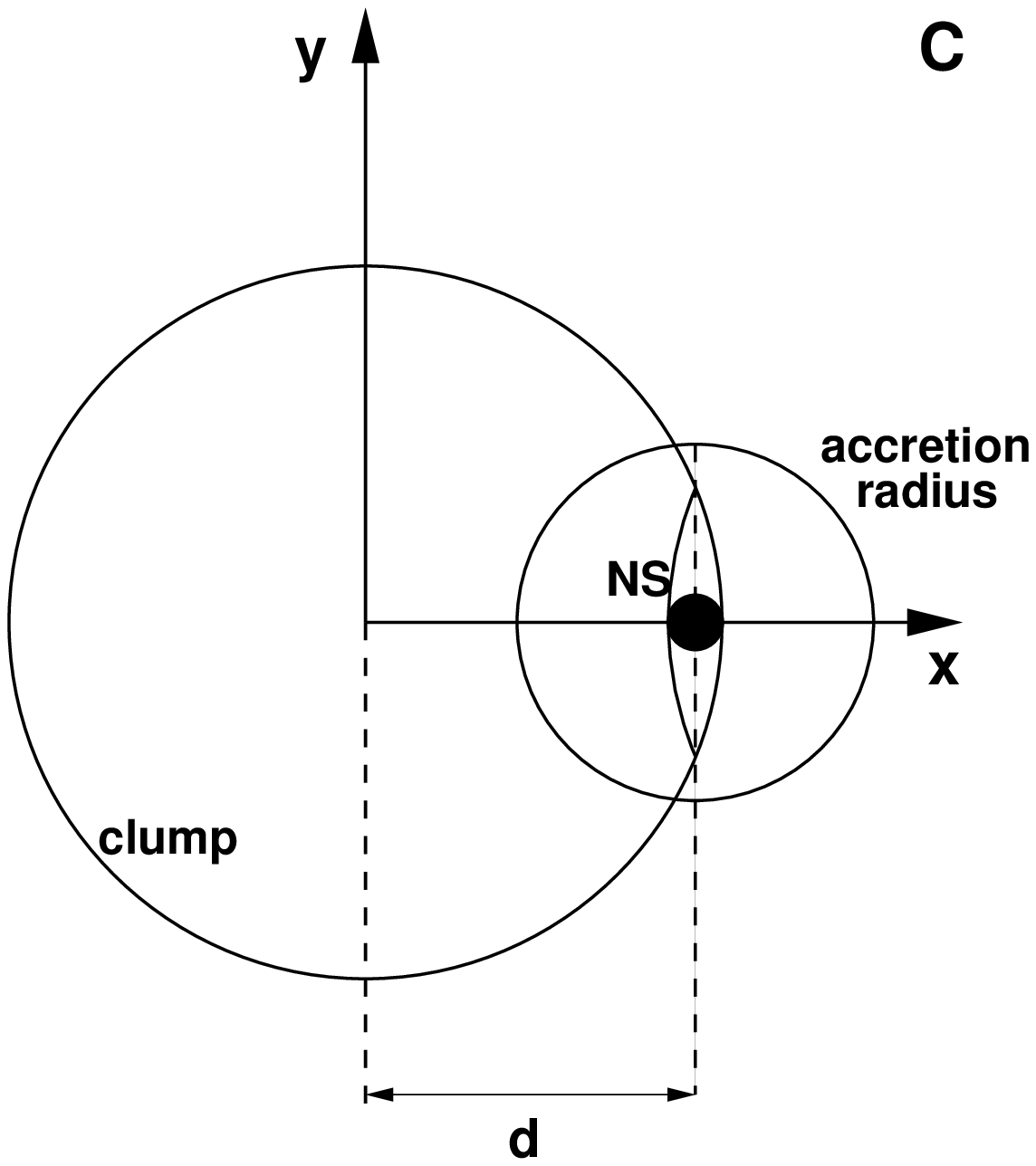} &
\includegraphics[height=4.5cm]{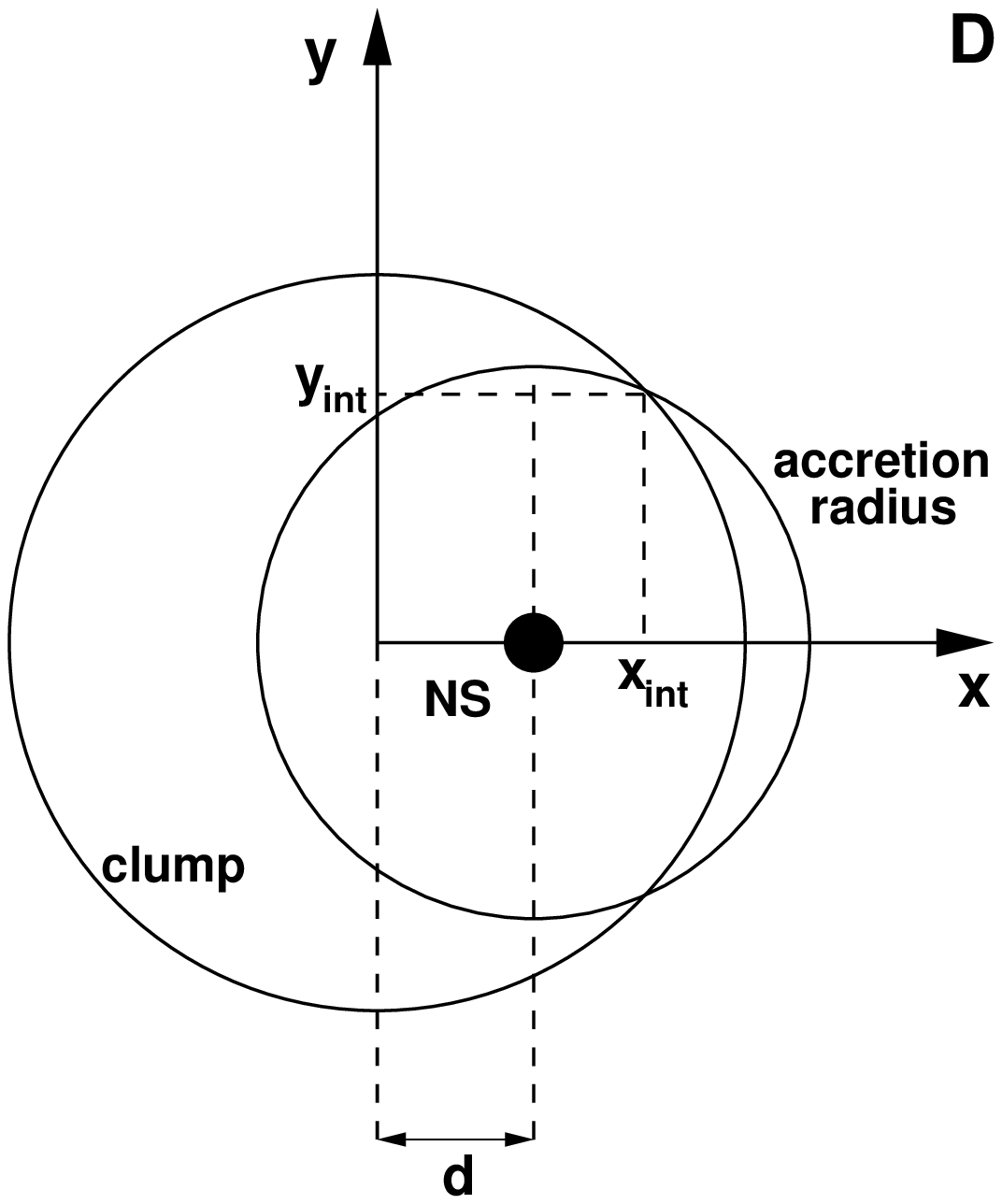}
\end{tabular}
\end{center}
\caption{Schematic view of different possibilities for the
accretion of clumps of radius $R_{\rm cl}$.
  A: $R_{\rm cl} \leq R_{\rm a}$ and $d \neq 0$ (Eq. \ref{Rcl minore Ra});
  B: $R_{\rm cl} > R_{\rm a}$ and $R_{\rm cl}-d \geq R_{\rm a}$
  (Eq. \ref{Rcl maggiore Ra and accrescimento totale});
  C: $R_{\rm cl} -d< R_{\rm a}$ and $\sqrt{R_{\rm cl}^2-d^2} < R_{\rm a}$
  (Eq. \ref{Rcl minore Ra});
  D: $R_{\rm cl} -d< R_{\rm a}$ and $\sqrt{R_{\rm cl}^2-d^2} > R_{\rm a}$
  (Eq. \ref{Rcl maggiore Ra and sqrt(Rcl^2 - d^2) maggiore Ra}).}
\label{figura A B C D}
\end{figure*}
The Bondi-Hoyle-Lyttleton accretion theory, that is based on a
homogeneous wind, requires an important modification to properly
take into account the presence of inhomogeneity in a clumpy wind.
In the homogeneous case the  wind particles are deflected by the
gravitational field of the neutron star, and collide with the
particles having the symmetric trajectory in a cylindrical region
with axis along the relative wind direction. The collisions
dissipate the kinetic energy perpendicular to this axis, and only
the particles with a parallel kinetic energy component lower than
the gravitational potential energy are accreted. The application
of this accretion mechanism to an inhomogeneus wind can lead to a
partial accretion of the clump: when the distance $d$ between the
neutron star and the projection of the centre of the clump on the
accretion cross section is smaller than the clump radius and $d
\neq 0$, only a fraction of the mass of the clump will be accreted
(see Figure \ref{figura accrescimento clump}).
In particular, if an incoming clump
that crosses the accretion cross-section $\pi R_{\rm a}^2$,
is smaller than the accretion radius
($R_{\rm cl} < R_{\rm a}$) and $d < R_{\rm cl}$,
the accretion cross-section in Equation (\ref{M_accr})
must be replaced by%\footnote{where:
\begin{equation} \label{Rcl minore Ra}
\Sigma = 4 \int_d^{R_{\rm cl}} \sqrt{R_{\rm cl}^2 - x^2} dx
\end{equation}
and the mass accretion rate is
$\dot{M}_{\rm accr} = \rho_{\rm cl}(r) v_{\rm rel}(r) \times \Sigma$
(see Figure \ref{figura A B C D},A).
If $R_{\rm cl} \ge R_{\rm a}$ we have three cases:
\begin{itemize}
\item when $R_{\rm cl} - d \ge R_{\rm a}$ (Figure \ref{figura A B C D},B),
the accretion cross-section is given by:
\begin{equation} \label{Rcl maggiore Ra and accrescimento totale}
\Sigma  =  \pi R_{\rm a}^2
\end{equation}
\item when $R_{\rm cl} -d < R_{\rm a}$ and $\sqrt{R_{\rm cl}^2
-d^2} < R_{\rm a}$ (Figure \ref{figura A B C D},C), the accretion
cross-section is given by Equation (\ref{Rcl minore Ra}).
\item when $R_{\rm cl} -d < R_{\rm a}$ and $\sqrt{R_{\rm cl}^2 -d^2} > R_{\rm a}$
(Figure \ref{figura A B C D},D),
the accretion cross-section is given by:
\begin{eqnarray} \label{Rcl maggiore Ra and sqrt(Rcl^2 - d^2) maggiore Ra}
\Sigma & = & 4  \left [  \int_{x_{int}}^{R_{\rm cl}} \sqrt{R_{\rm cl}^2 -
    x^2}dx - (x_{int} - d)y_{int} \right. + \nonumber \\
       & + & \left. \int_{y_{int}}^{R_{\rm a}}  \sqrt{R_{\rm a}^2 - y^2}dy \right ]
\end{eqnarray}
\end{itemize}
The number density of clumps $n_{\rm cl}$ obeys the equation of
continuity $\dot{N}_{\rm cl}=4 \pi r^2 n_{\rm cl}(r) v_{\rm
cl}(r)$, where $\dot{N}_{\rm cl}$ is the rate of clumps emitted by
the OB supergiant (see Equation \ref{Npunto}). Thus:
\begin{equation} \label{clump number density}
n_{\rm cl}(r) = \frac{\dot{N}_{\rm cl}}{4 \pi r^2 v_{\rm cl}(r)} \mbox{ \ \ clumps cm}^{-3}
\end{equation}
The rate of clumps accreted by the neutron star is
given by:
\begin{equation} \label{Naccr}
\dot{N}_{\rm accr} = n_{\rm cl}(r) v_{\rm rel}(r) \times (\pi R_{\rm a}^2) \mbox{ \ \ clumps s}^{-1}
\end{equation}
When the neutron star accretes only the  inter-clump wind, the
X$-$ray luminosity variations are due to changes in its distance
from the OB companion due to orbital eccentricity. The
corresponding luminosity (see Equations \ref{accretion radius} and
\ref{M_accr}) is:
\begin{equation} \label{Lx interclump dettagliata}
L_{\rm x, wind}(\phi) = \frac{G M_{\rm NS}}{R_{\rm NS}}\dot{M}_{accr} = \frac{(G M_{\rm NS})^3}{R_{\rm NS}} \frac{ 4 \pi \rho_{\rm w}(r)}{[(v_{\rm rel}^2(r) + c_s^2]^{3/2} }
\end{equation}
where $\phi$ is the orbital phase, and $\rho_{\rm w}(r)$
is given by the Equation (\ref{legge_densita_vento_omogeneo}).

When the neutron star accretes a clump its X$-$ray luminosity is
given by:
\begin{equation} \label{L_x clump dettagliata}
L_{\rm x, cl} (\phi) = \frac{G M_{\rm NS}}{R_{\rm NS}} \frac{M_{\rm cl}}{\frac{4}{3}\pi R_{\rm cl}^3} v_{\rm rel} \times \Sigma
\end{equation}
where $R_{\rm cl}(r)$ is given by the Equation (\ref{legge_Rcl_r}),
and $\Sigma$ by the Equations (\ref{Rcl minore Ra}) -- (\ref{Rcl maggiore Ra and sqrt(Rcl^2 - d^2) maggiore Ra}).
If two or more clumps are accreted at the same time,
the X$-$ray luminosity at the peak of the flare produced
by the accretion is given by the sum of the luminosities
(Equation \ref{L_x clump dettagliata})
produced by the accretion of each clump.

%---------------------------------------------
\section{Application of the clumpy wind model}
\label{Application of the clumpy wind model}
%---------------------------------------------

In this section we investigate how the X-ray luminosity and
variability properties depend on the different clumpy wind
parameters and orbital configurations.

\begin{figure}
\begin{center}
\includegraphics[width=8cm]{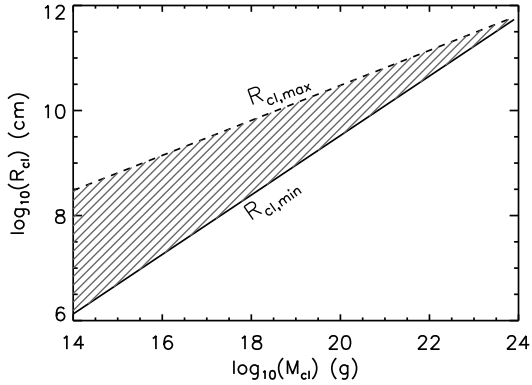}
\end{center}
\caption{Upper (dashed line) and lower-limit
  (solid line) for the clump radius at $r=R_{\rm s}$.
  They have been obtained from Equations (\ref{R_cl_min}) and (\ref{R_cl_max})
  assuming the following parameters for the
  supergiant: $M_{\rm OB}=30$~M$_{\odot}$, $R_{\rm OB}=23.8$~R$_{\odot}$,
  $\dot{M}_{\rm tot}=10^{-6}$~M$_{\odot}$~yr$^{-1}$,
  $v_{\infty}=1700$~km~s$^{-1}$, $\beta=1$, $v_0=10$~km~s$^{-1}$,
  $\dot{M}_{\rm cl}/\dot{M}_{\rm wind}=0.7$.}
\label{relazione_Mcl_Rcl_Mloss025}
\end{figure}

As an example, we  consider a binary consisting of an O8.5I star
with $M_{\rm OB}=30$~M$_{\odot}$, $R_{\rm OB}=23.8$~R$_{\odot}$
\citep{Vacca-et-al.-1996}, and a neutron star with $M_{\rm
NS}=1.4$~M$_{\odot}$, $R_{\rm NS}=10$~km. The parameters for the
supergiant wind are the following: $\dot{M}_{\rm
tot}=10^{-6}$~M$_{\odot}$~yr$^{-1}$ \citep{Puls-et-al.-1996},
$v_{\infty}=1700$~km~s$^{-1}$, $\beta=1$, $v_0=10$~km~s$^{-1}$,
and the force multiplier parameters are
$k=0.375$, $\alpha=0.522$, and $\delta=0.099$
\citep{Shimada-et-al.-1994}. The corresponding upper and lower
limits to the clump radius, derived from Equations
(\ref{R_cl_min}) and (\ref{R_cl_max}), are shown in Figure
\ref{relazione_Mcl_Rcl_Mloss025}.

%

%---------------------------------------------
\subsection{The effect of the mass distribution}
\label{The effect of the mass distribution law}
%---------------------------------------------

To study the effects of the clump masses  we computed the
distributions of the flares luminosity and durations for different
values of $\zeta$, $f$, $M_{\rm a}$ and $M_{\rm b}$ considering
for simplicity a circular orbit with  $P_{\rm orb}=10$~d.  We
first neglected the clump radii distribution
assuming that clumps which start from the sonic radius have radii
given by Equation (\ref{R_cl_min}), and follow the expansion law
(Equation \ref{legge_Rcl_r}).
We found that when $M_{\rm a}$ and/or $M_{\rm b}$ increase, the
number of clumps produced by the supergiant decreases (see
Equation \ref{Ndot emitted}), resulting in  a smaller number of
X$-$ray flares. The average flare luminosity, the average flare duration,
the number of flares and the shapes of the luminosity and flare duration
distributions do not change much for different values of
$M_{\rm a}$ and $M_{\rm b}$.
\begin{figure*}
\begin{center}
\begin{tabular}{@{}c@{}c@{}}
\includegraphics[width=9cm]{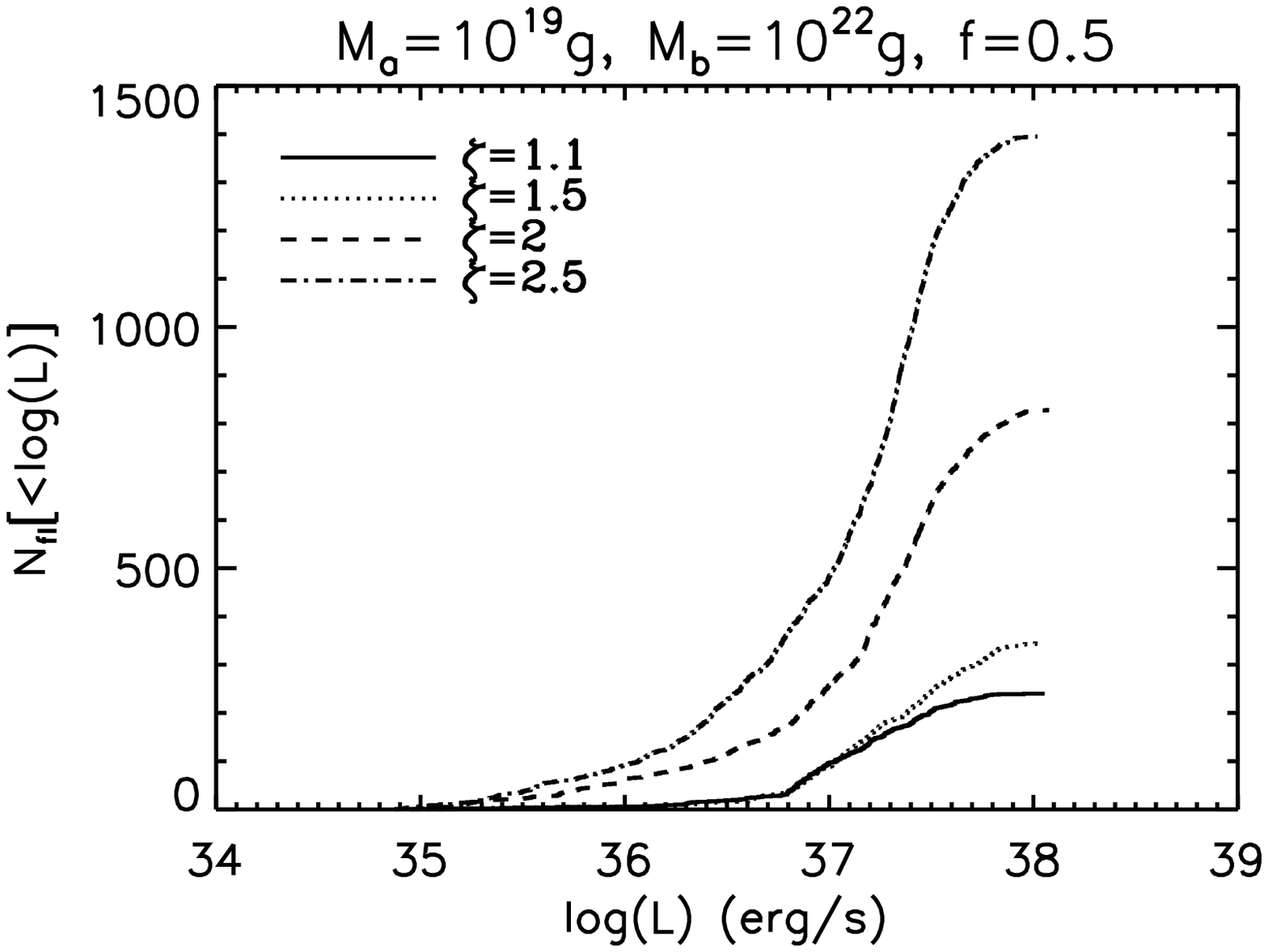} &
\includegraphics[width=9cm]{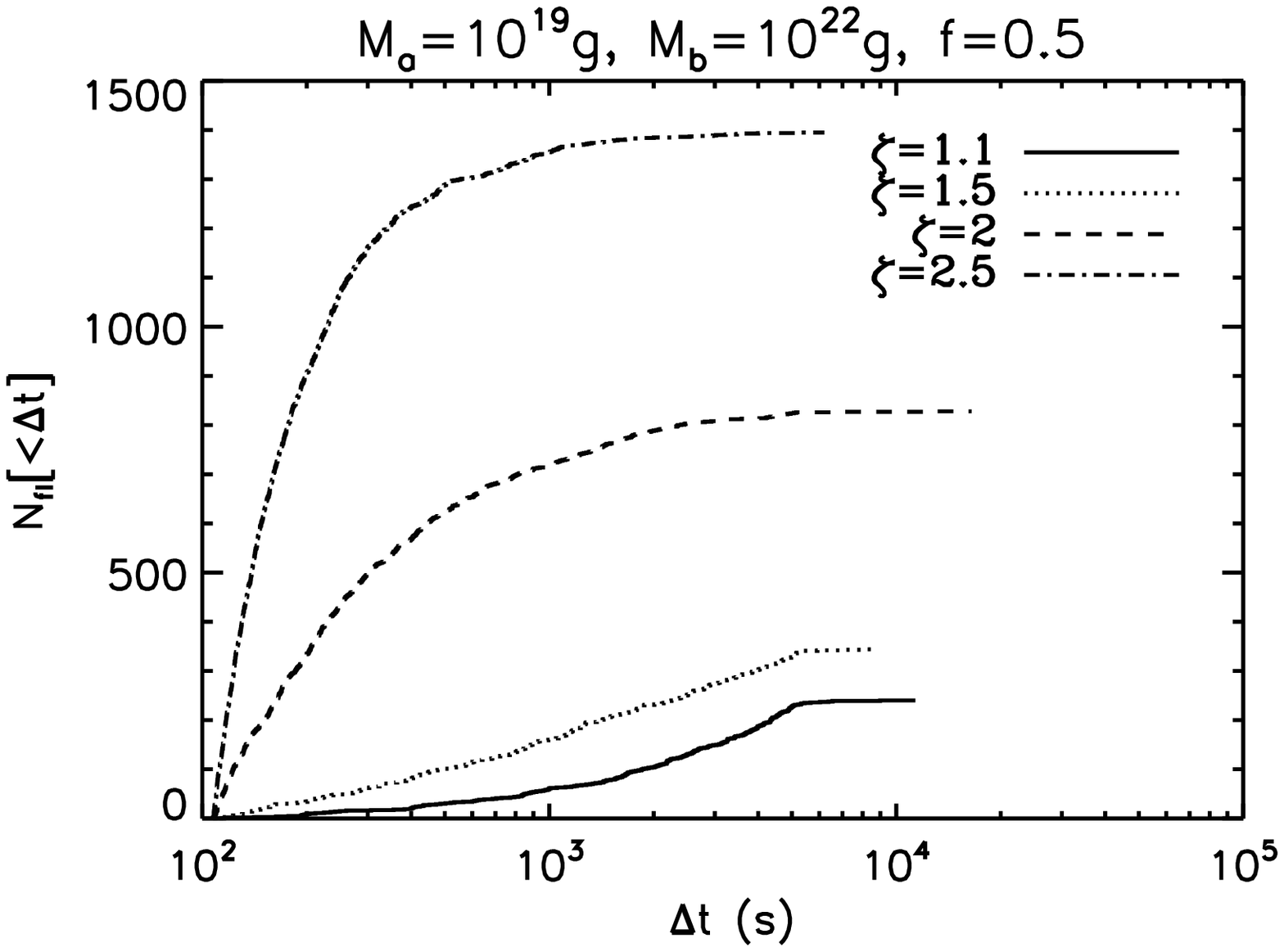} \\
\includegraphics[width=9cm]{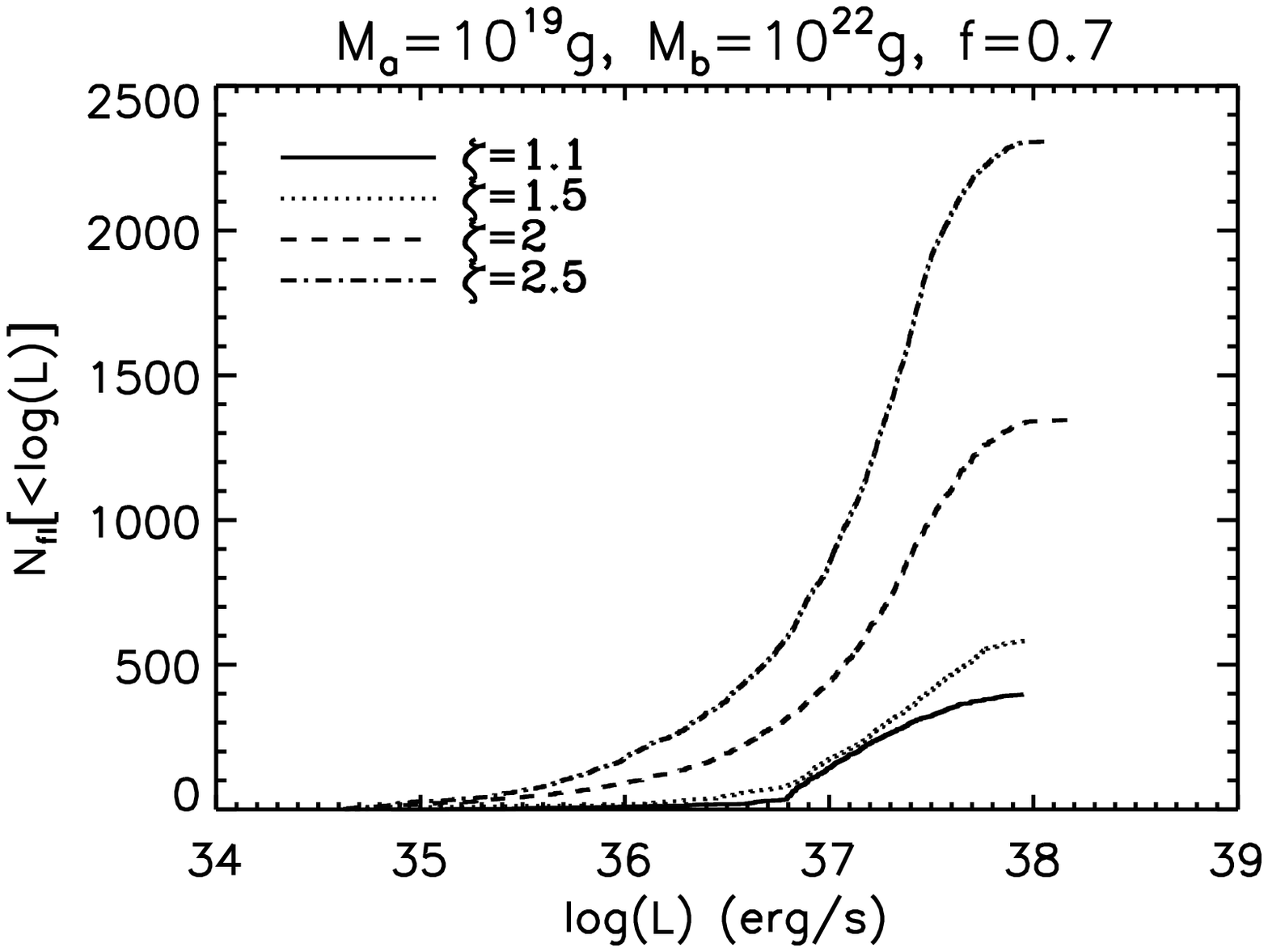} &
\includegraphics[width=9cm]{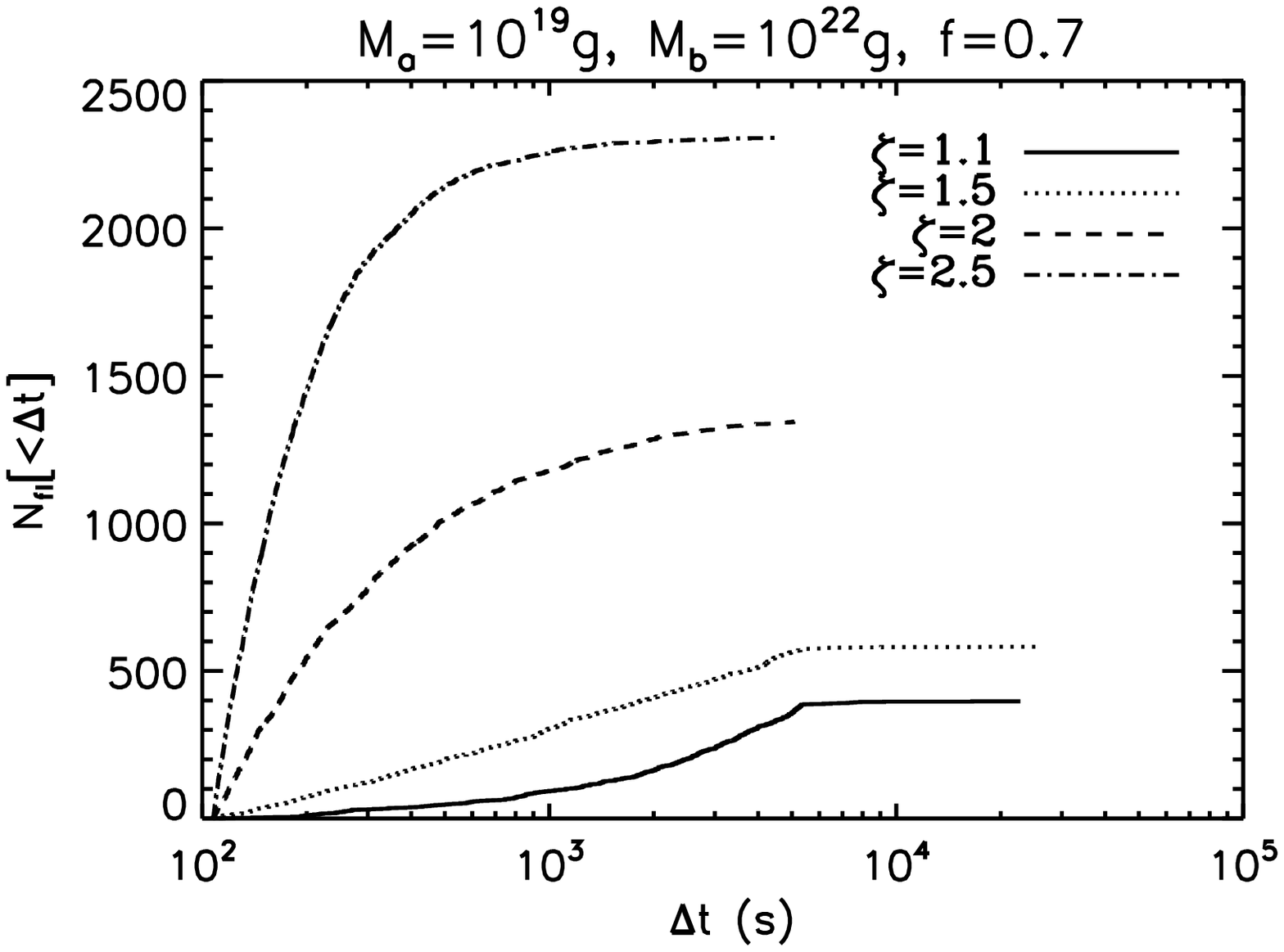} \\
\includegraphics[width=9cm]{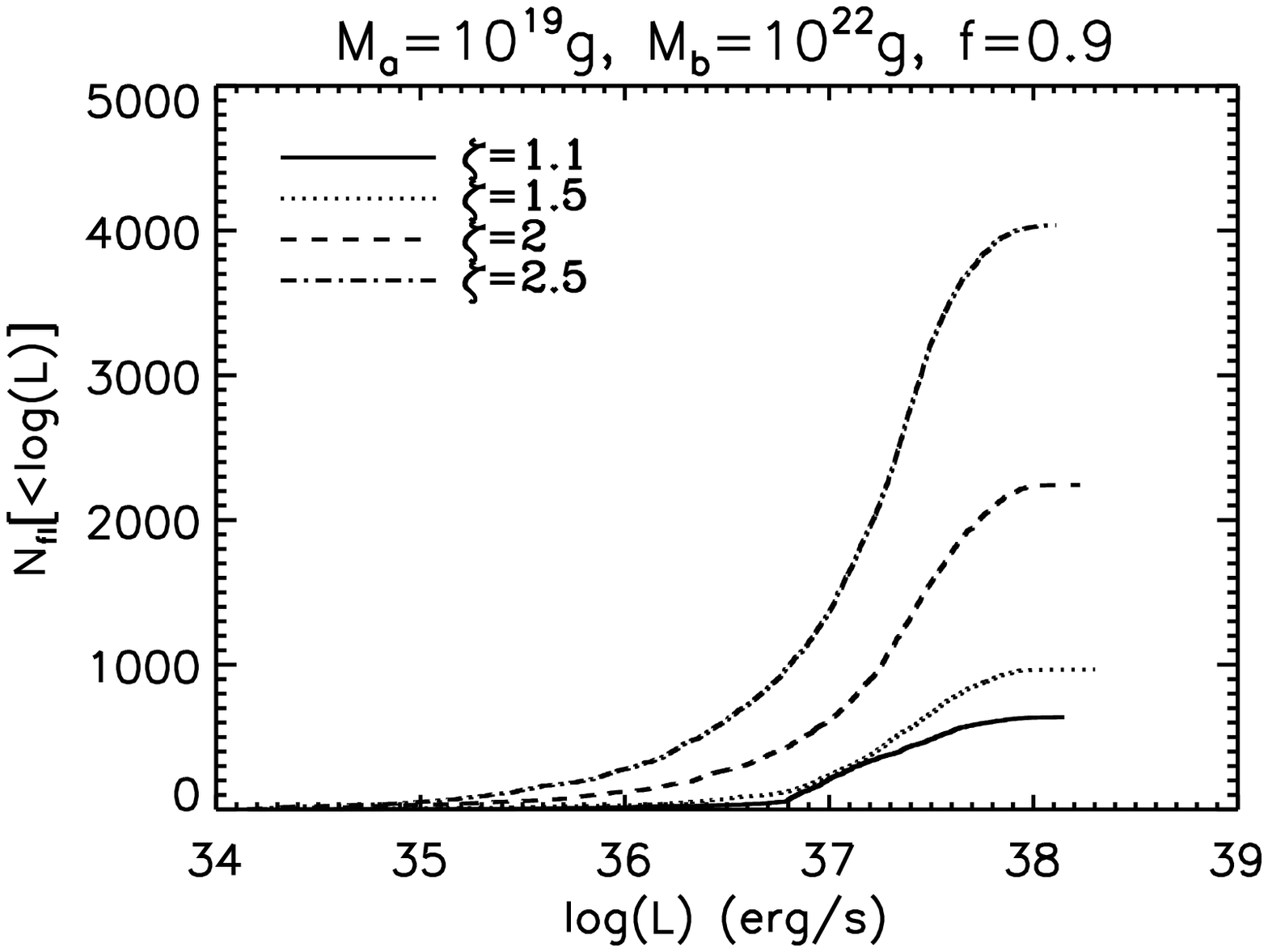} &
\includegraphics[width=9cm]{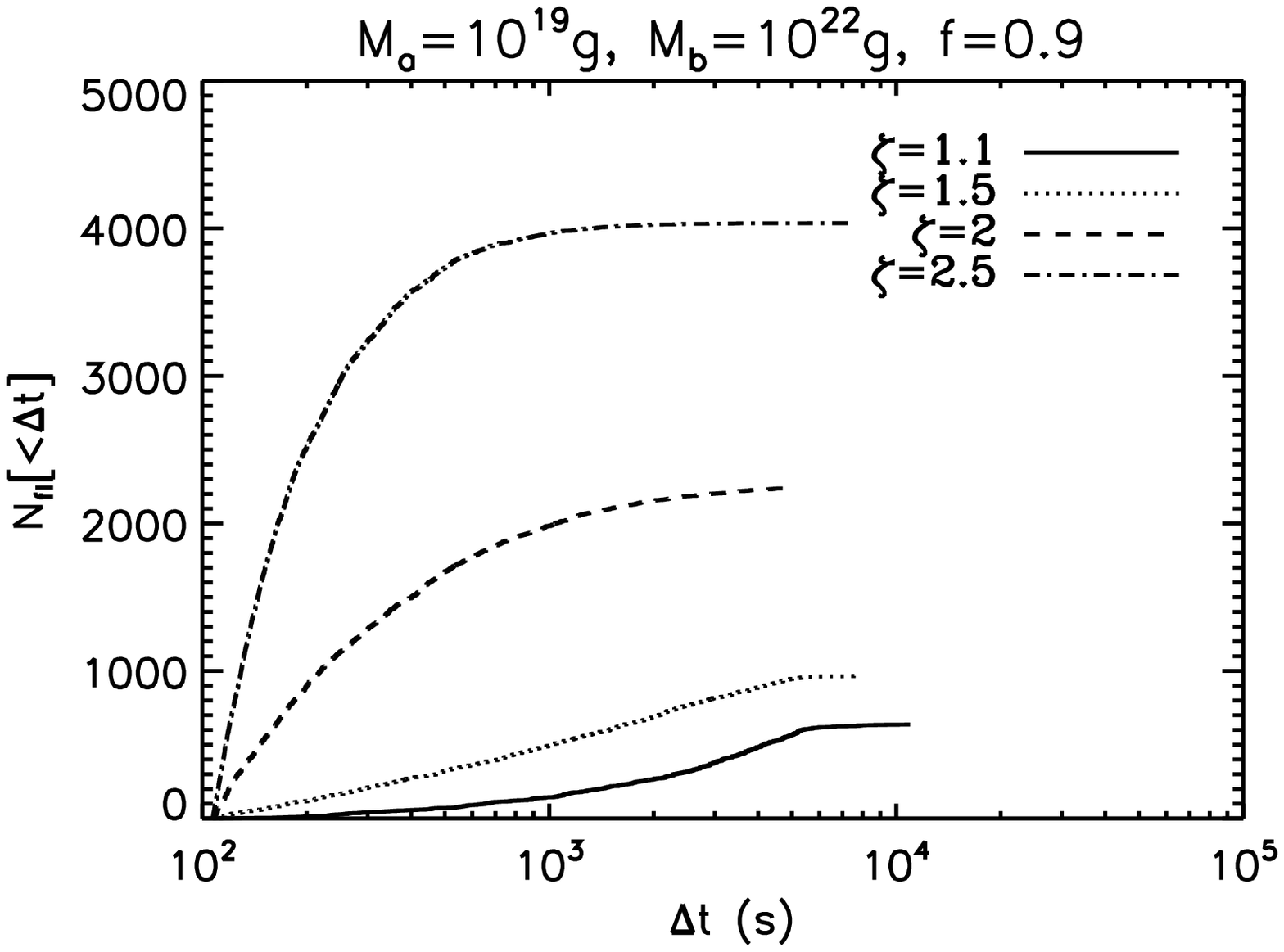}
\end{tabular}
\end{center}
\caption{Expected integral distributions of the flare luminosities
(left panels)      and  durations (right panels) for
         different values of $\zeta$ and $f=\dot{M}_{\rm cl} / \dot{M}_{\rm tot}$.
         The binary system parameters are:
         $M_{\rm OB}=30$~M$_{\odot}$, $R_{\rm OB}=23.8$~R$_{\odot}$,
         $M_{\rm NS}=1.4$~M$_{\odot}$, $R_{\rm NS}=10$~km, $P_{\rm orb}=10$~d, $e=0$.
         The parameters for the supergiant wind are:
         $\dot{M}_{\rm tot}= 10^{-6}$~M$_{\odot}$~yr$^{-1}$,
         $v_{\infty}=1700$~km~s$^{-1}$, $\beta=1$, $v_0=10$~km~s$^{-1}$,
         $M_{\rm a} = 10^{19}$~g and $M_{\rm b}= 10^{22}$~g
         and $f=0.5$, $0.7$, $0.9$.
         The time interval for each histogram
         corresponds to 100 days.}
\label{figure zeta}
\end{figure*}
In Figure \ref{figure zeta} we show the dependence of the
distributions  of flare luminosities and durations on $\zeta$ and $f$
(for $M_{\rm a}=10^{19}$~g, $M_{\rm b}=10^{22}$~g). If $\zeta$
increases, the number of clumps and their density increases (see
Equations \ref{Ndot emitted} and \ref{R_cl_min}), implying a higher number of flares, a
shift to higher luminosities and to shorter flare durations.

\begin{figure*}
\begin{center}
\begin{tabular}{@{}c@{}c@{}}
\includegraphics[width=9cm]{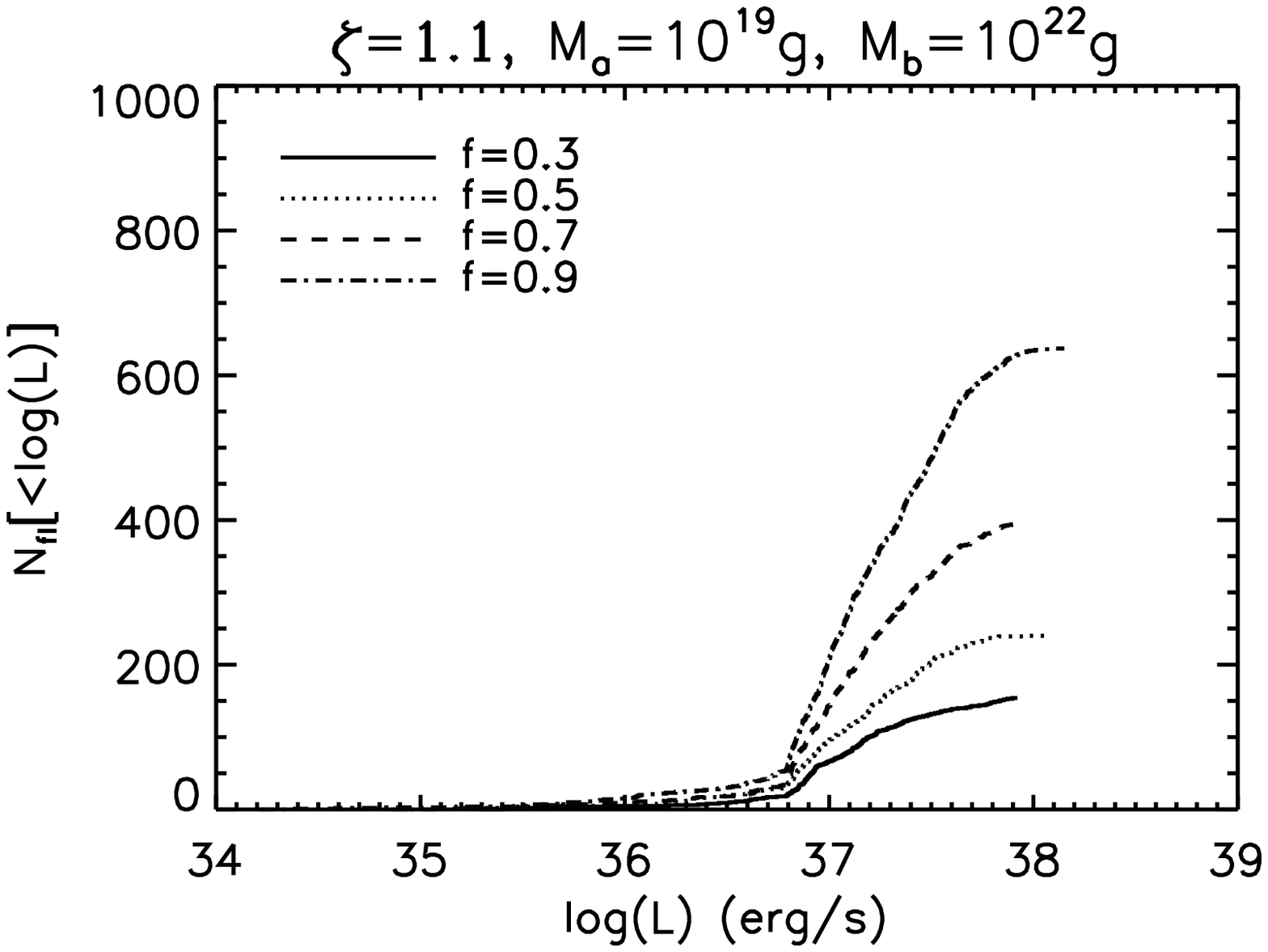} &
\includegraphics[width=9cm]{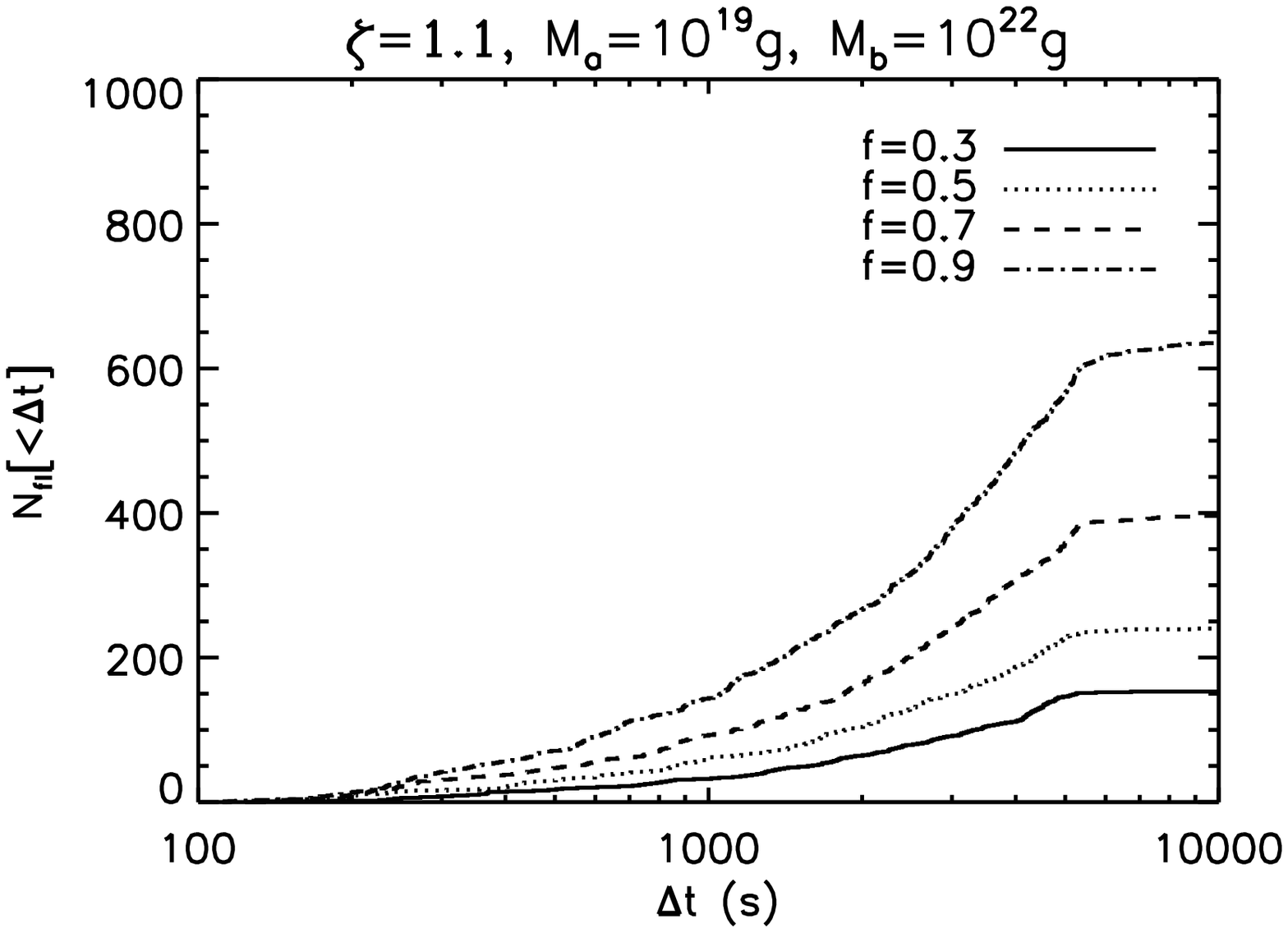} \\
\includegraphics[width=9cm]{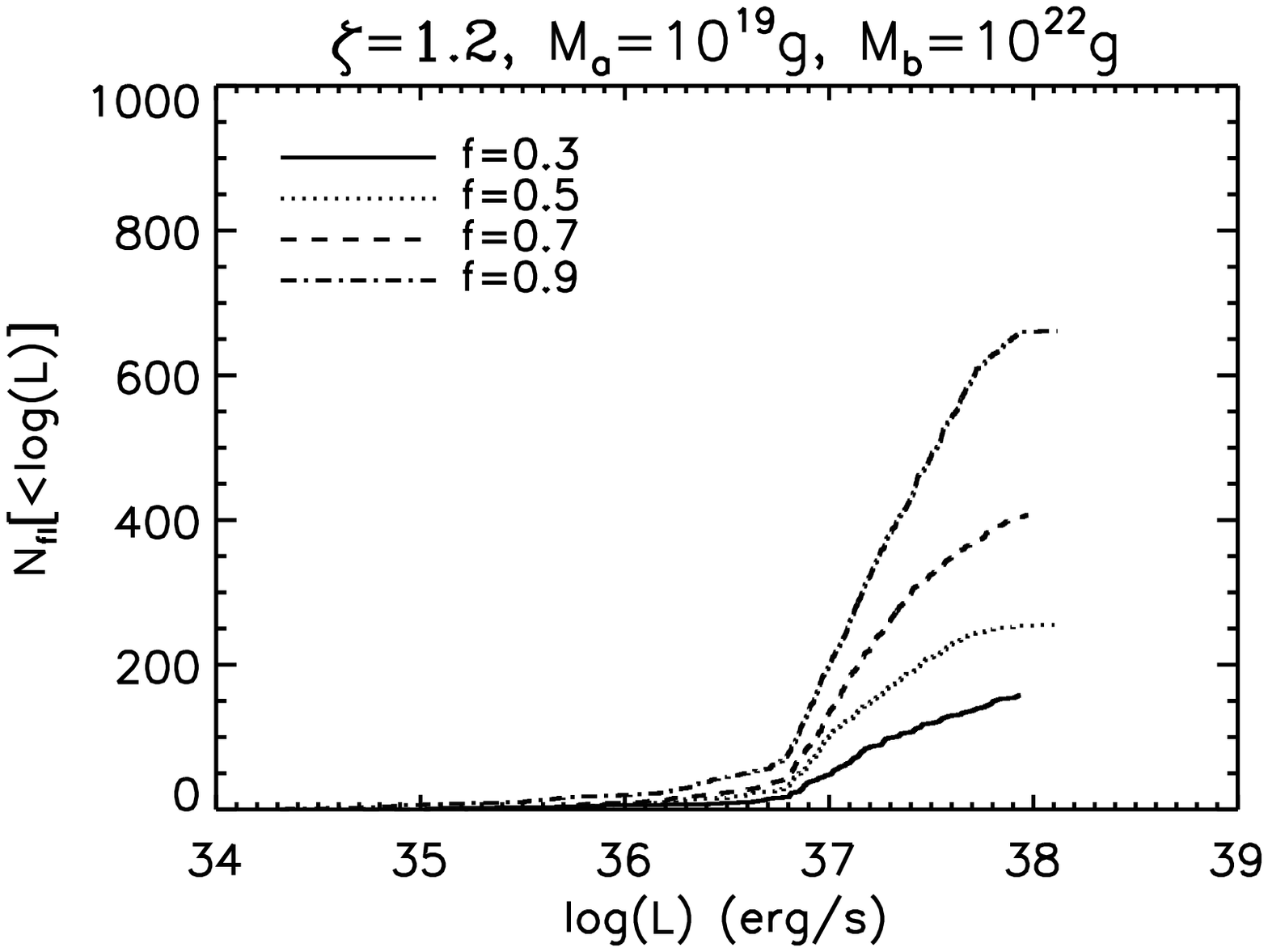} &
\includegraphics[width=9cm]{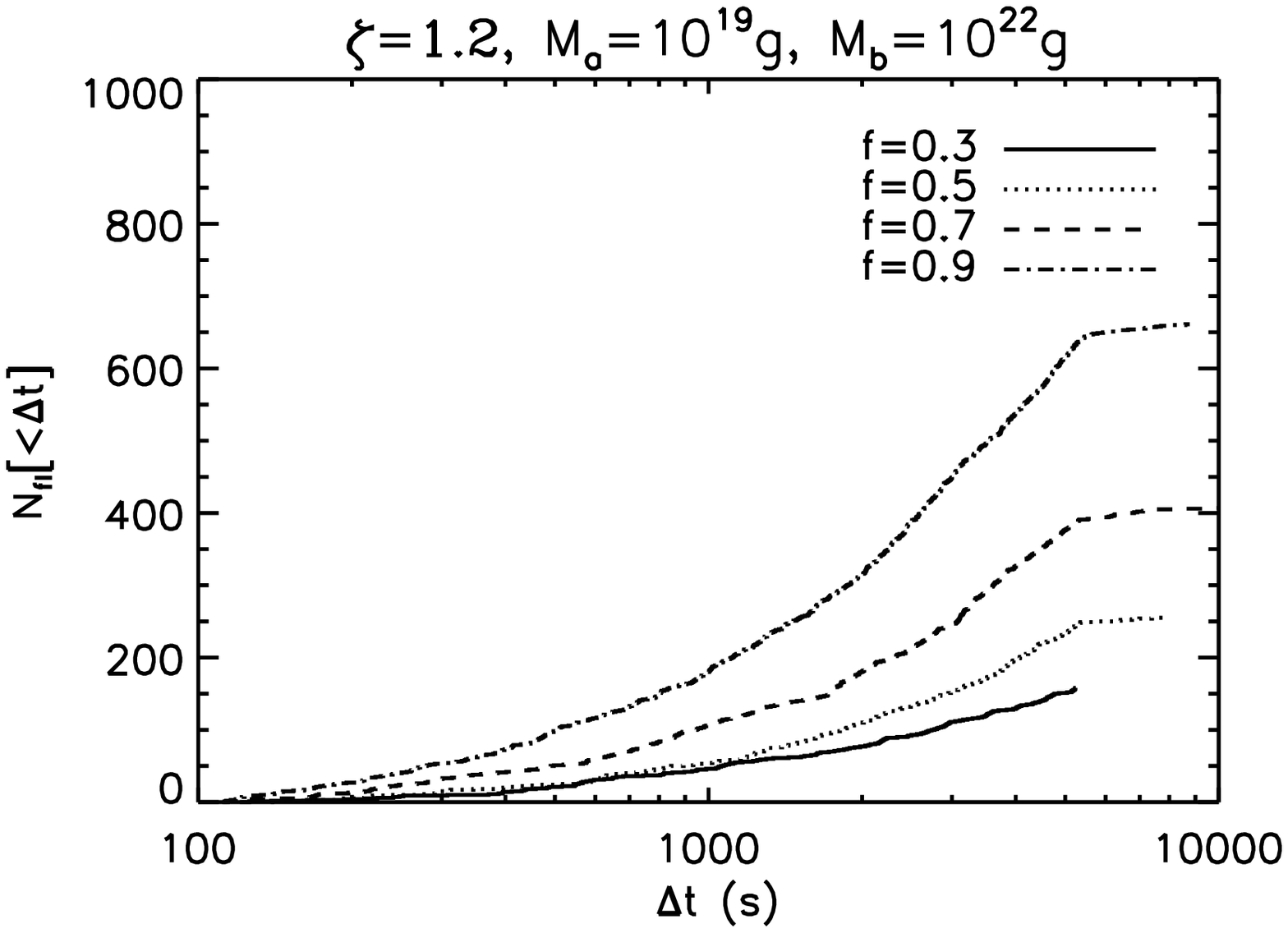} \\
\includegraphics[width=9cm]{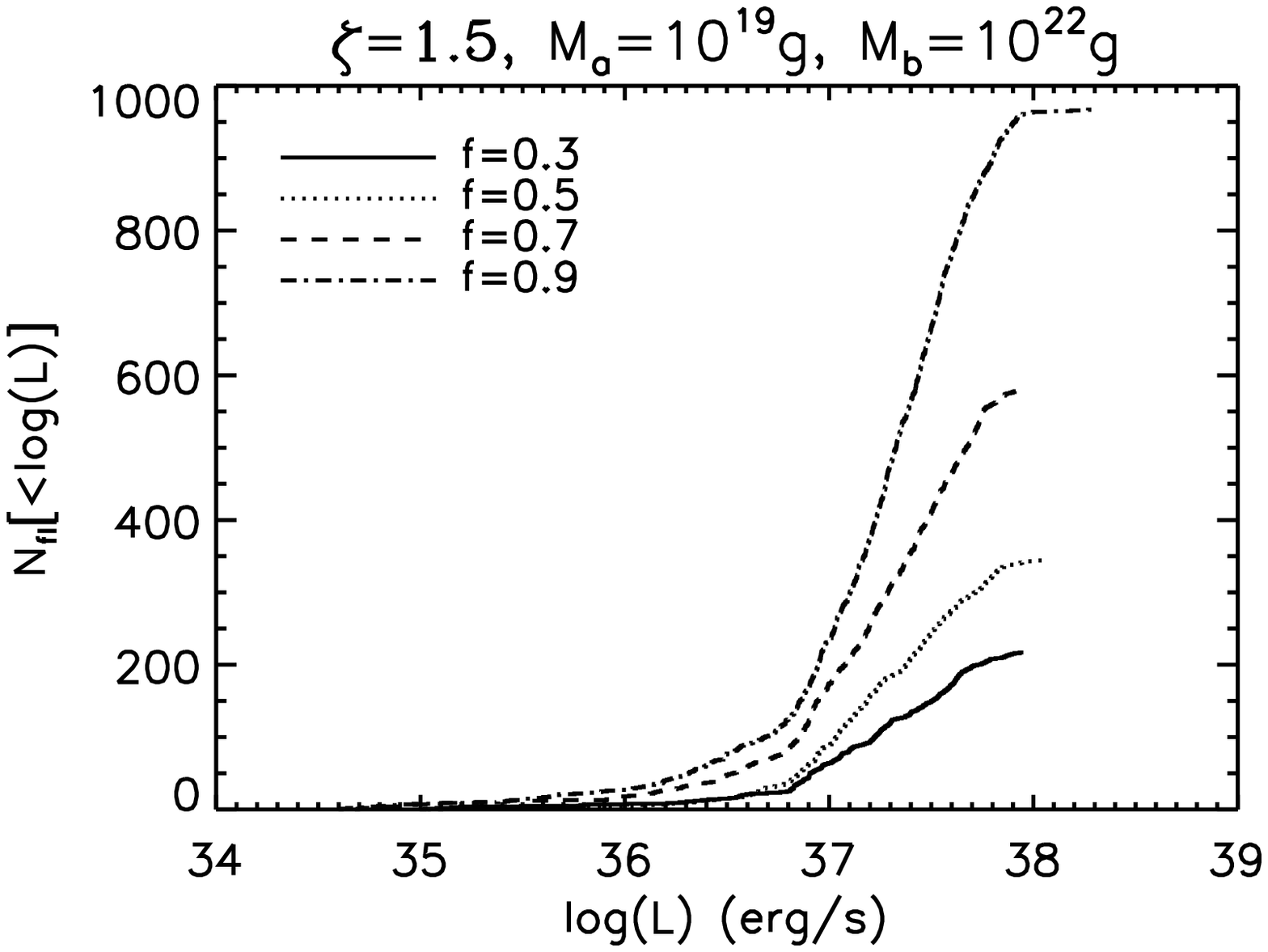} &
\includegraphics[width=9cm]{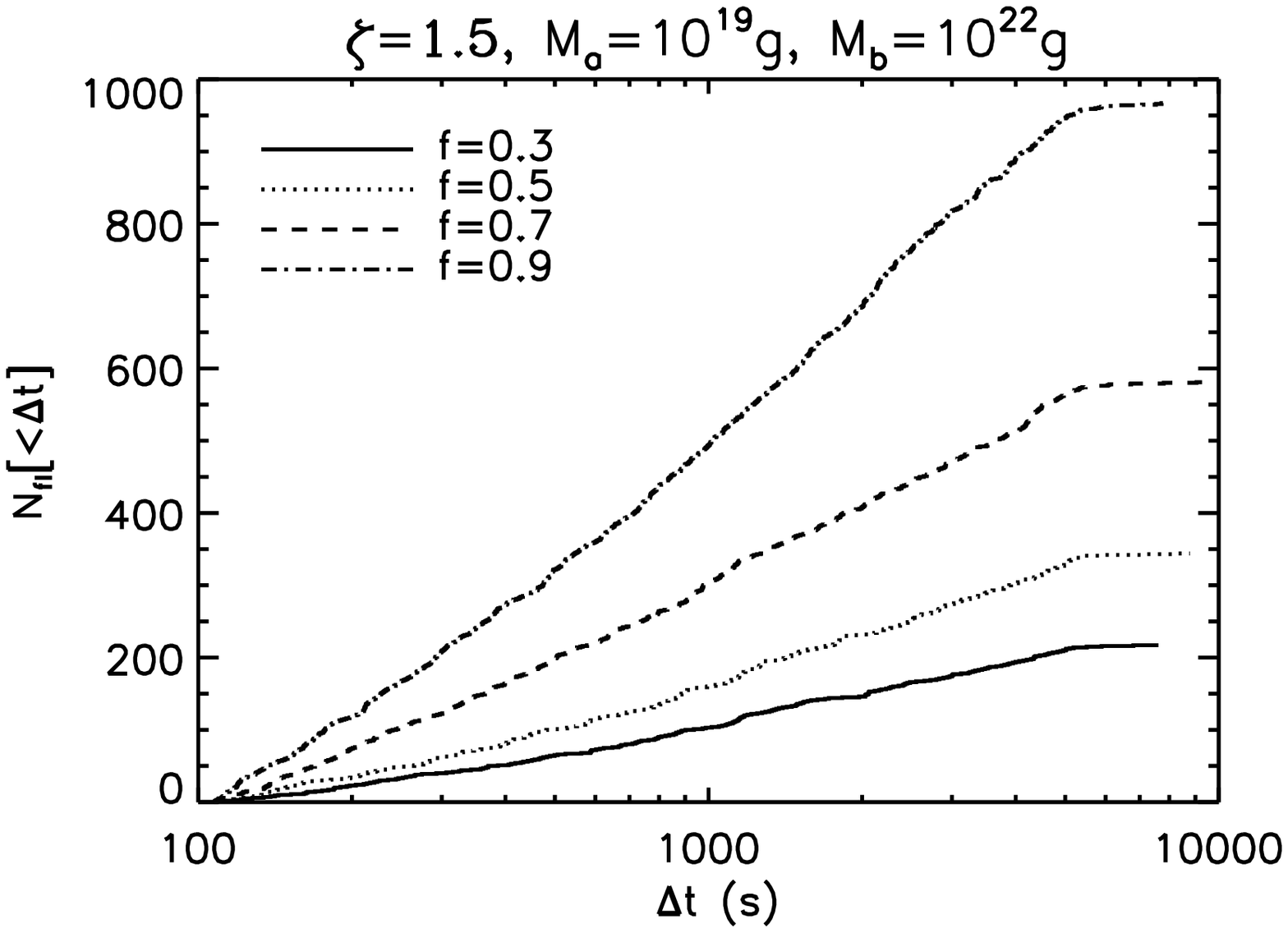}
\end{tabular}
\end{center}
\caption{Expected integral distributions of the flare luminosities
(left panels)  and  durations (right panels) for different values
         of $f=\dot{M}_{\rm cl}/\dot{M}_{\rm wind}$ and $\zeta$.
The binary system parameters are in the caption of Figure \ref{figure zeta}.
         The time interval corresponds to 100 days.}
\label{figure f}
\end{figure*}

Figure \ref{figure f} shows the effect of changing the fraction of
wind mass in the form of clumps $f$
for different values of $\zeta$. When $f$ increases, the
supergiant produces more clumps (see Equations \ref{Ndot emitted}), 
thus the number of flares and their average luminosity
increase, while their average  duration remains unchanged.
%
%

%---------------------------------------------
\subsection{The effect of the radii distribution}
\label{The effect of the radii distribution law}
%---------------------------------------------

In Section \ref{Section Clumpy stellar winds properties} we showed
that, for any given mass, the clump dimensions are constrained
within the limits  given by Equations (\ref{R_cl_min}) and
(\ref{R_cl_max}). Here we show the effect of different assumptions
on the radii distributions laws. We considered both  a power law
(Equation \ref{distrib R_cl}) and  a truncated normal distribution 
(Equation \ref{distrib R_cl gauss}), described by  the parameters $\gamma$
(or $N_{\sigma}$).
\begin{figure*}
\begin{center}
\begin{tabular}{@{}c@{}c@{}}
\includegraphics[width=9cm]{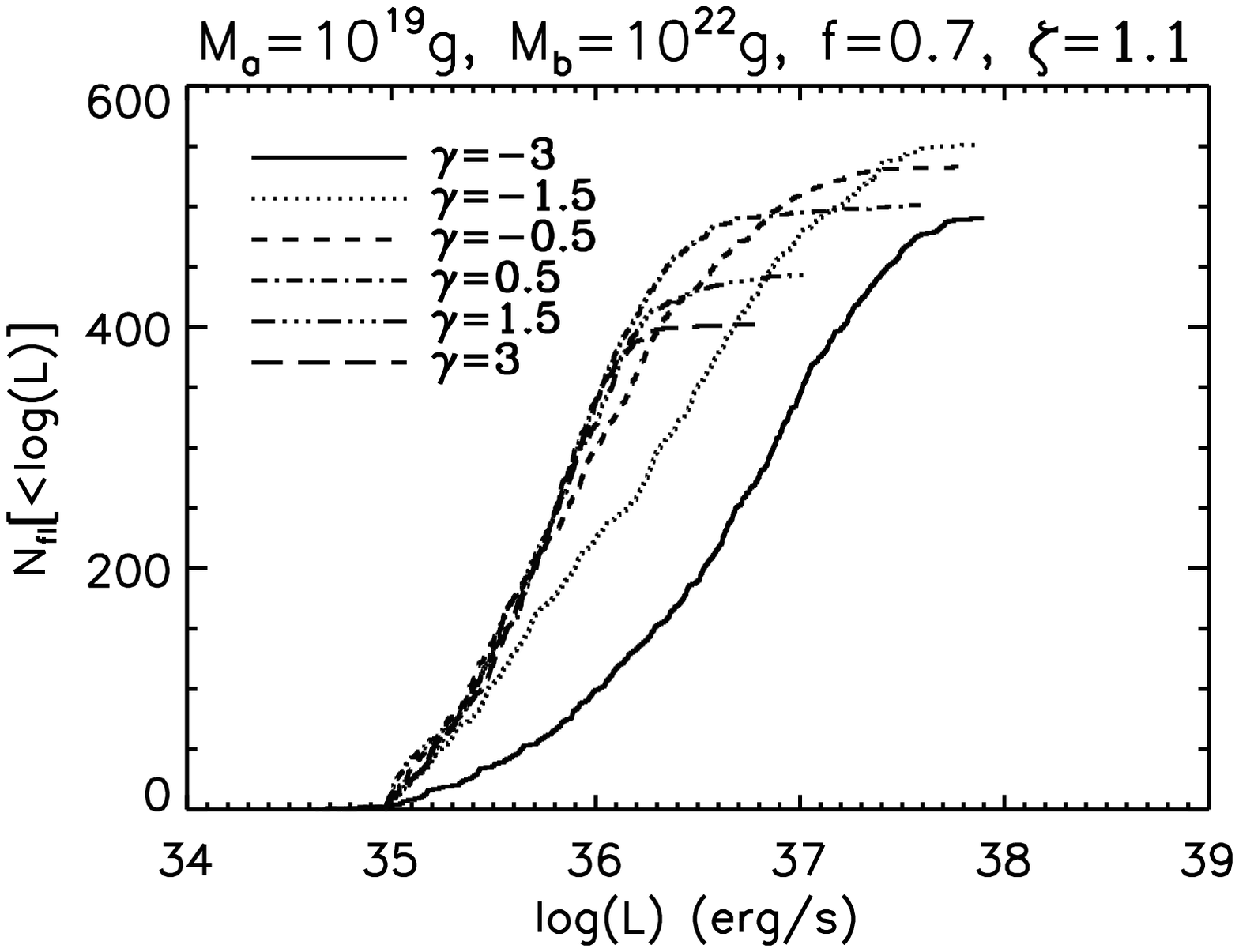} &
\includegraphics[width=9cm]{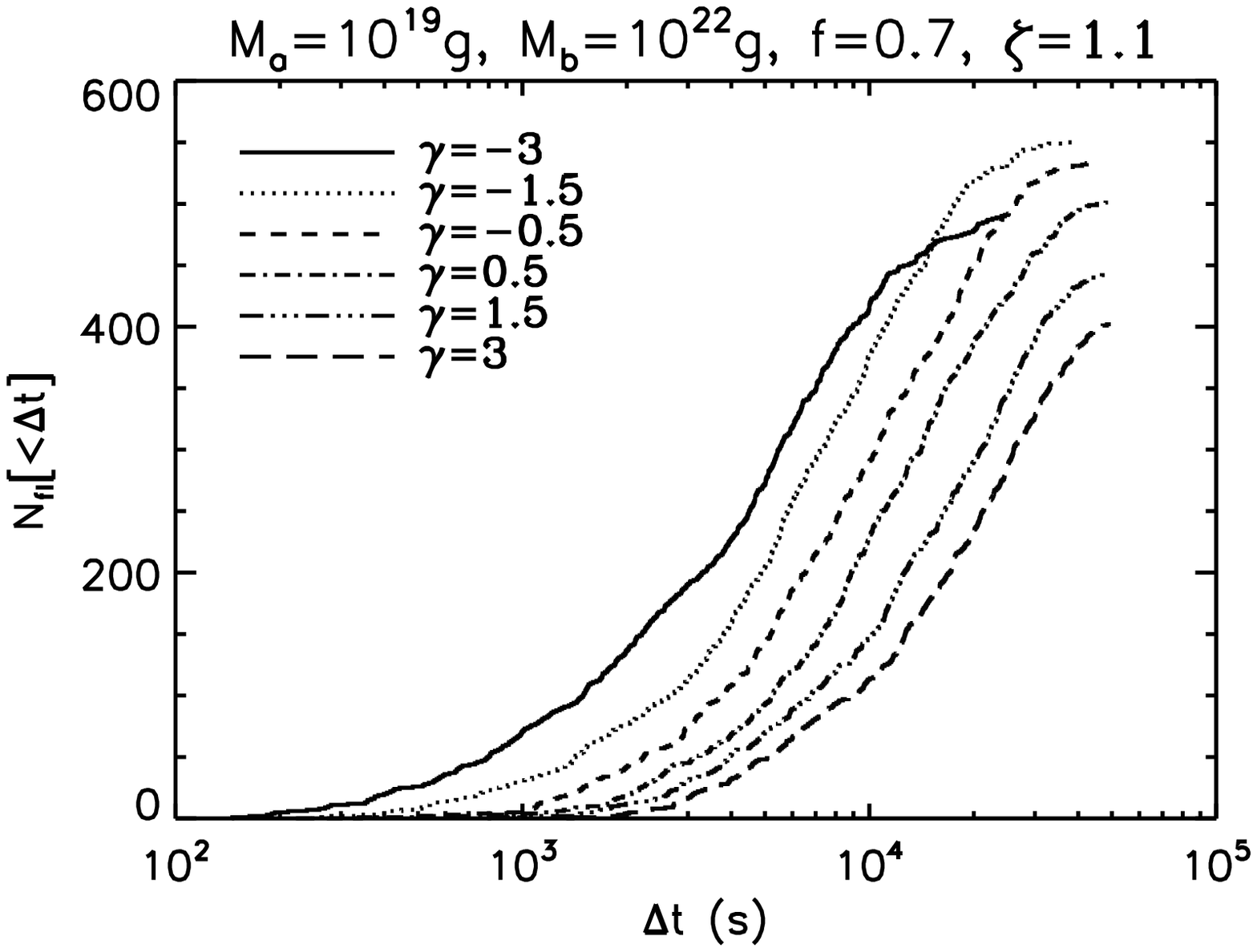} \\
\includegraphics[width=9cm]{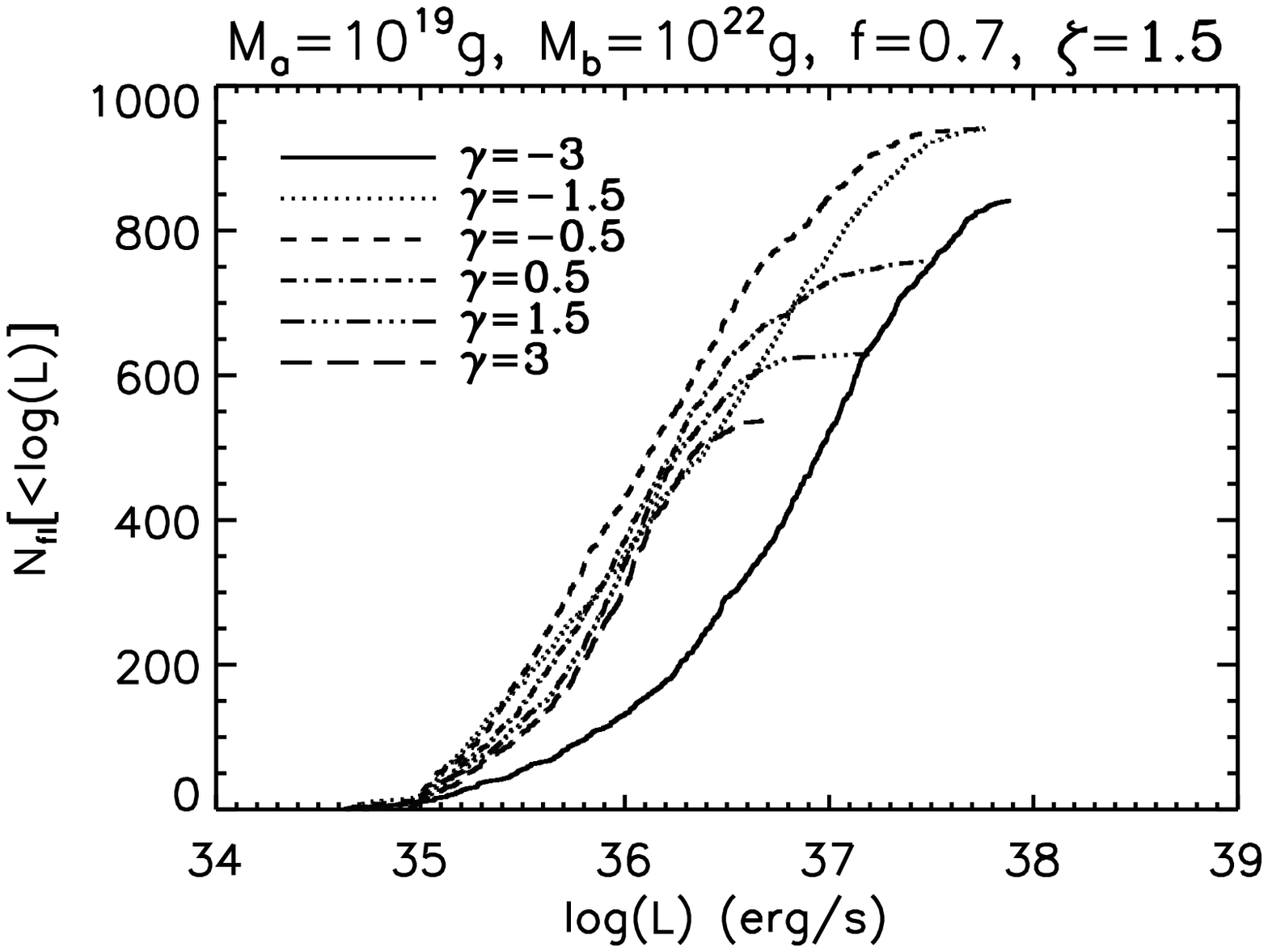} &
\includegraphics[width=9cm]{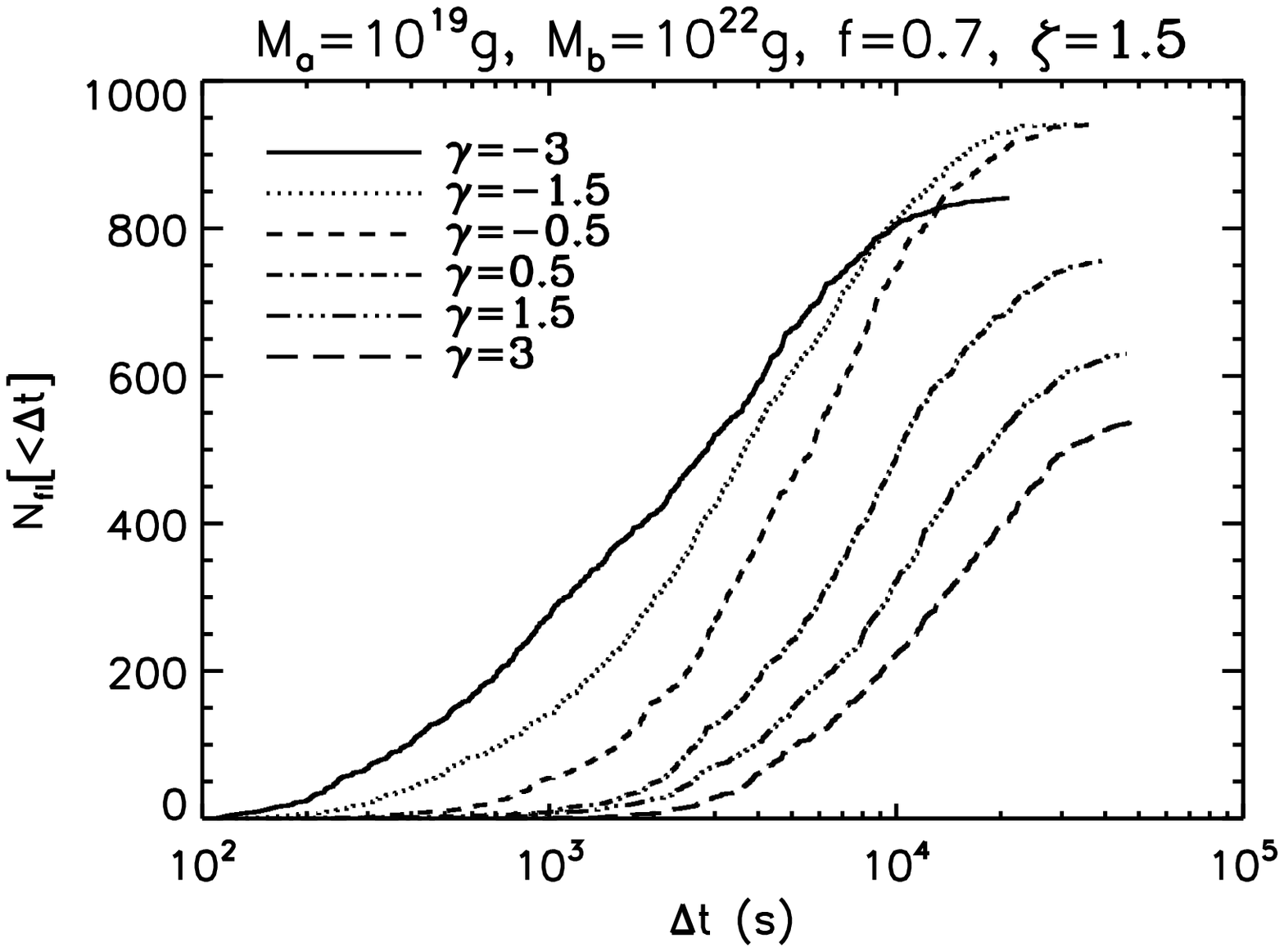} \\
\includegraphics[width=9cm]{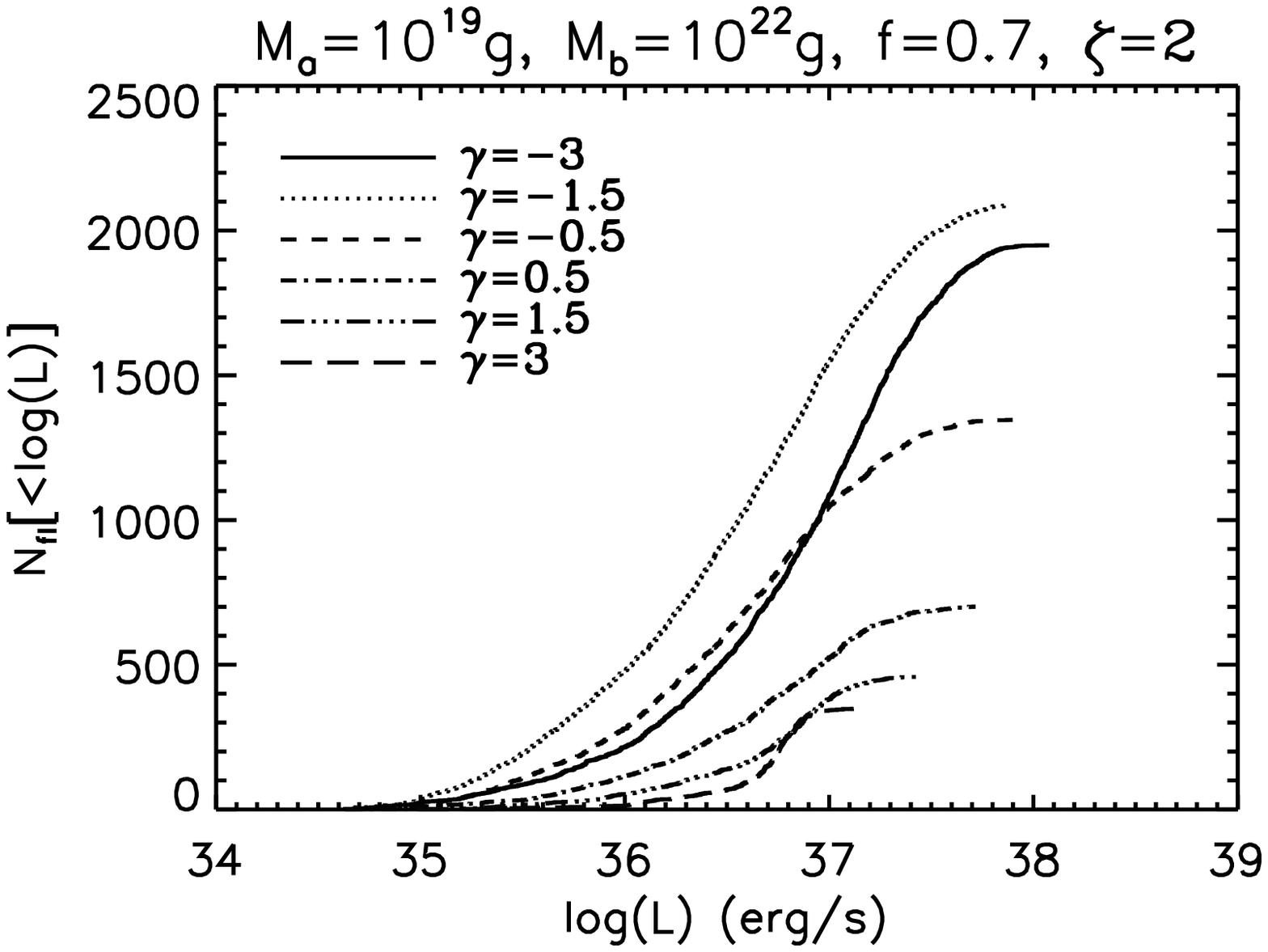} &
\includegraphics[width=9cm]{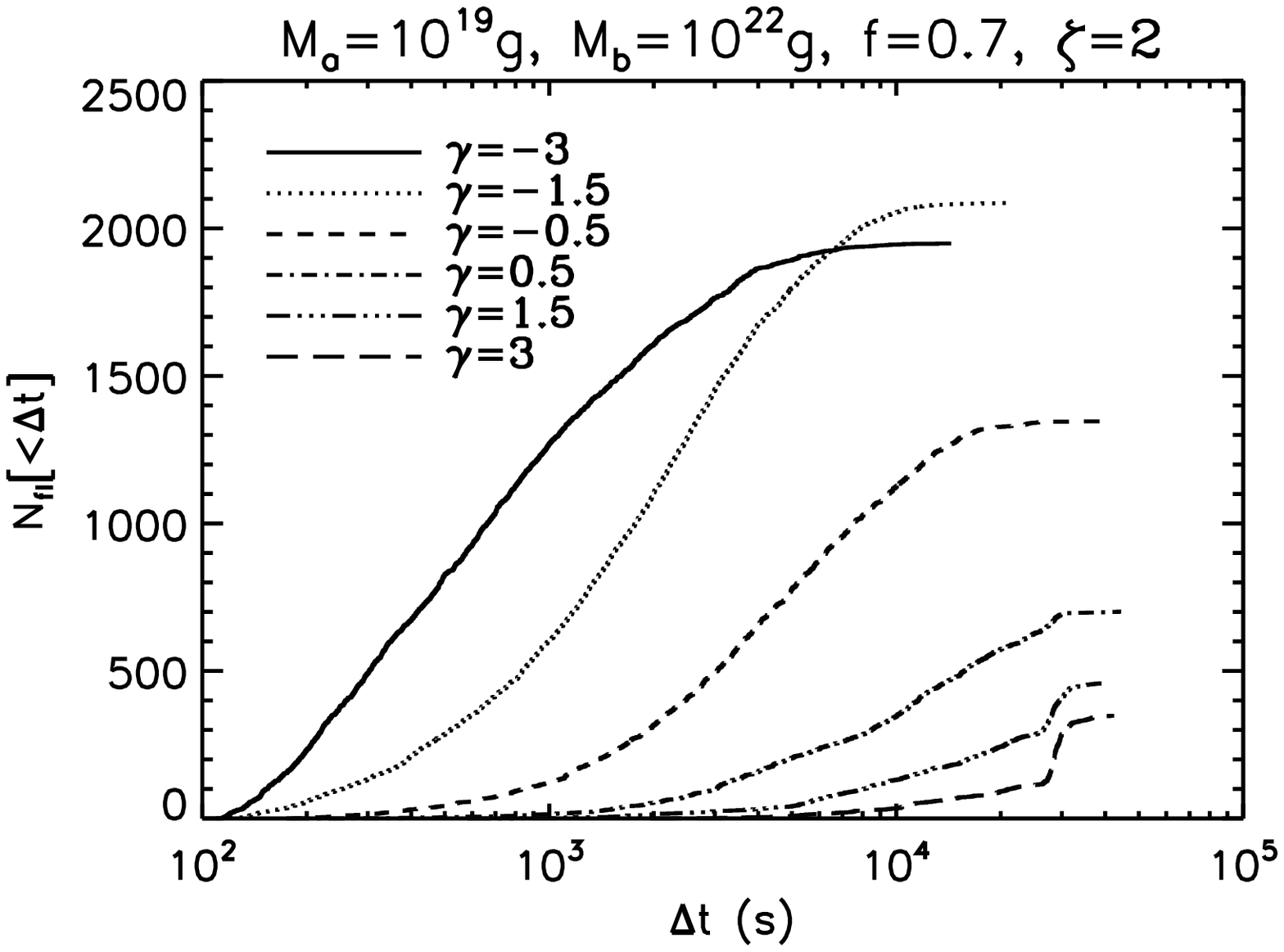}
\end{tabular}
\end{center}
\caption{Expected integral distributions of the flare luminosities
(left panels)  and  durations (right panels) for
         different values of $\gamma$ and $\zeta$ of the Equations (\ref{distrib R_cl})
         and (\ref{Npunto}).
The binary system parameters are in the caption of Figure \ref{figure zeta}.
         The time interval for each histogram
         corresponds to 100 days.}
\label{figure gamma}
\end{figure*}
When $\gamma$ increases, the number of clumps with larger density decreases,
thus the average  luminosity of the flares decreases and their
average   duration increases (see Figure \ref{figure gamma}). We
also found that when $\zeta$ increases, the flare distributions
with a positive $\gamma$ follow a different behaviour than those
with a negative value:  for $\gamma<0$, the flare distributions
behave as described above (i.e. the number of flares increases
with $\zeta$), while this does not happen for  $\gamma>0$. This is
due to the fact that in this case the clumps are larger, thus
there is a high  probability that two or more clumps overlap
thus  reducing the number of flares.

\begin{figure*}
\begin{center}
\begin{tabular}{@{}c@{}c@{}}
\includegraphics[width=9cm]{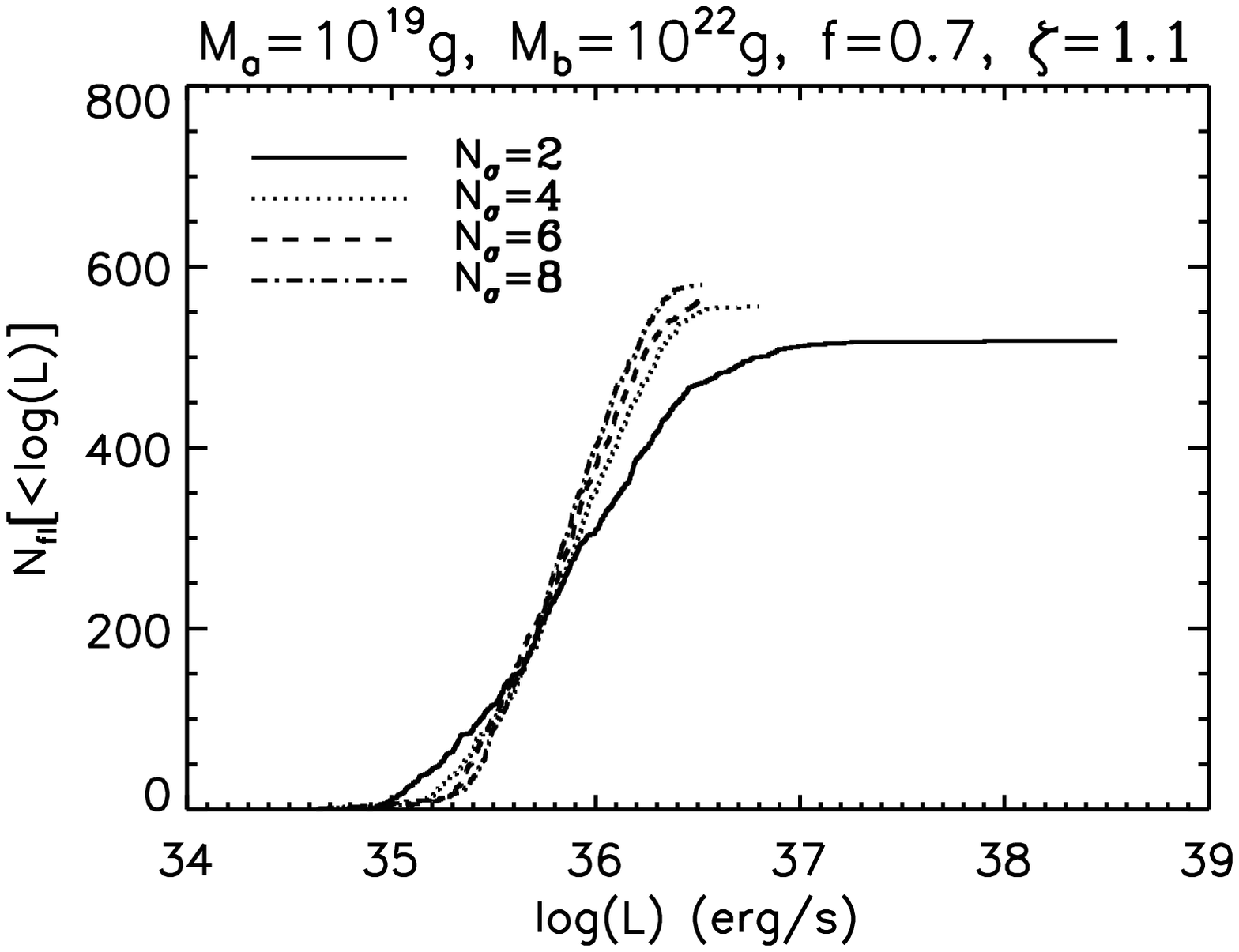} &
\includegraphics[width=9cm]{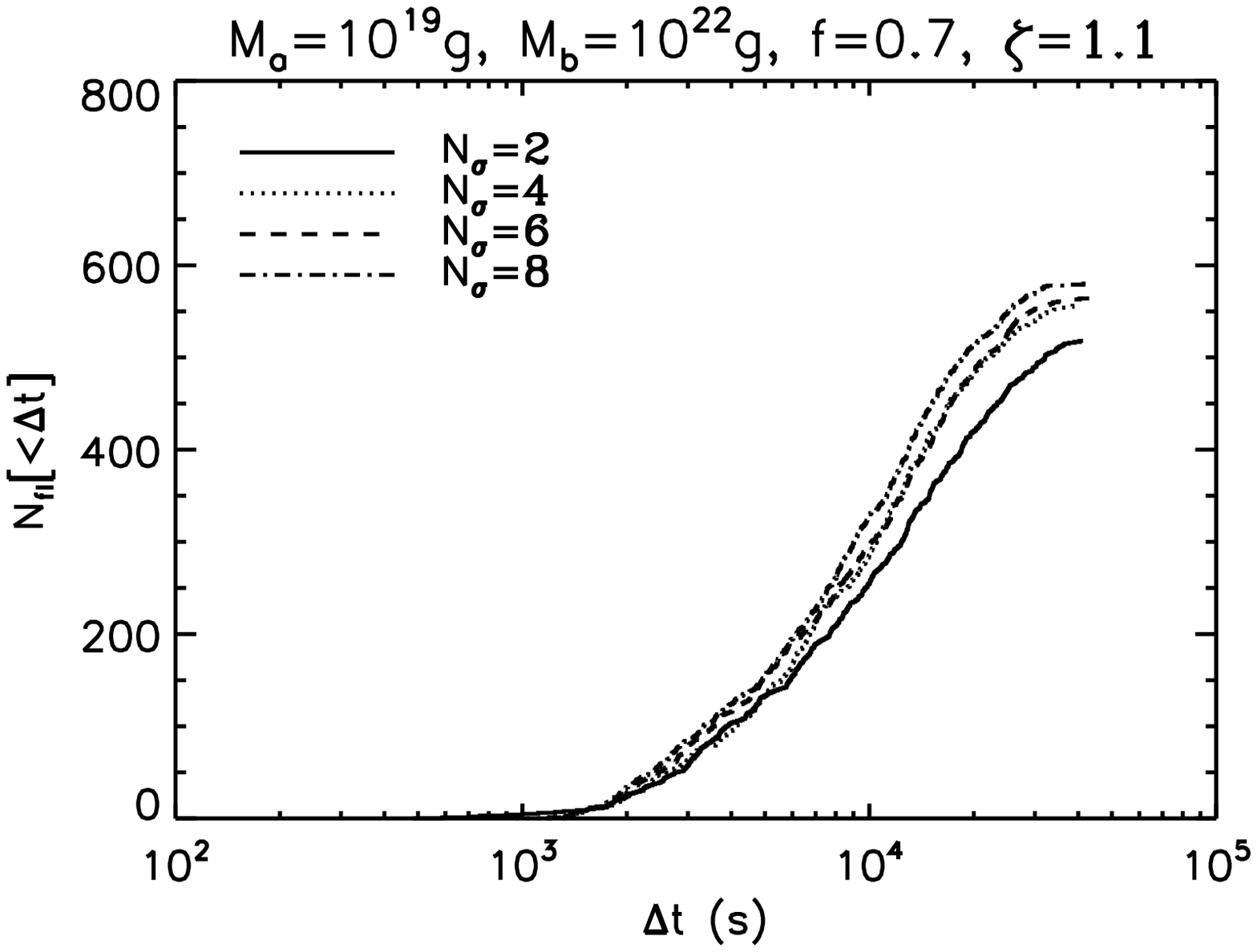} \\
\includegraphics[width=9cm]{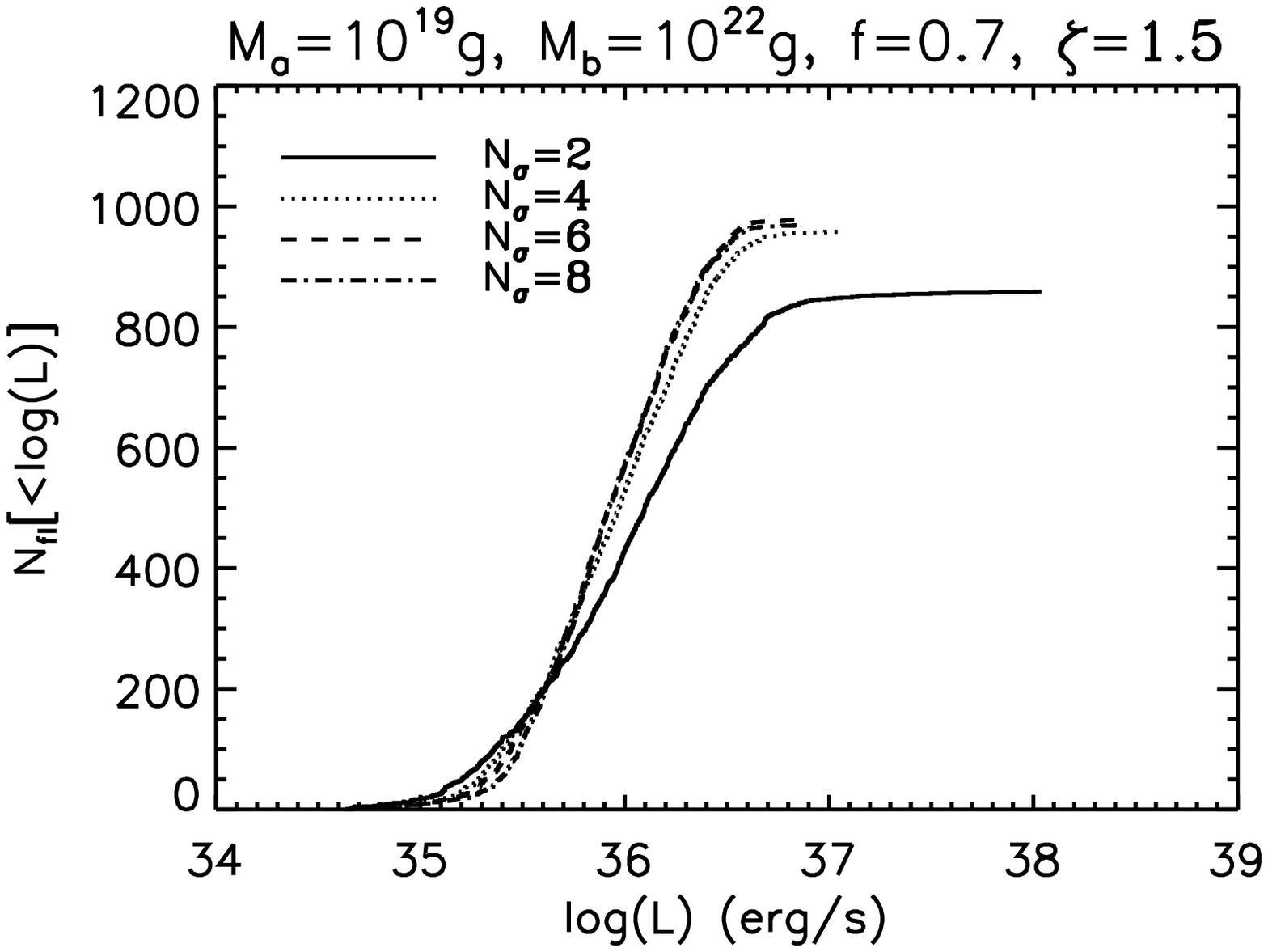} &
\includegraphics[width=9cm]{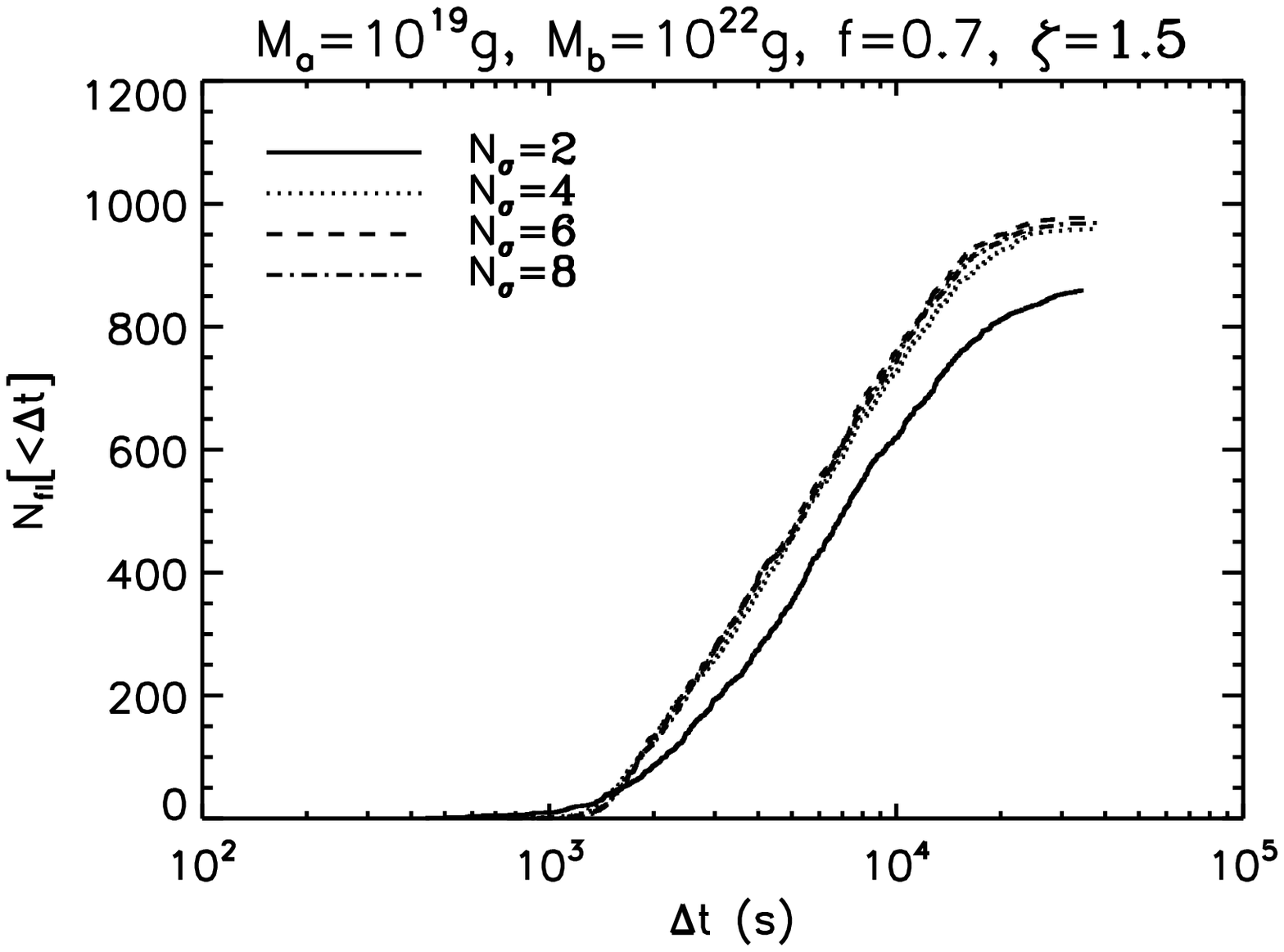} \\
\includegraphics[width=9cm]{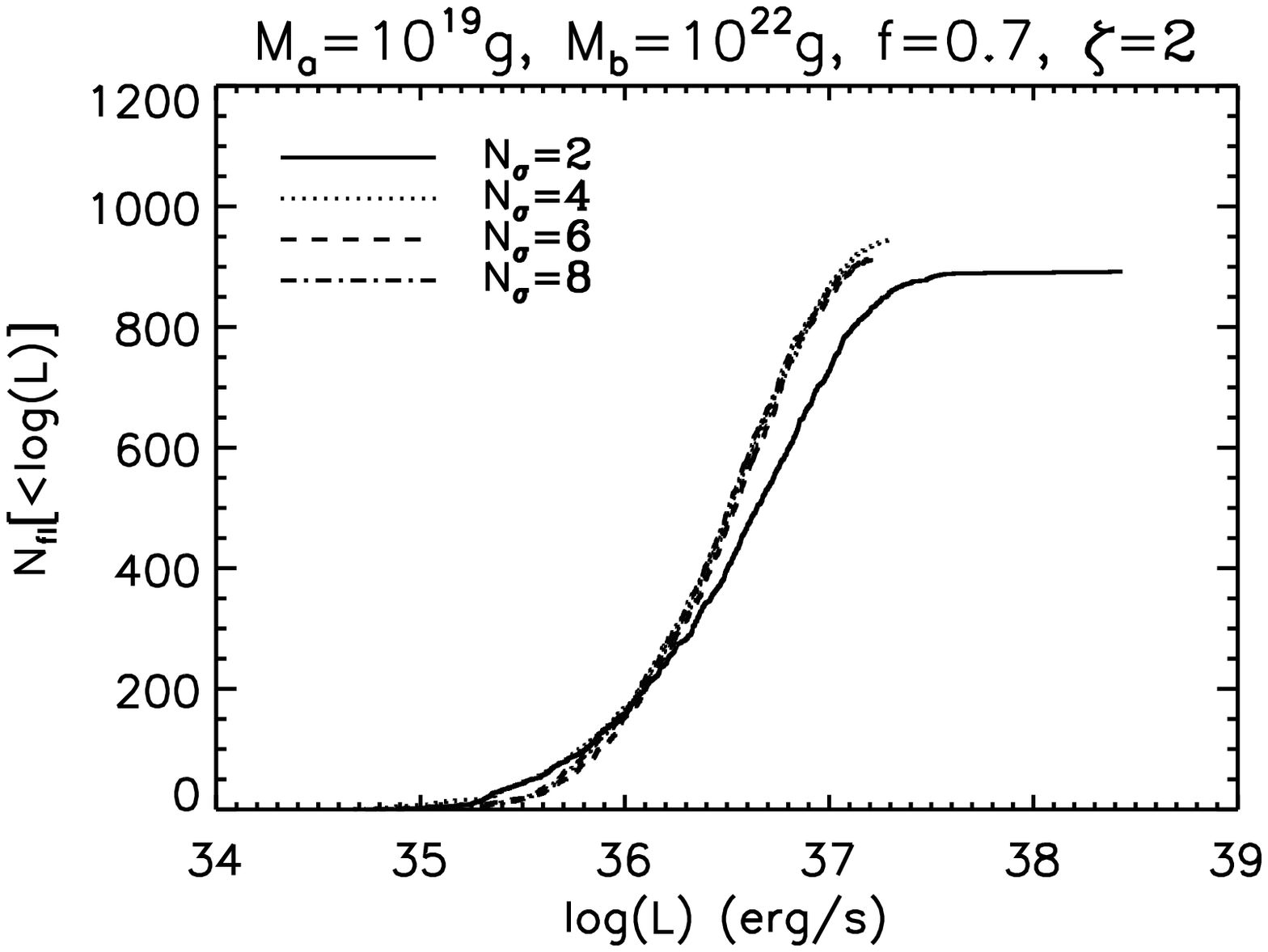} &
\includegraphics[width=9cm]{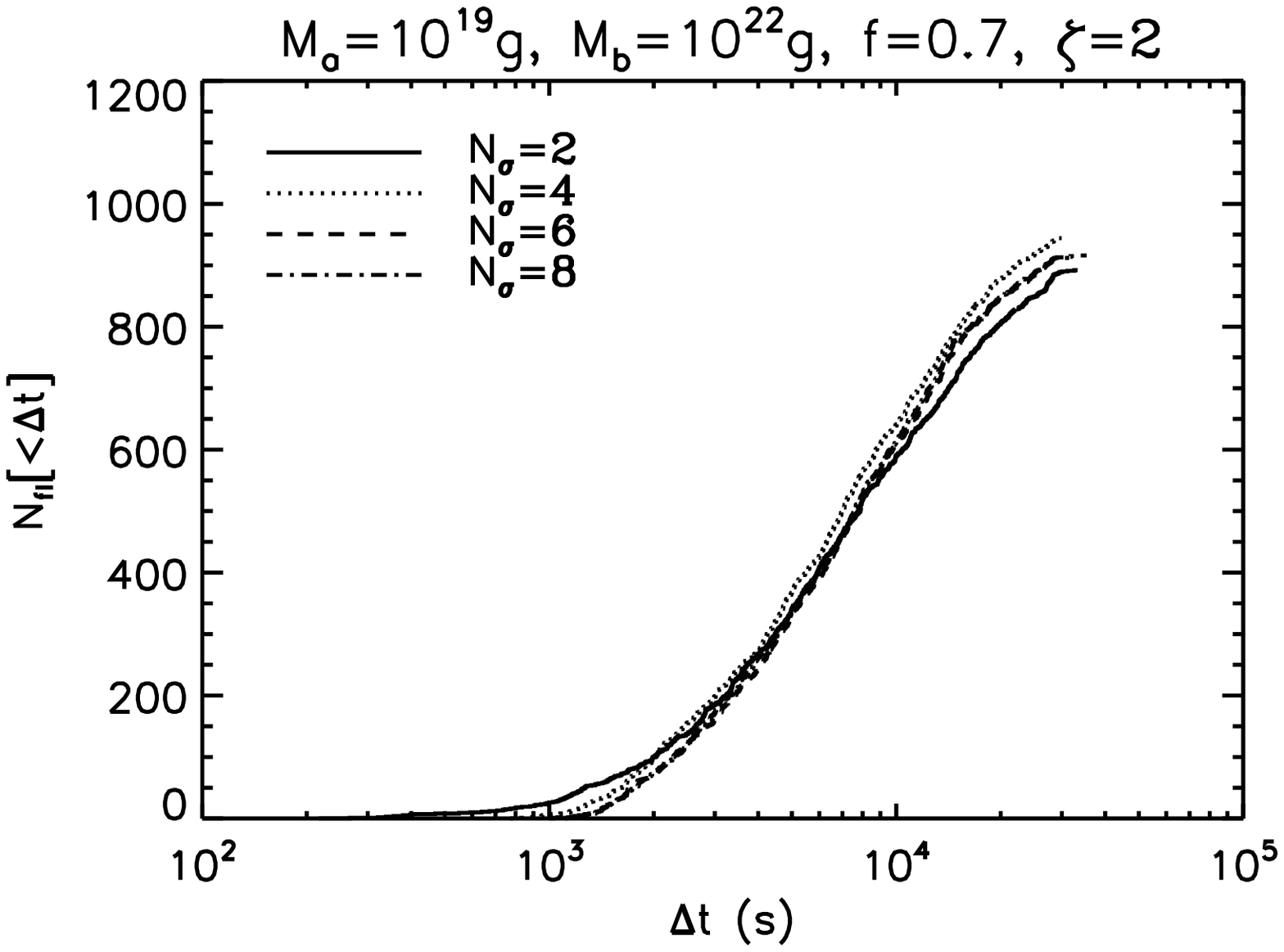}
\end{tabular}
\end{center}
\caption{Expected integral distributions of the flare luminosities
(left panels)  and durations (right panels) for
         different values of $N_{\sigma}$ and $\zeta$ of the Equations (\ref{distrib R_cl gauss})
         and (\ref{Npunto}).
The binary system parameters are in the caption of Figure \ref{figure zeta}.
         The time interval for each histogram
         corresponds to 100 days.}
\label{figure Nsigma}
\end{figure*}
For the case of a normal distribution of clump radii,  we
calculated the distributions of flare luminosities and durations
for different values of $N_{\sigma}$
(see Figure \ref{figure Nsigma}).
When $N_{\sigma}$ increases, the number of clumps with larger and
smaller radii is reduced,
resulting in  a narrower flare luminosity distribution.
When $\zeta$ increases, the flare distributions have the same
behaviour described above for the case  $\gamma>0$.

We then tried another test, increasing the orbital period (e.g.
from 10 days to 100 days), finding that the shape of the integral
distributions in Figure \ref{figure gamma} remains similar, except
for the number of X$-$ray flares, which decreases.

%---------------------------------------------
\subsection{The effect of the mass-loss rate}
\label{The effect of the mass-loss rate}
%---------------------------------------------

\begin{figure*}
\begin{center}
\begin{tabular}{@{}c@{}c@{}}
\includegraphics[width=9cm]{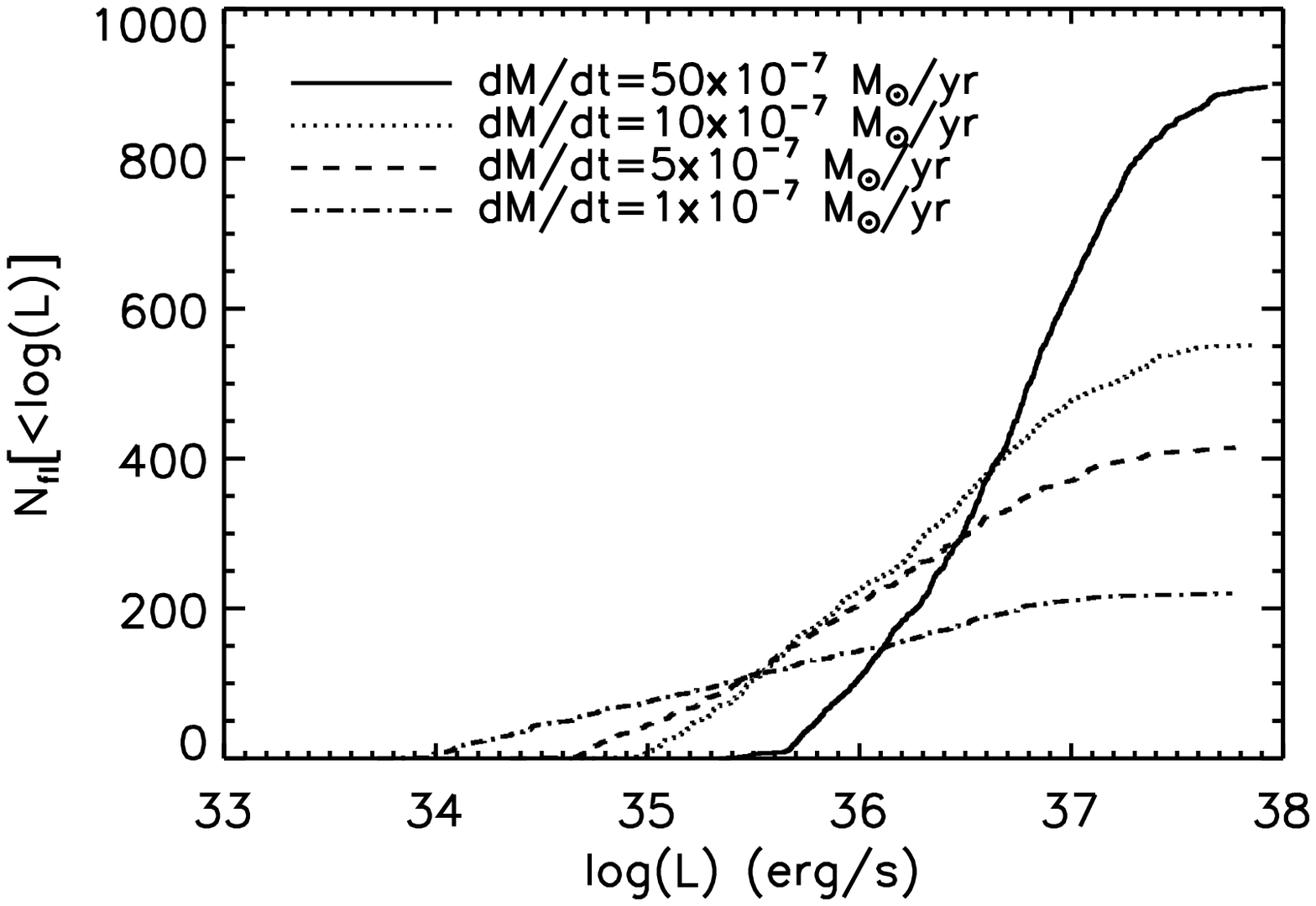} &
\includegraphics[width=9cm]{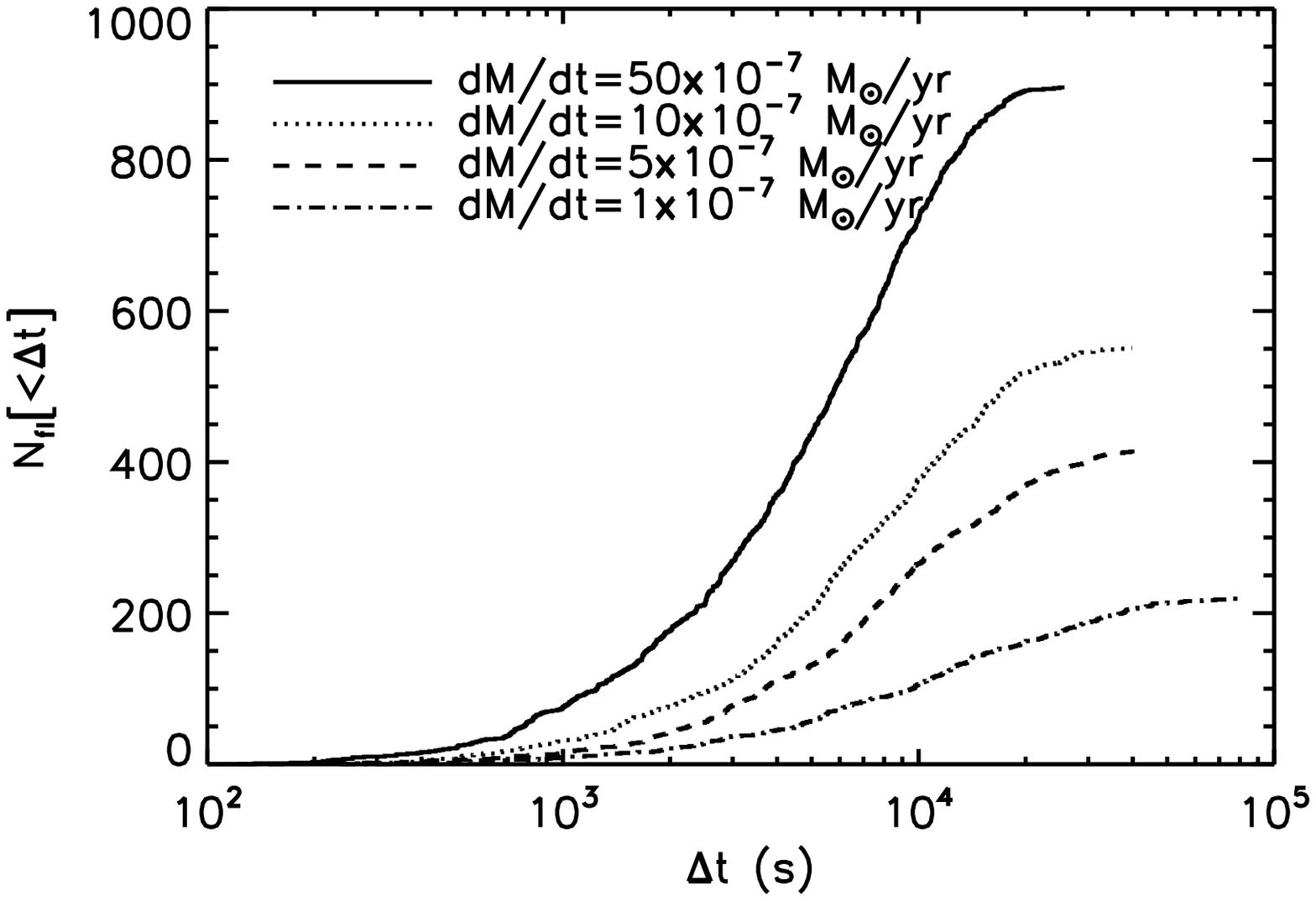}
\end{tabular}
\end{center}
\caption{Expected integral distributions of the flare luminosities
and  durations for different values of the mass-loss rate and
for an assumed orbital period of 10~days and a circular orbit.
$\zeta=1.1$, $\gamma=-1.5$, $f=0.7$.
The other binary system parameters are in the caption of Figure \ref{figure zeta}.
         The time interval corresponds to 100 days.}
\label{figure Mloss}
\end{figure*}

The effects of different wind mass-loss rates $\dot{M}_{\rm tot}$
are shown in  Figure \ref{figure Mloss}. The mass-loss rate is
usually derived observationally from the strength of the H$\alpha$
emission line, since this gives smaller  uncertainties than the
method based on  UV P-Cygni lines \citep{Kudritzki-and-Puls-2000}.
Since the H$\alpha$ line opacity depends on $\rho^2$, the presence
of wind inhomogeneities  leads to an over-estimate of the
mass-loss rate.
In particular, the mass-loss rates from O stars derived from
smooth-wind models measurements with the H$\alpha$ method need to
be reduced by a factor 3 to 10 if the wind is clumpy [see
\citet{Lepine-and-Moffat-2008} and \citet{Hamann-et-al.-2008}]. In
Figure \ref{figure Mloss} we show that if $\dot{M}_{\rm tot}$
decreases, also the number of flares decreases due to the reduced
number of clumps. Also the average luminosity of the flares is
reduced because the number density of clumps decreases resulting
in a smaller probability that two or more clumps overlap.

%---------------------------------------------
\subsection{The effect of the orbital parameters}
\label{The effect of the orbital parameters}
%---------------------------------------------

\begin{figure*}
\begin{center}
\begin{tabular}{@{}c@{}c@{}}
\includegraphics[width=9cm]{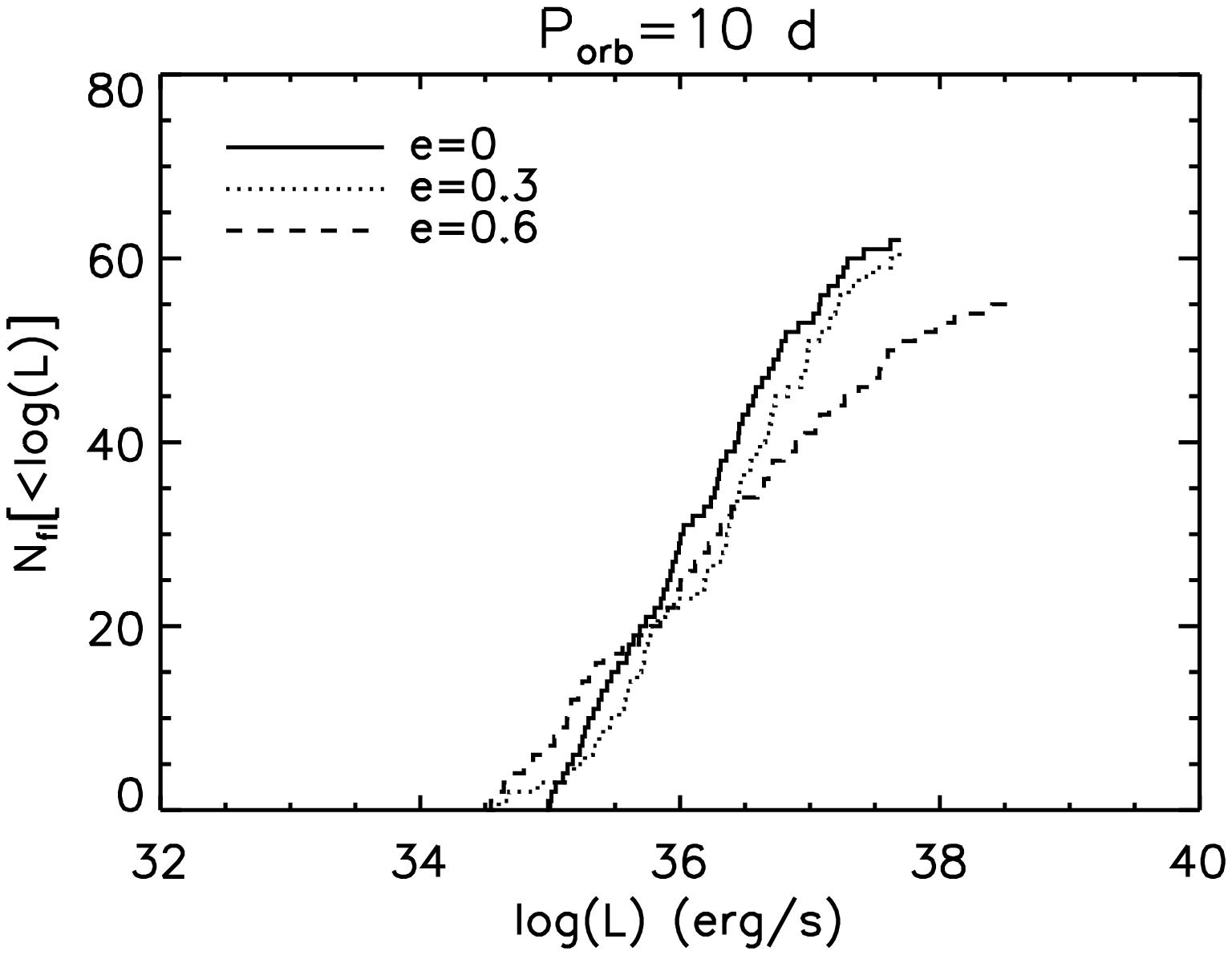} &
\includegraphics[width=9cm]{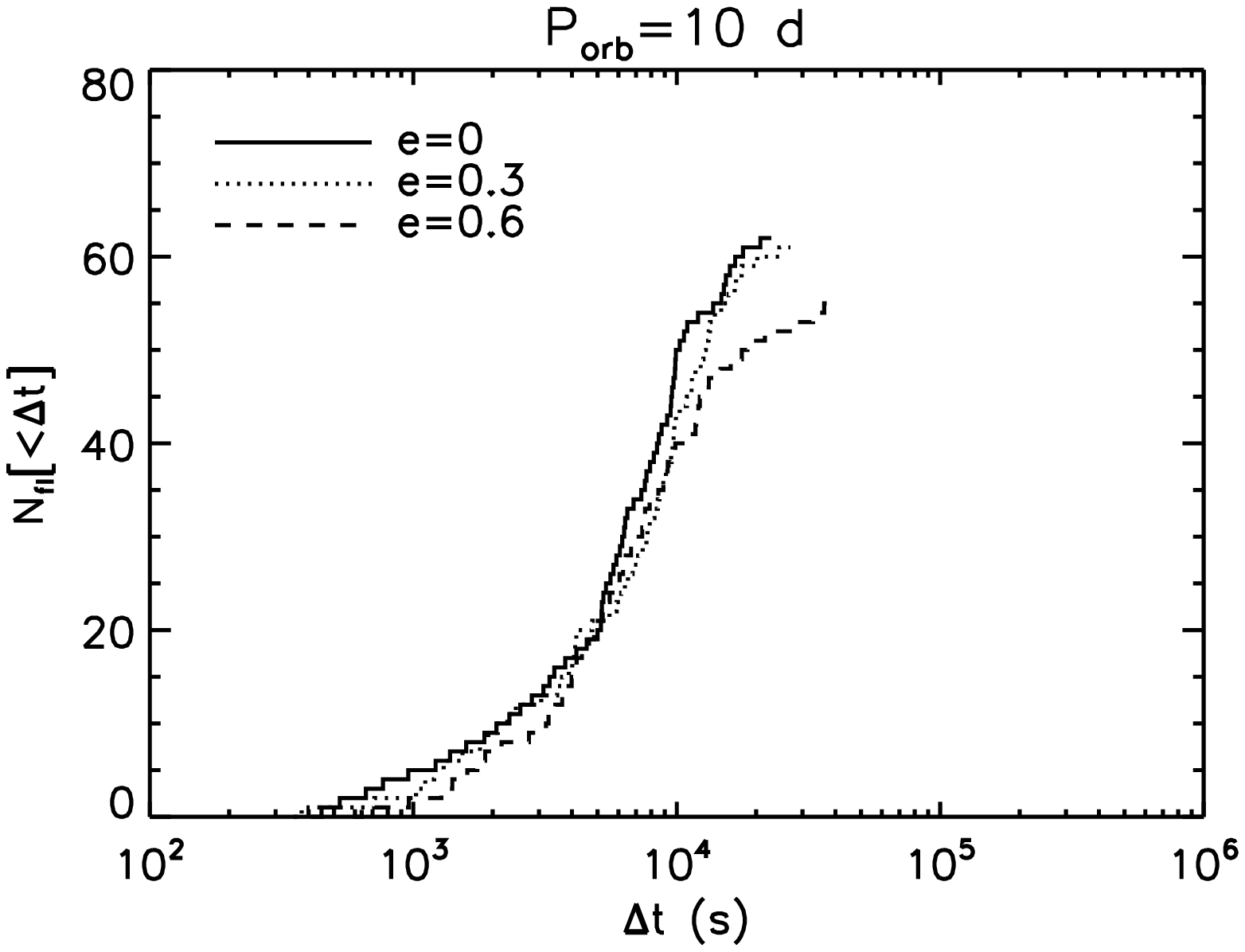} \\
\includegraphics[width=9cm]{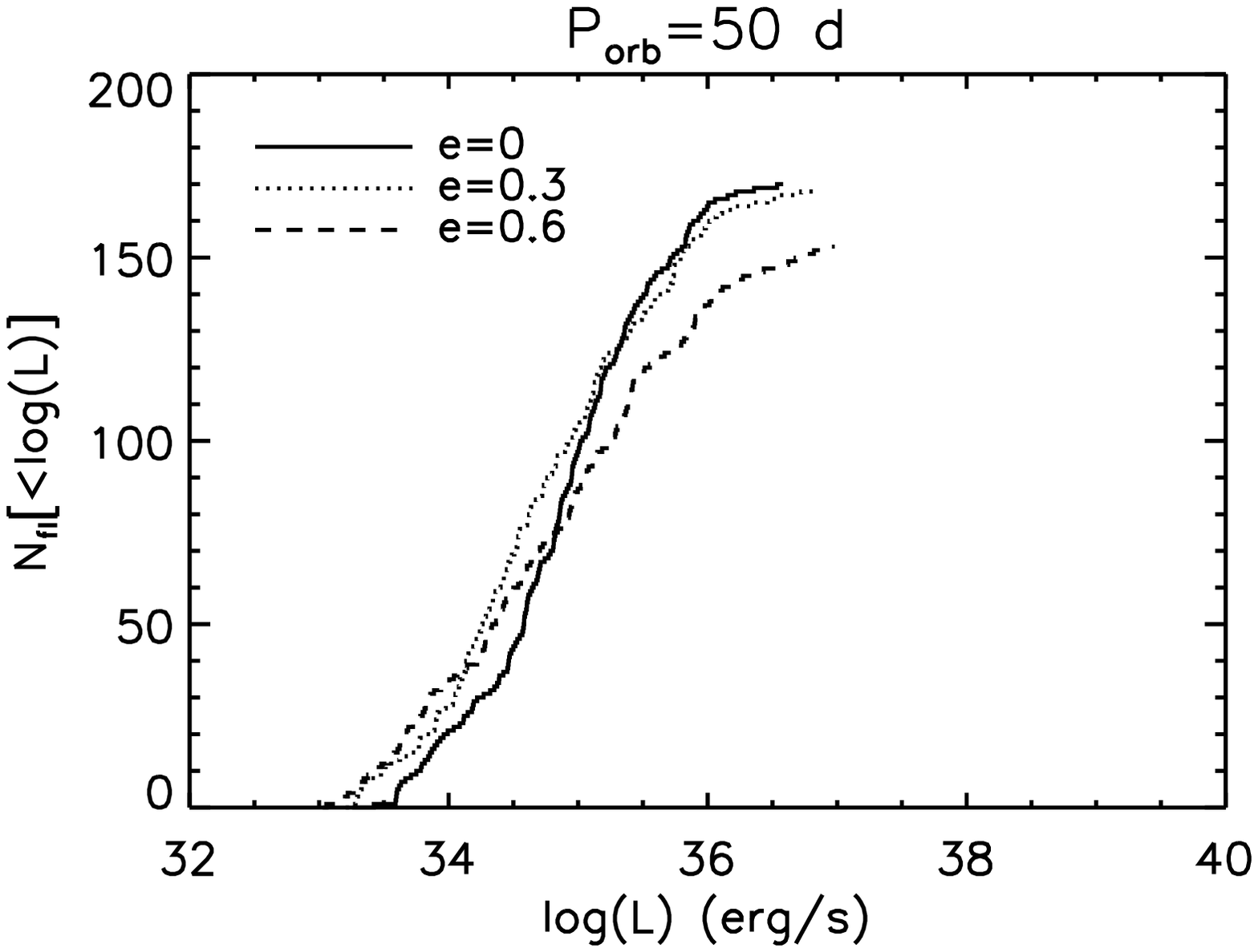} &
\includegraphics[width=9cm]{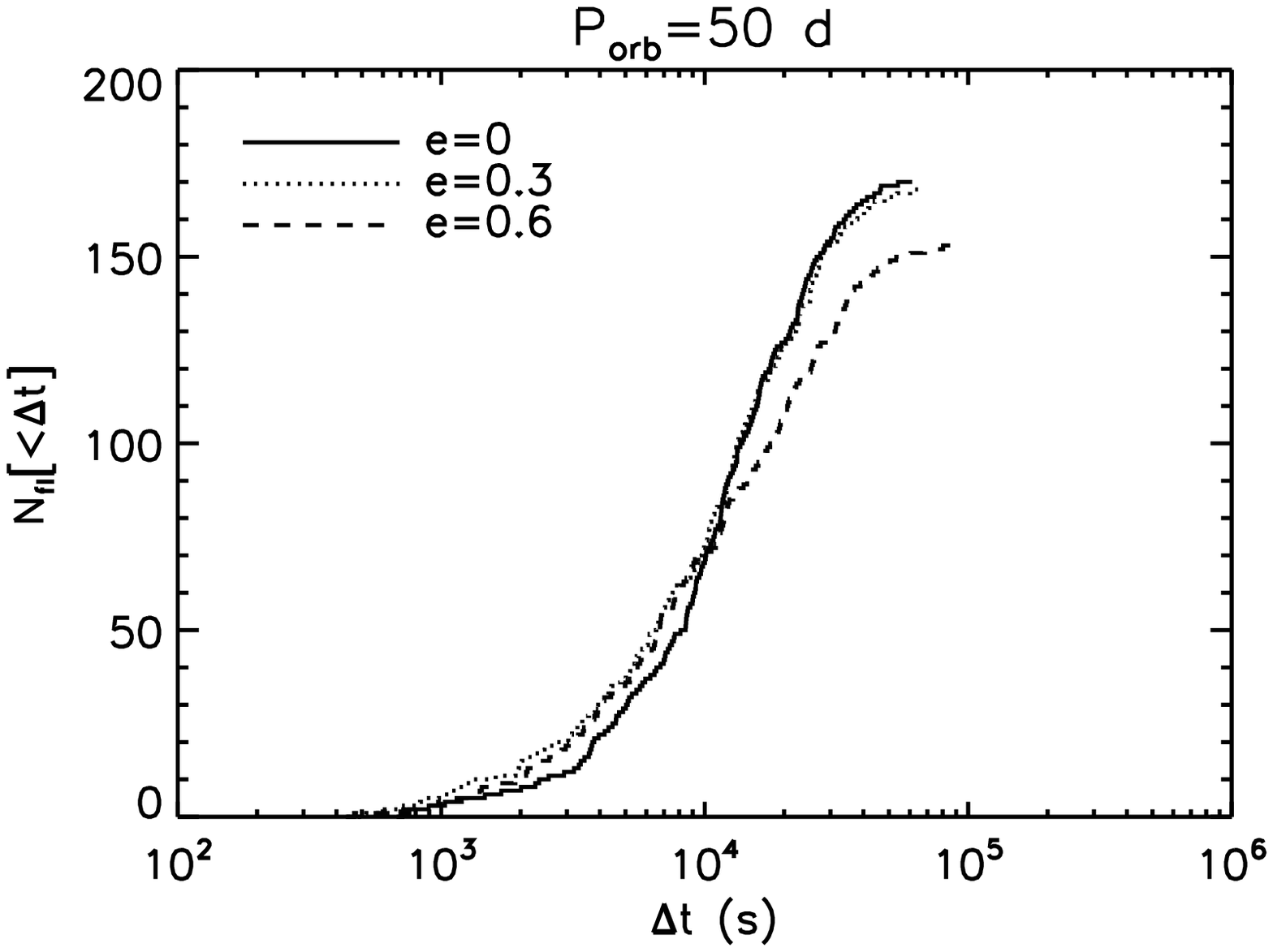} \\
\includegraphics[width=9cm]{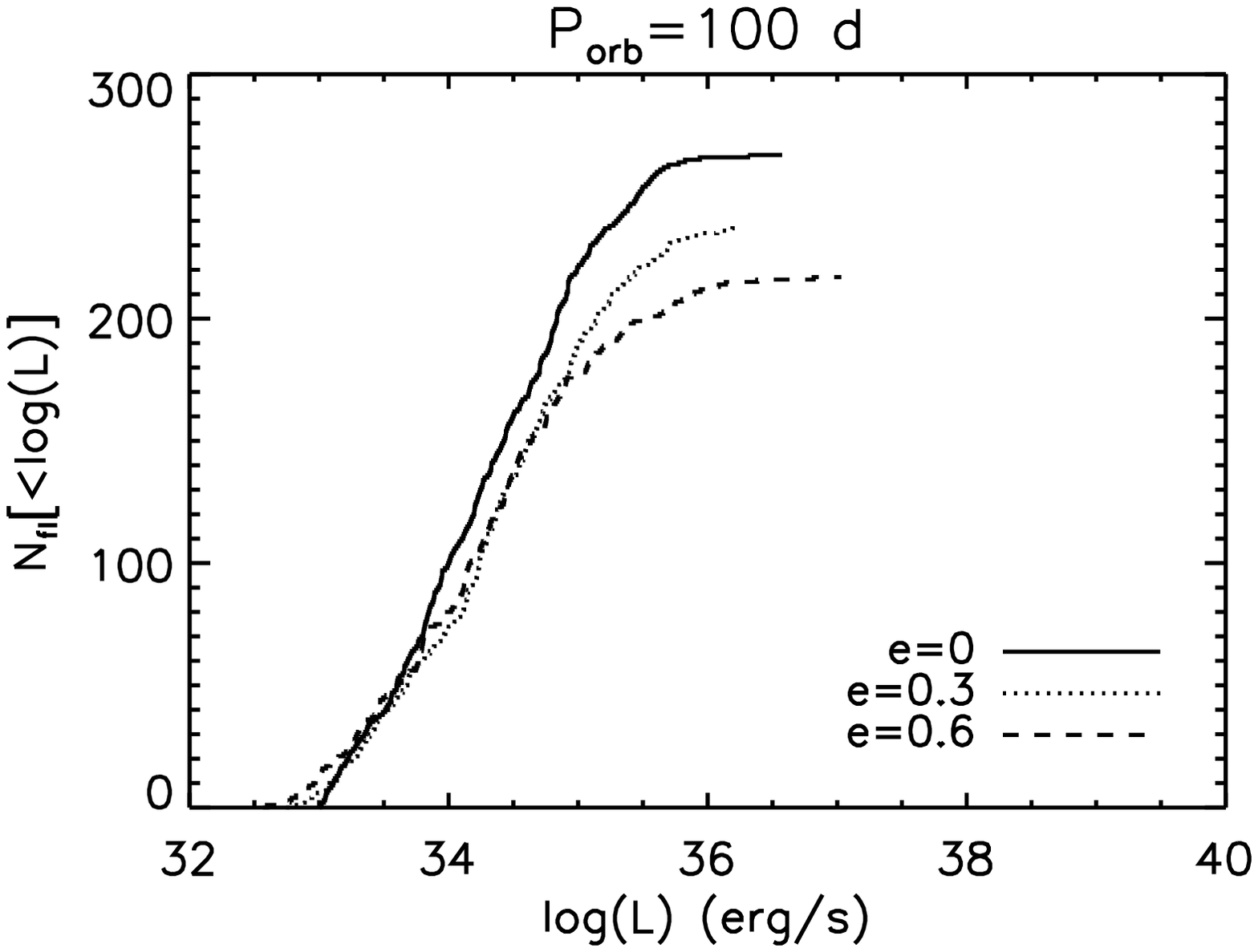} &
\includegraphics[width=9cm]{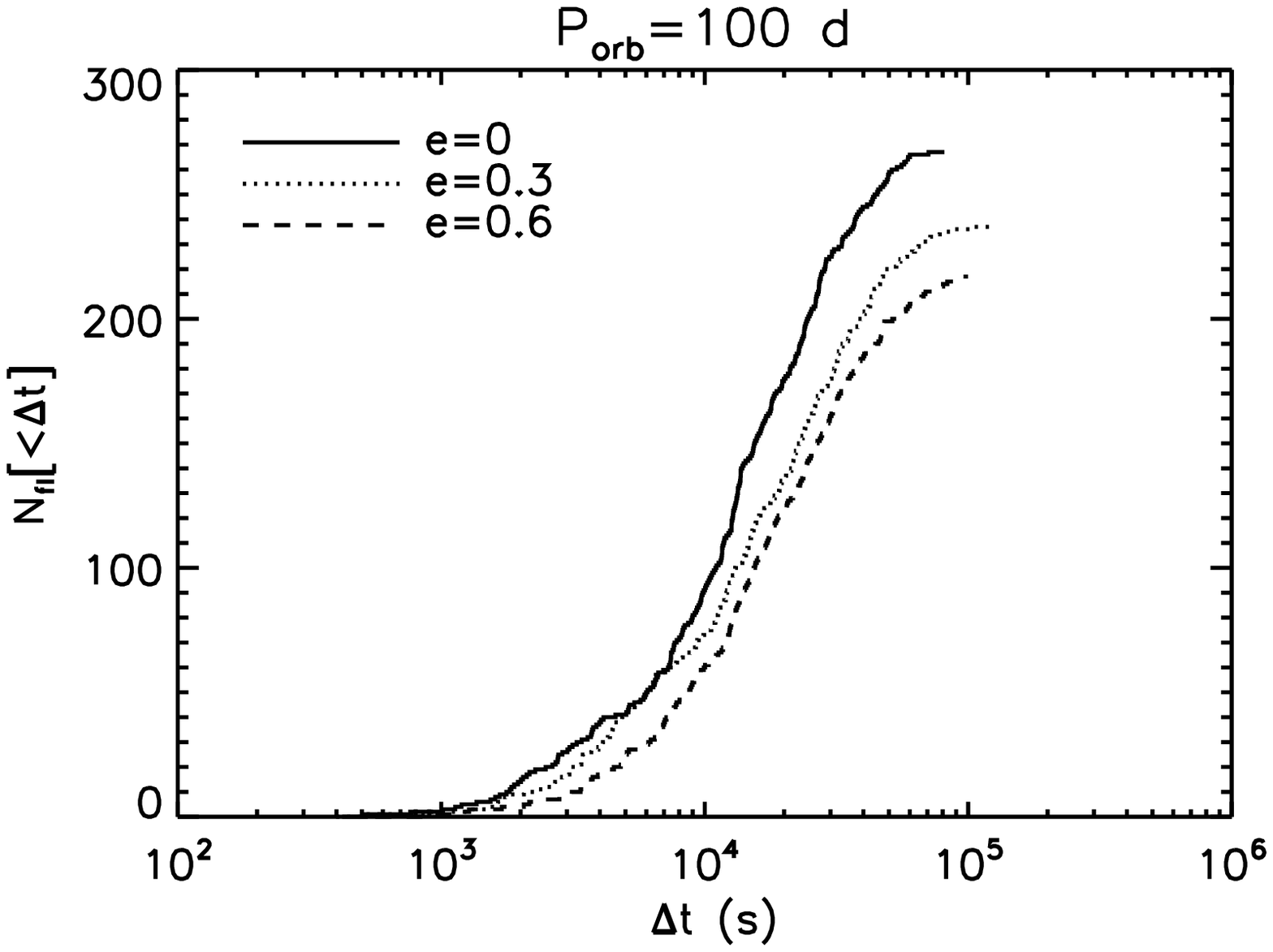}
\end{tabular}
\end{center}
\caption{Expected integral distributions of the flare luminosities
for different orbital periods and eccentrities of the binary system.
$\zeta=1.1$, $\gamma=-1.5$ and $f=0.7$.
The other binary system parameters are in the caption of Figure \ref{figure zeta}.
         The time interval for each histogram
         corresponds to the orbital period.}
\label{figure Porb-e}
\end{figure*}

In Figure \ref{figure Porb-e} we show the effect of changing the
orbital period $P_{\rm orb}$ and the eccentricity $e$. We assumed
$\zeta=1.1$, $\gamma=-1.5$, $f=0.7$.
When the orbital period increases, the number of flares emitted by
the neutron star decreases (see Equation \ref{clump number density}) and the
neutron star accretes clumps with a smaller density (see Equation
\ref{legge_Rcl_r}), implying a shift to lower luminosities and
higher flare durations (see Figure \ref{figure Porb-e}). When the
eccentricity increases, the neutron star accretes clumps with a
higher density range (in general  clumps are denser when they are
closer to the supergiant), thus   the luminosity range of the
flares increases,  as shown in Figure \ref{figure Porb-e}.

%----------------------------------------------------------
\section{Comparison with the HMXB Vela~X$-$1}
\label{Study of the HMXB Vela X-1}
%----------------------------------------------------------

Vela~X$-$1 (4U~1900$-$40) is a bright eclipsing  X-ray binary
($P_{orb}$= $8.964$~d, $e\sim 0.09$) formed by the B0.5~Ib
supergiant HD~77581 \citep{Brucato_and_Kristian-1972} (M =
$23$~M$_{\odot}$, R= $30$~R$_{\odot}$,
\citep{van-Kerkwijk-et-al.1995}) and a pulsar with spin period
$\sim283$~s and  mass $1.9$~M$_{\odot}$.
This source shows significant X--ray variability on short
time-scales, with flares lasting from $\sim 500$~s to $\sim
40000$~s [\citet{Haberl-et-al.-1994}; \citet{Kreykenbohm-et-al.-2008}].

Recently \citet{Kreykenbohm-et-al.-2008} analyzed \emph{INTEGRAL}
observations of Vela~X$-$1 obtained during a phase of high flaring
activity, finding two kinds of flares: brief and bright flares
softer than longer flares. They also found several off-states,
during which the source is not detected (at least by \emph{INTEGRAL}).
\citet{Kreykenbohm-et-al.-2008} proposed that the short flares are
caused by the flip-flop instability, while the long ones are due
to the accretion of clumps ejected by the supergiant. The
off-states are explained as due to the onset of the propeller
effect when the neutron star crosses the lower density inter-clump
medium.

In this Section we apply our wind model to Vela X-1, assuming
the wind parameters derived by \citet{Searle-et-al.-2008} and
reported in Table \ref{3 HMXBs}.

\begin{table*}
\begin{center}
\caption{Parameters of the HMXBs studied in this paper.
         Values labeled with ($^*$) are taken from
         \citet{Searle-et-al.-2008}, values labeled with ($^{\dagger}$)
         are taken from \citet{Lefever-et-al.-2007}.}
\label{3 HMXBs}
\begin{tabular}{lccc}
\hline
                    &          Vela~X$-$1           & 4U~1700$-$377 & IGR~J11215$-$5952 \\
\hline
Type                &            SGXB                &         SGXB         &                SFXT            \\
Spectral Type       &           B0.5~Ib              &     O6.5~Iaf$^+$     &               B0.7~Ia           \\
$M_{\rm OB}$         &       $23$~M$_{\odot}$          &    $58$~M$_{\odot}$   &           $23$~M$_{\odot}$ ($^*$) \\
$R_{\rm OB}$         &       $30$~R$_{\odot}$          &   $21.9$~R$_{\odot}$  &          $35.1$~R$_{\odot}$ ($^*$)\\
$T_{\rm eff}$        &     $2.5 \times 10^4$~K ($^*$) &   $3.5 \times 10^4$~K &       $2.36 \times 10^4$~K ($^*$) \\
$\log (L/L_{\odot})$ &            $5.58$  ($^*$)      &        $5.82$        &                $5.5$     ($^*$)  \\
$\dot{M}_{\rm tot}$           & $0.7 - 1.2 \times 10^{-6}$~M$_{\odot}$~yr$^{-1}$ ($^*$)& $9.5 \times 10^{-6}$~M$_{\odot}$~yr$^{-1}$ & $1 - 2.5 \times 10^{-6}$~M$_{\odot}$~yr$^{-1}$ \\
$v_{\infty}$         & $\sim 1520 - 1600$~km~s$^{-1}$ ($^*$) &    $1700$~km~s$^{-1}$ &   $\sim 1000 - 1400$~km~s$^{-1}$ ($^*$) ($^{\dagger}$)\\
$\beta$             &        $\sim 1 - 1.5$  ($^*$)  &         $1.3$        &           $\sim 1 - 1.5$        \\
$M_{\rm NS}$         &       $1.9$~M$_{\odot}$         &   $2.44$~M$_{\odot}$  &          $1.4$~M$_{\odot}$       \\
$P_{\rm orb}$        &           $8.964$~d            &       $3.412$~d      &              $164.5$~d          \\
$e$                 &         $e\sim 0.09$           &          $0$         &                 --              \\
$P_{\rm spin}$       &         $\sim 283$~s            &          --         &          $186.78 \pm 0.3$~s     \\
distance            &        $\sim 2.0$~kpc          &       $1.9$~kpc      &               $8$~kpc           \\
\hline
\end{tabular}
\end{center}
\end{table*}

\begin{figure}
\begin{center}
\includegraphics[width=4cm, angle=-90]{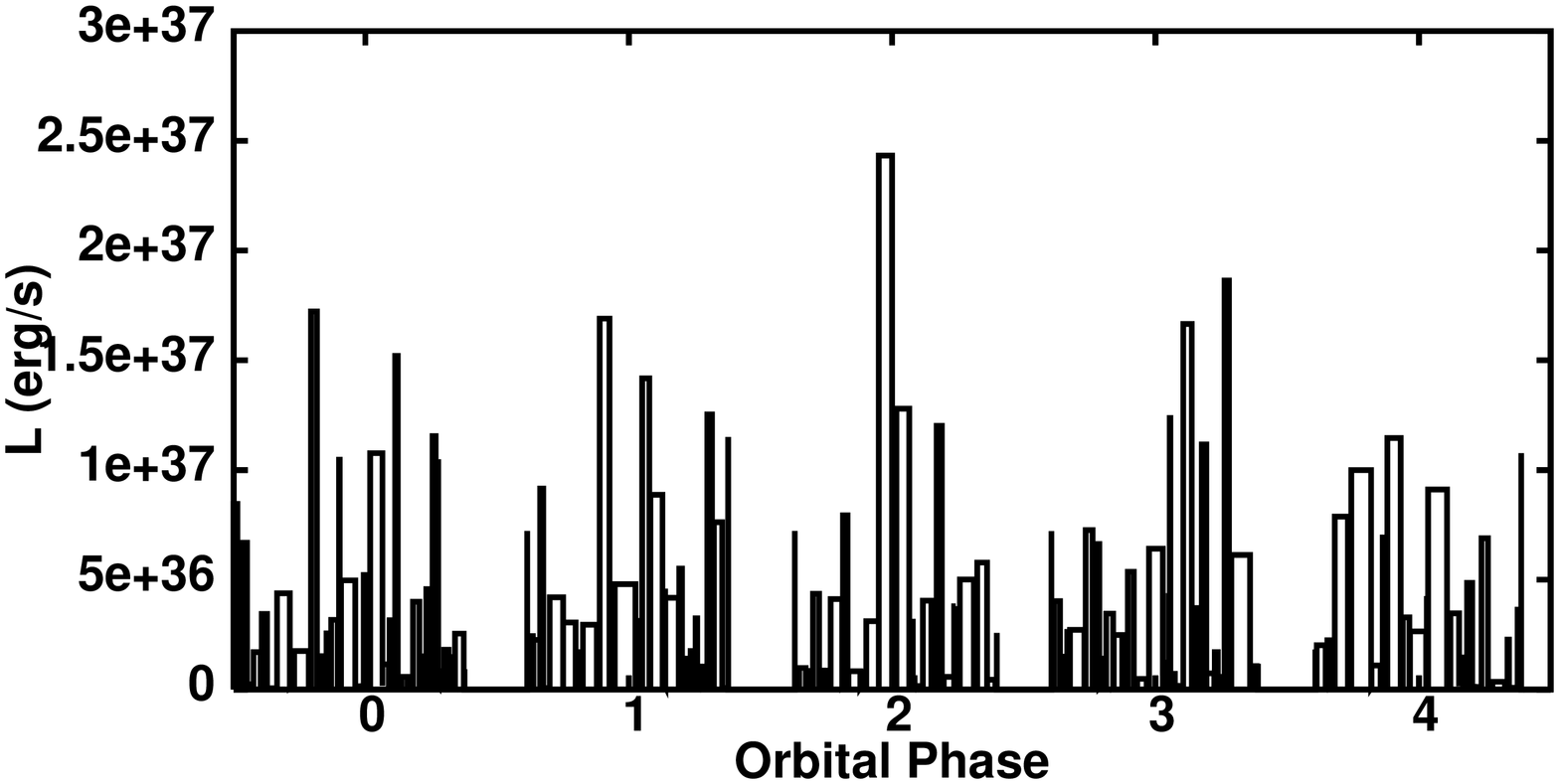}
\includegraphics[width=4cm, angle=-90]{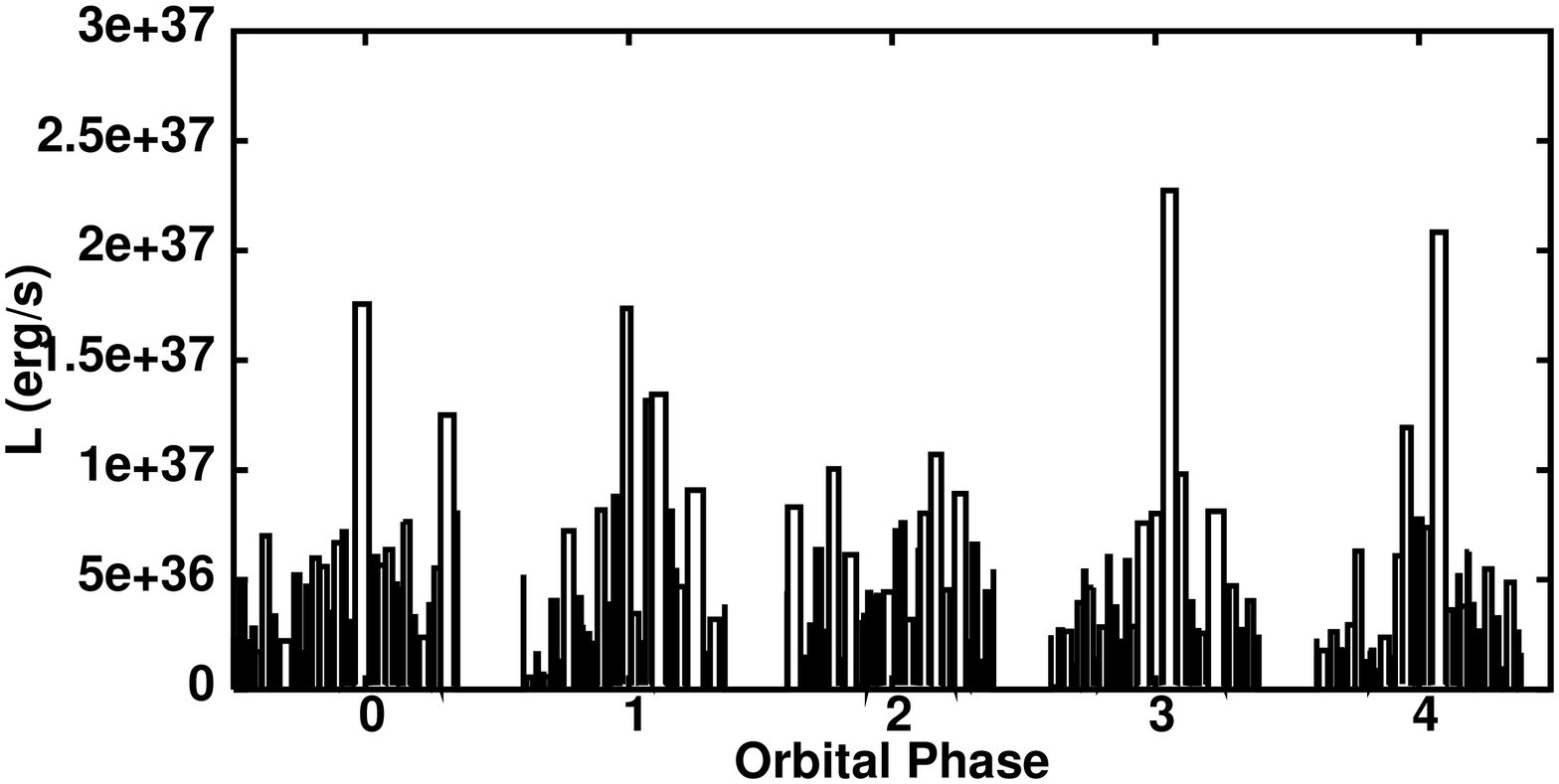}
\includegraphics[width=4cm, angle=-90]{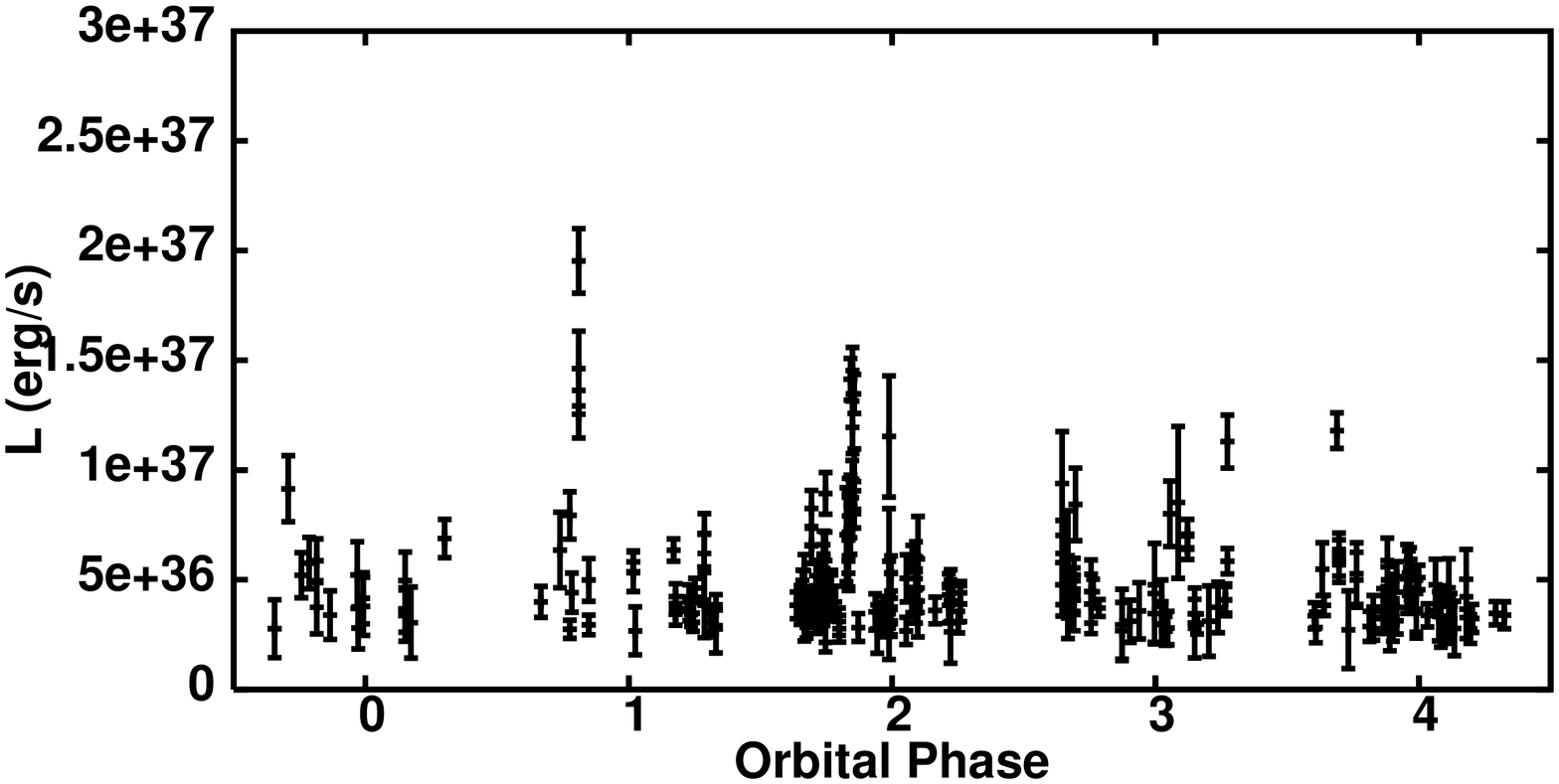}
\end{center}
\caption{Comparison of the Vela~X$-$1 light curve,
      as   observed with ASM/\emph{RXTE} (lower panel),
         with that calculated with our clumpy wind model
         for the  parameters reported in Table \ref{3 HMXBs},
         and $\dot{M}_{\rm tot}=2.1 \times 10^{-7}$~M$_{\odot}$~yr$^{-1}$,
         $\beta=1$, $v_{\infty}=1600$~km~s$^{-1}$,
         $f=\dot{M}_{\rm cl}/\dot{M}_{\rm wind}=0.75$, $M_{\rm a}=5 \times 10^{18}$~g, $M_{\rm b}=5 \times 10^{21}$~g,
         $\zeta=1.1$, $N_{\sigma}=5$ (middle panel), and $\gamma=-1$ (upper panel),
         $k=0.375$, $\alpha=0.522$, $\delta=0.099$.
         Orbital phase $\phi=0$ corresponds to 53750~MJD.
         Note that Vela~X$-$1 is not continuously  observed with
         ASM/\emph{RXTE}. Therefore
         it is possible that some flares have been missed.
}
\label{lcr_VelaX-1_vento_clump}
\end{figure}
In Figure \ref{lcr_VelaX-1_vento_clump}  we  compare  the light
curve measured with the ASM/\emph{RXTE} instrument with that
calculated with our  clumpy wind model assuming a spherical
symmetry for the outflowing wind. The
ASM/\emph{RXTE} count rate, measured in the $2-10$~keV range,  has
been converted to the $1-100$~keV luminosity using  the average
spectral parameters obtained by \citet{Orlandini-et-al.-1998}
and the distance of $2$~kpc \citep{Sadakane-et-al.-1985}.
The observed light curve of Vela~X$-$1 is well reproduced by our
clumpy wind model for 
$\dot{M}_{\rm tot}= 2.1 \times 10^{-7}$~M$_{\odot}$~yr$^{-1}$, 
$M_{\rm a} = 5 \times 10^{18}$~g and $M_{\rm b}= 5 \times 10^{21}$~g,
$\zeta=1.1$, $f=0.75$ and $\gamma=-1$.
Acceptable light curves were also obtained with $\gamma=1$ 
and $\dot{M}_{\rm tot}=4 \times 10^{-7}$~M$_{\odot}$~yr$^{-1}$,
and with a normal distribution law for the clump radii, with $N_{\sigma}=5$.
We point out that the average luminosity
observed by ASM/\emph{RXTE} out of the flares is $\approx 3-4
\times 10^{36}$~erg~s$^{-1}$. This luminosity is obtained in our
model with the accretion of numerous clumps with low density.
%

%----------------------------------------------------------
\section{Comparison with the HMXB 4U~1700$-$377}
\label{Study of the SGXB 4U 1700-377 }
%----------------------------------------------------------

4U~1700$-$377 \citep{Jones-et-al.-1973} is a $3.412$~day eclipsing HMXB
composed of a compact object, (a neutron star or a black hole),
and the O6.5~Iaf$^{+}$ star HD~153919, located at a distance of
$1.9$~kpc \citep{Ankay-et-al.-2001}. Despite extensive searches,
no  X$-$ray pulsations have been found in this system. Therefore
the X$-$ray mass function cannot be determined and the  system
parameters ($M_{\rm OB}$, $R_{\rm OB}$, $M_{x}$,) cannot be derived
directly. They  have been estimated from the radial velocity curve
of the supergiant and from the duration of the X$-$ray eclipse, by
making assumptions about possible values of the radius of the O
star and the orbital inclination. Several studies indicate that
the mass of the compact object is larger than $2$~M$_{\odot}$
[\citet{Rubin-et-al.-1996}; \citet{Clark-et-al.-2002}]. The
similarity of the X$-$ray spectrum to other pulsars suggest that
the compact object of 4U~1700$-$377 is a neutron star
\citep{White-et-al.-1983}, but  the presence of a low-mass black
hole cannot be excluded \citep{Brown-et-al.-1996}. 
\citet{Reynolds-et-al.-1999} reported the
presence of a possible cyclotron feature at $\sim37$~keV observed
with \emph{BeppoSAX}. If confirmed, this would demonstrate that
4U~1700$-$377 hosts a neutron star with  a magnetic field of about
$2.3 \times 10^{12}$~G.

The X$-$ray light curve of 4U~1700$-$377  is characterized by a
strong flaring activity with variations as large as a factor of
$10-100$ on time scales from minutes to hours
[\citet{Haberl-et-al.-1989}; \citet{White-et-al.-1983};
\citet{Rubin-et-al.-1996}]. We assumed the most recent set of
system parameters of 4U~1700$-$377, obtained by means of a
detailed NLTE (Non-Local Thermal Equilibrium) line-driven wind
model analysis of HD~153919 and a Monte Carlo simulation for the
determination of the masses of both components
\citep{Clark-et-al.-2002}. These authors found that the supergiant
has a luminosity $\log(L/L_{\odot})=5.82 \pm 0.07$, 
an effective temperature $T_{\rm eff} \approx 35000$~K, 
radius $R_{\rm OB} \approx 21.9$~R$_{\odot}$,
mass $M_{\rm OB} \approx 58$~M$_{\odot}$, mass loss rate $\dot{M} =
9.5 \times 10^{-6}$~M$_{\odot}$~yr$^{-1}$, and a mass for the
compact object $M_{\rm x}=2.44$~M$_{\odot}$ (Table \ref{3 HMXBs}).

The X$-$ray spectrum of 4U~1700$-$377 is well described by an
absorbed power law with  high-energy cutoff
\citep{van-der-Meer-et-al.-2005}. The spectrum above $20$~keV was
studied using different satellites (e.g. BATSE detector on board
the \emph{CGRO}, \emph{INTEGRAL}) and can be modelled using a
thermal bremsstrahlung model with $kT \sim 25$~keV
\citep{Rubin-et-al.-1996} or with a thermal Comptonization model
\citep{Orr-et-al.-2004}.

The analysis of the $0.5-12$~keV spectrum with XMM-\emph{Newton}
during the eclipse, the egress, and a low-flux interval led
\citet{van-der-Meer-et-al.-2005} to suggest that the low-flux
interval is probably due to a lack of accretion such as expected
in a structured and inhomogeneous wind. Moreover,
\citet{van-der-Meer-et-al.-2005} proposed that the fluorescence
line from near-neutral iron detected in all spectra is produced by
dense clumps. They also observed recombination lines during the
eclipse which indicate the presence of ionized zone around the
compact object.

We  analyzed the public archival \emph{INTEGRAL} data of
4U~1700$-$377, using all the  IBIS/ISGRI observations obtained
from 2003 March 12 to 2003 April 22, and from 2004 February 2 to
2004 March 1. These data correspond to a net exposure time of
$\sim 5.2$~days (excluding the eclipse phase). We reduced the data
using OSA~7.0, and extracted the light curve in the energy range
$15-60$~keV, finding a total of $123$ flares. For each flare we
extracted the spectrum in the range $22-100$~keV. All the spectra
could be well fit by a thermal Comptonization model ({\sc comptt}
in {\sc xspec}).  Based on the spectral results, we computed the
$1-200$~keV  luminosity of each flare. All of them have a
luminosity greater than $5.8 \times 10^{36}$~erg~s$^{-1}$ (for
lower luminosities it is difficult to evaluate the flare duration
and then to distinguish the flares from the average level of the
X$-$ray emission). For each flare we have measured two parameters:
the peak luminosity and the flare duration. We then applied our
clumpy wind model to the \emph{INTEGRAL} observations of
4U~1700$-$377.
\begin{figure*}
\begin{center}
\includegraphics[width=7cm]{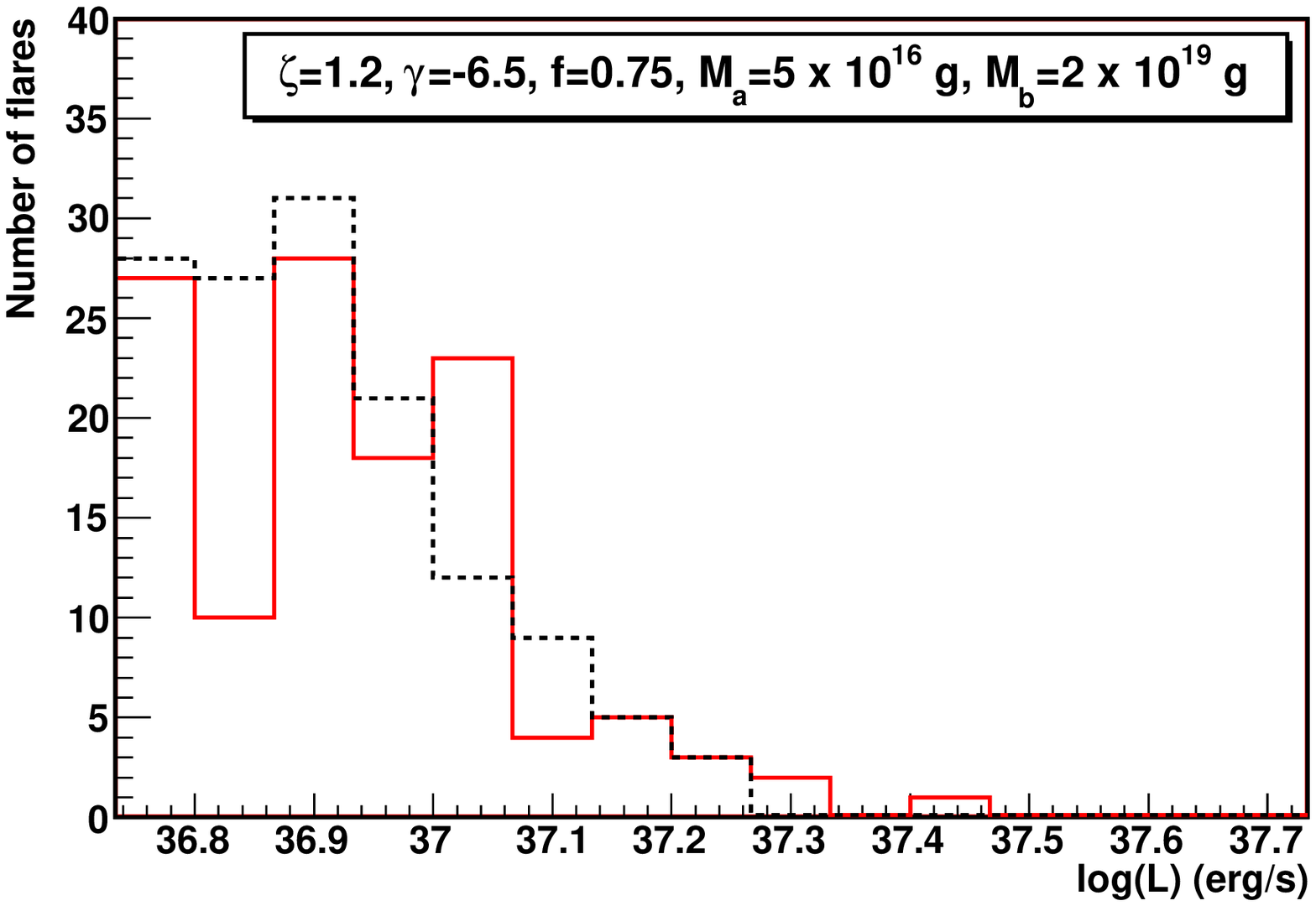}
\includegraphics[width=7cm]{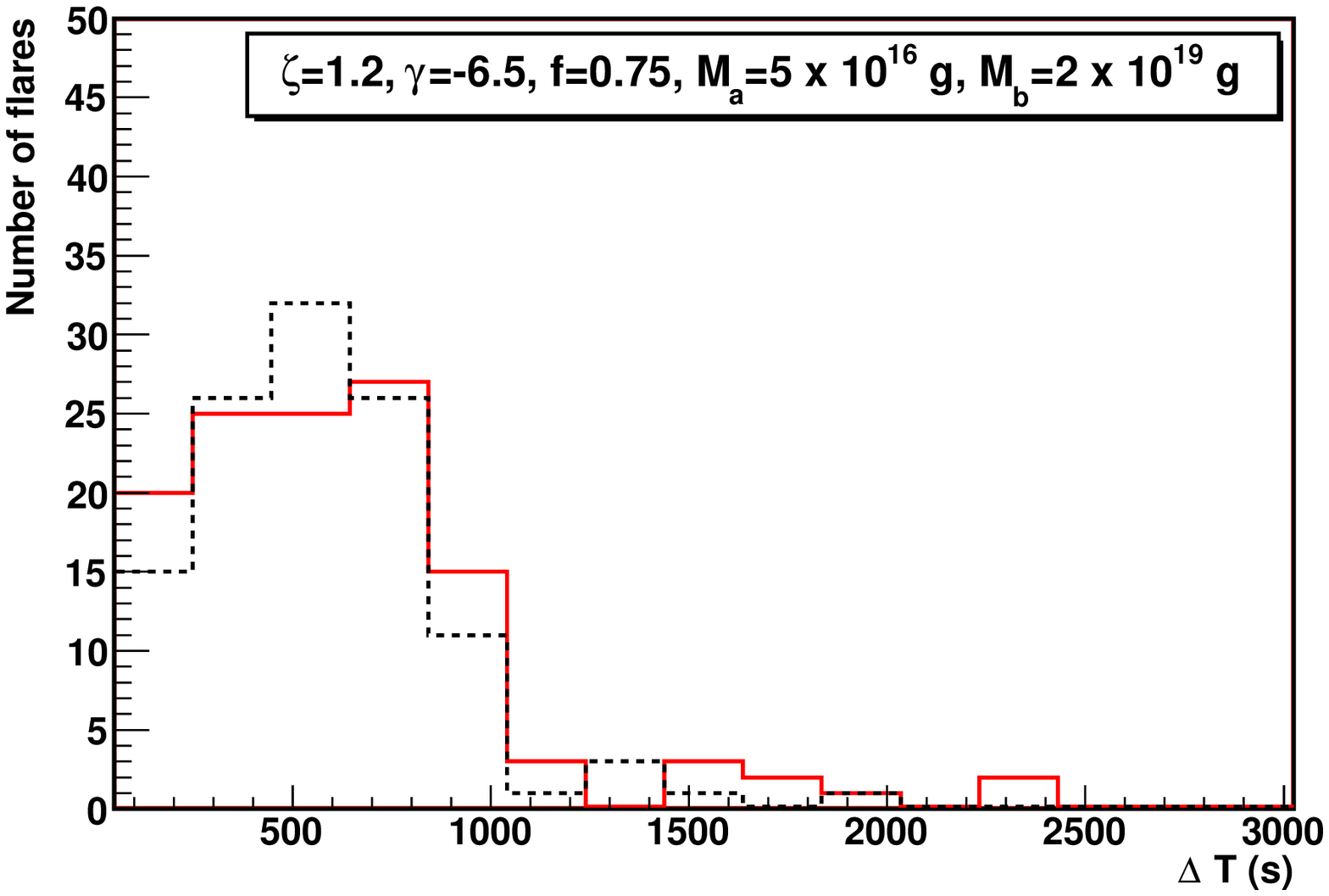}
\end{center}
\caption{Comparison between observed (solid line) and calculated
(dashed line) distributions
         of the flare luminosities and  durations
         for  4U 1700$-$377.
         The binary system parameters are:
         $M_{\rm OB}=58$~M$_{\odot}$, $R_{\rm OB}=21.9$~R$_{\odot}$,
         $M_{\rm NS}=2.44$~M$_{\odot}$, $R_{\rm NS}=10$~km.
         The parameters for the supergiant wind are:
         $\dot{M}_{\rm tot}= 1.3 \times 10^{-6}$~M$_{\odot}$~yr$^{-1}$,
         $v_{\infty}=1700$~km~s$^{-1}$, $\beta=1.3$, $v_0=10$~km~s$^{-1}$,
         $M_{\rm a} = 5 \times 10^{16}$~g and $M_{\rm b}= 2 \times 10^{19}$~g,
         $\zeta=1.2$, $\gamma=-6.5$ and $f=0.75$.
       }
\label{confronto_istogrammi_4U1700}
\end{figure*}
We first compared the observed  distributions of the flare
luminosities and durations with those computed adopting the system
parameters reported in Table \ref{3 HMXBs}. We  assumed  for the
computed distribution a time interval equal to the exposure time
of the 4U~1700$-$377 observations considered here. As shown in
Figure \ref{confronto_istogrammi_4U1700} the flare properties are
well reproduced with our clumpy wind model for  $\dot{M}_{\rm
tot}= 1.3 \times 10^{-6}$~M$_{\odot}$~yr$^{-1}$, $M_{\rm a} = 5
\times 10^{16}$~g and $M_{\rm b}= 2 \times 10^{19}$~g,
$\zeta=1.2$, $\gamma=-6.5$ and $f=0.75$. 
We found that the numbers of  observed (123) and calculated flares (116)
are in good agreement.
\begin{figure}
\begin{center}
\includegraphics[width=4cm, angle=-90]{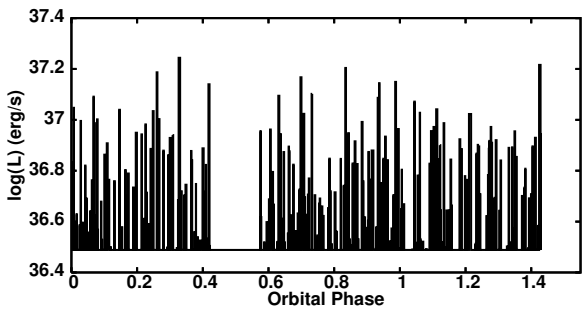}
\end{center}
\caption{Computed light curve of 4U 1700$-$377. The off-state at
orbital phase         $0.42 \la \phi \la 0.58$ is due to the
eclipse.
         The stellar parameters are reported in the caption of  Figure
         \ref{confronto_istogrammi_4U1700}.}
\label{lcr_calculated_4U1700}
\end{figure}
The light curve of 4U~1700$-$377 calculated with our clumpy wind
model is shown in Figure \ref{lcr_calculated_4U1700}.

%----------------------------------------------------------
\section{Comparison with the SFXT IGR~J11215$-$5952}
\label{Study of the SFXT IGR J11215-5952}
%----------------------------------------------------------

The SFXT IGR~J11215$-$5952 was discovered in April 2005 with
\emph{INTEGRAL} \citep{Lubinski-et-al.-2005}. It is associated
with HD~306414, a  B0.7~Ia star located at an estimated
distance of $\sim 8$~kpc \citep{Negueruela-et-al.-2007}. 
\emph{RXTE} observations showed a pulse period $P_{\rm spin}=186.78 \pm 0.3$~s
[\citet{Smith-et-al.-2006}; \citet{Swank-et-al.-2007}]. This is
the first SFXT for which a  periodicity in the outbursts
recurrence time was  discovered \citep{Sidoli-et-al.-2006}.
Subsequent observations \citep{Romano-et-al.-2009} showed that the
true periodicity is about 164.5 days, i.e. half of the originally
proposed value. This periodicity is very likely due to the orbital
period of the system.

For a distance of $8$~kpc the peak fluxes of the outbursts
correspond to a luminosity of $\sim 5 \times 10^{36}$~erg~s$^{-1}$
[$5-100$~keV, \citet{Romano-et-al.-2007b}]. 
\emph{Swift} monitoring of this source revealed that the outburst
(lasting a few days) %15d
is composed by many flares (lasting from minutes to a few hours),
and before and after the whole outburst the source is fainter than
$10^{33}$~erg~s$^{-1}$ \citep{Sidoli-et-al.-2007}.

\citet{Sidoli-et-al.-2007} showed that
accretion from a spherically symmetric homogeneous wind
could not reproduce the observed light curve and therefore
proposed a model based on an anisotropic wind characterized by a
denser and slower equatorial component, periodically
crossed by the neutron star along its orbit. 
However this result was based on the old determination 
of the orbital period (329 days).
Therefore, before applying our clumpy wind model, we checked the
spherically  symmetric homogeneous wind with   $P_{\rm
orb}=164.5$~d, different eccentricity values, and  the set of
stellar parameters derived by \citet{Searle-et-al.-2008} and
\citet{Lefever-et-al.-2007} and reported in Table \ref{3 HMXBs}.
We assumed a terminal velocity $v_{\infty}$ ranging from
$1000$~km~s$^{-1}$ to $1400$~km~s$^{-1}$, and $\dot{M}$ ranging
from $1 \times 10^{-6}$~M$_{\odot}$~yr$^{-1}$ to $2.5 \times
10^{-6}$~M$_{\odot}$~yr$^{-1}$.
In all cases we found that the duration of the  X$-$ray outburst
observed with \emph{Swift} is  shorter than that of the calculated
light curves. 

\begin{figure*}
\begin{center}
\includegraphics[width=9cm, angle=-90]{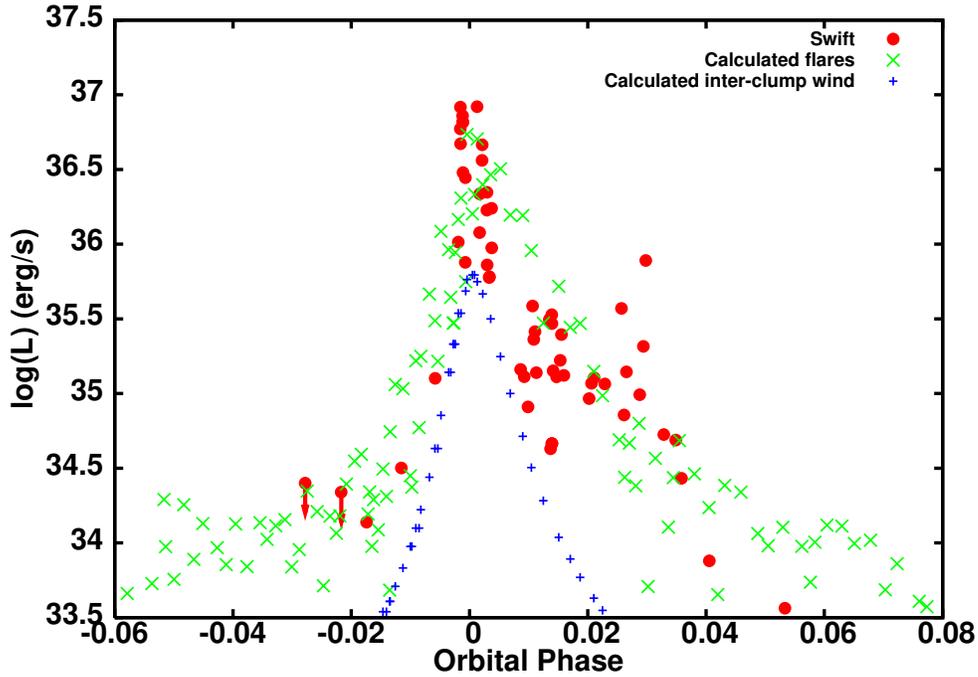}
\end{center}
\caption{Comparison of the IGR~J11215$-$5952 light curve of the February 2007 outburst
         observed with \emph{Swift}/XRT
         with the calculated X$-$ray light curve of our clumpy wind model
         (presented in Section \ref{Section Clumpy stellar winds properties}) with
         the stellar parameters reported in Table \ref{3 HMXBs},
         and $P_{\rm orb}=164.5$~d, $e=0.89$, $\dot{M}_{\rm tot}=2 \times 10^{-7}$~M$_{\odot}$~yr$^{-1}$,
         $\beta=1$, $v_{\infty}=1400$~km~s$^{-1}$,
         $f=\dot{M}_{\rm cl}/\dot{M}_{\rm wind}=0.75$, $M_{\rm a}=10^{17}$~g, $M_{\rm b}=10^{20}$~g,
         $\zeta=1.1$, $\gamma=6$,
         $k=0.709$, $\alpha=0.470$, $\delta=0.089$.
}
\label{lcr_igrj11215_vento_clump}
\end{figure*}

A better agreement with the observations could be obtained with
our clumpy wind model, especially for what concerns the flaring
variability during the outburst phase. However, also in this case
the calculated light curve always produces an outburst longer than
the observed one. This is shown in Fig. \ref{lcr_igrj11215_vento_clump}, where
the  green symbols corresponds to the accretion of a dense clump
(producing a flare), while the blue symbols indicate
the lower luminosity level produced by the accretion of the
inter-clump matter. We found that the probability to observe a
flare, rather than the
inter-clump luminosity level,  is 90\%. %%

In order to improve the agreement between the observed and the
calculated light curve, we introduced an anisotropic outflow similar to that
proposed by \citet{Sidoli-et-al.-2007}. In this modified model
we introduce a denser clumpy wind component in the equatorial
plane, with a thickness $2h$, a terminal velocity $v_{\rm ed}$ and
a mass loss rate $\dot{M}_{\rm ed}$, together with a 
spherically symmetric clumpy wind component (polar wind) 
with terminal velocity $v_{\rm pw}$ and a mass loss rate $\dot{M}_{\rm pw}$. 
We linked the mass loss rate from the equatorial outflow with the mass loss
rate from the polar wind, by means of the factor $f_{\rm ed}$:
\begin{equation} \label{f_ED}
\dot{M}_{\rm ed} = f_{\rm ed} \dot{M}_{\rm pw}
\end{equation}
Both wind components are clumpy and obey laws
described in Section \ref{Section Clumpy stellar winds properties}.
We assume that the second wind component has a Gaussian density profile perpendicular
to the equatorial plane of the supergiant.
In this framework we have considered an orbital period $P_{\rm orb}=164.5$~d
and a high eccentricity in order to produce only one outburst per orbit.
\begin{figure*}
\begin{center}
\includegraphics[width=9cm, angle=-90]{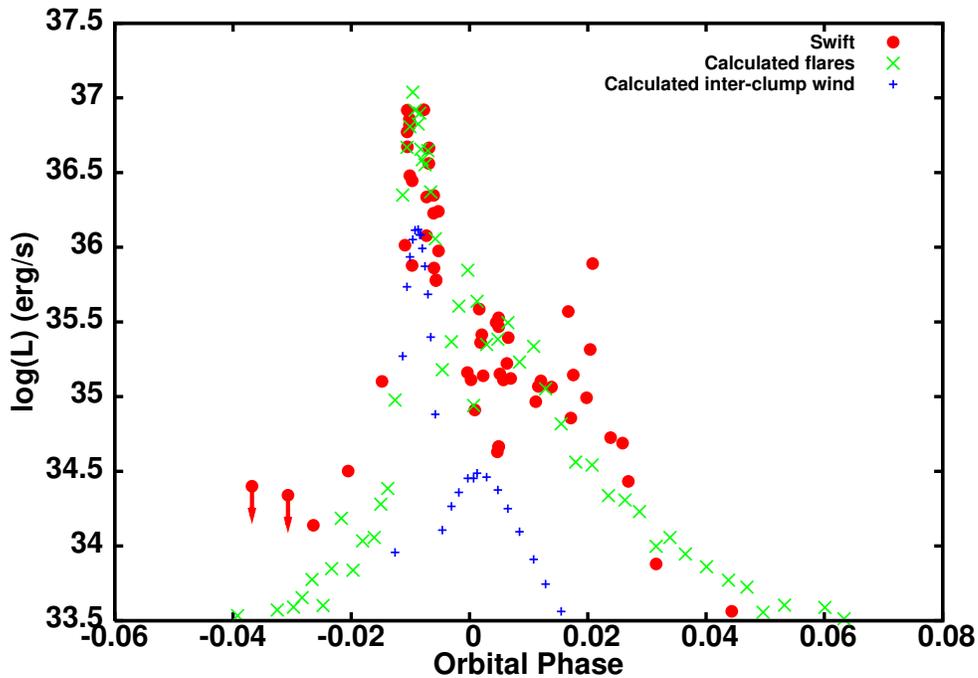}
\end{center}
\caption{Comparison of the IGR~J11215$-$5952 light curve of the February 2007 outburst
         observed with \emph{Swift}/XRT
         with the calculated X$-$ray light curve of our clumpy wind model
         with equatorially enhanced wind component
         (presented in Section \ref{Section Clumpy stellar winds properties}). We used
         the stellar parameters reported in Table \ref{Parameters IGR J11215 equatorially enhanced wind component}.
       }
\label{lcr_igrj11215_vento_clump_disco}
\end{figure*}
The comparison between the \emph{Swift} light curve and that
predicted with this model is shown in Figure
\ref{lcr_igrj11215_vento_clump_disco}. A good agreement with the
data is obtained with  the parameters reported in Table
\ref{Parameters IGR J11215 equatorially enhanced wind component}.

\begin{table}
\begin{center}
\caption{Parameters for the calculated X$-$ray light curves of IGR~J11215$-$5952
         of our clumpy wind model with equatorially enhanced wind component
         (see Figure \ref{lcr_igrj11215_vento_clump_disco}).
         $i_0$ is the angle of inclination of the orbital plane with respect
         the equatorial wind component; $i_p$ is the angle between the orbital plane
         intersection with the equatorial wind component and the direction of the periastron.}
\label{Parameters IGR J11215 equatorially enhanced wind component}
\begin{tabular}{lc}
\hline
Parameter            & Value                                  \\
\hline
$\beta$              & $1$                                    \\
$\dot{M}_{\rm pw}$    & $1 \times 10^{-7}$~M$_{\odot}$~yr$^{-1}$ \\
$v_{\infty, \ pw}$     & $1500$~km~s$^{-1}$                     \\
$i_0$                & $20^{\circ}$                            \\
$i_p$                & $70^{\circ}$                            \\
$h_{\rm ed}$          & $2. \times 10^{11}$~cm                  \\
$f=\dot{M}_{\rm cl}/\dot{M}_{\rm wind}$ & $0.75$                 \\
$M_{\rm a}$           & $5 \times 10^{18}$~g                    \\
$M_{\rm b}$           & $10^{20}$~g                             \\
$\zeta$               & $1.1$                                 \\
$\gamma$              & $6$                                   \\
$f_{\rm ed}$           & $100$                                 \\
$v_{\infty, \ ed}$      & $1000$~km~s$^{-1}$                     \\
\hline
\end{tabular}
\end{center}
\end{table}

%---------------------------------------------
\section{Discussion and Conclusions}
\label{Discussion and Conclusions}
%---------------------------------------------

We have developed a clumpy wind model (where the dynamical problem is not treated) and explored the resulting
effects on an accreting compact object in order to explain the
observed behavior of the SFXTs and the SGXBs. Compared to previous
attempts to  explain the SFXTs outbursts in the context of clumpy
winds  [\citet{Walter-and-Zurita-Heras-2007};
\citet{Negueruela-et-al.-2008}],
we introduced a distribution for the masses and initial dimensions
of the clumps.
We described the subsequent expansion of the clumps
(Equation \ref{legge_Rcl_r})
taking into account realistic upper and lower limits for their
radius (Equations \ref{R_cl_min} and \ref{R_cl_max}).

This model, together with the theory of wind accretion modified
because of the presence of clumps, allow us a comparison with the
observed properties of both the light curves and luminosity
distributions of the flares in SGXBs and SFXTs.

From the calculated integral distributions (Section
\ref{Application of the clumpy wind model}), we found that the
observable characteristics of the flares, such as luminosity,
duration, number of flares produced, depend mainly on the orbital
period (Figure \ref{figure Porb-e}), the scaling parameter $\zeta$
of the power-law distribution for the clump formation rate
(Equation \ref{Npunto}), and the fraction of wind mass 
in the form of clumps ($f=\dot{M}_{\rm cl} /
\dot{M}_{\rm tot}$), as shown in Figures \ref{figure zeta} and
\ref{figure f}. Thus the variability properties of the different
systems do not depend only on the orbital parameters,
but are also significantly affected by the properties of the
clumps (in particular by the parameters $\zeta$, $\gamma$, $f$,
$M_{\rm a}$, $M_{\rm b}$).

We successfully applied our clumpy wind model to three different
high mass X-ray binaries: Vela~X$-$1, 4U~1700$-$377 and
IGR~J11215$-$5952. For the latter source, however, we had to
introduce a denser equatorial component (still with a clumpy
structure) in order to reproduce the flare duration.

\begin{figure}
\begin{center}
\includegraphics[width=6cm, angle=-90]{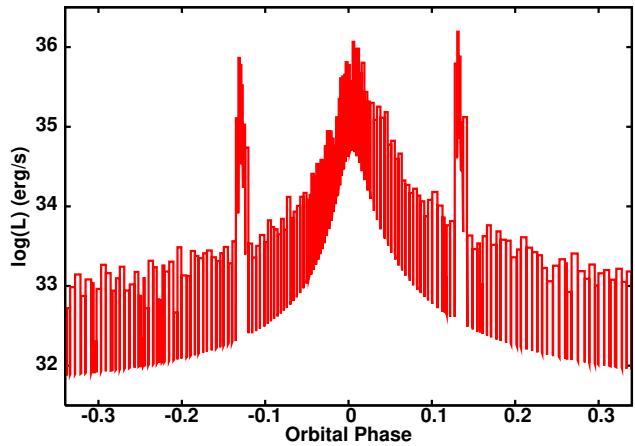}
\end{center}
\caption{Calculated light curve of a SFXT with a neutron star accreting
         an anisotropic clumpy wind. The parameters of the system are:
         $M_{\rm NS} = 1.4$~M$_{\odot}$, $M_{\rm OB} = 23$~M$_{\odot}$, $R_{\rm OB}=35.1$~R$_{\odot}$,
         $L_{\rm OB} = 3.16 \times 10^5$~L$_{\odot}$,
         $P_{\rm orb}=80$~d, $e=0.75$, $\dot{M}_{\rm tot} = 2 \times 10^{-7}$~M$_{\odot}$~yr$^{-1}$,
         $v_{\infty, pw} = 1500$~km~s$^{-1}$, $\beta=1$, $i_0=15^{\circ}$, $i_p=0^{\circ}$,
         $h=2. \times 10^{11}$~cm, $v_{\infty, ed} = 1000$~km~s$^{-1}$, $f_{\rm ed}=50$,
         $T_{\rm eff}=20000$~K, $k=0.709$, $\alpha=0.470$, $\delta=0.089$,
         $\zeta=1.1$, $\gamma=6$, $M_{\rm a}=5 \times 10^{18}$~g, $M_{\rm b}=10^{20}$~g.}
\label{figura_3_outburts}
\end{figure}
In Figure \ref{figura_3_outburts} we show an example
of a light curve of a generic SFXT
calculated in the case of anisotropic clumpy wind.
We assume for the generic SFXT properties similar to IGR~J11215$-$5952,
with an orbital period of $80$~d and an eccentricity $e=0.75$.
The orbital plane intersects the equatorial wind component at two phases
($\phi_1 \approx -0.12$, $\phi_2 \approx 0.12$) producing two outbursts.
The third outburst is produced at the periastron passage ($\phi_3 \approx 0$).
This implies that, if this explanation is correct,
up to 3 outbursts per orbit are possible.
\begin{table*}
\begin{center}
\caption{Summary of HMXBs studied in this paper. We give in table their name, their type (SGXB or SFXT)
         their spectral type, the nature of the compact object and their wind parameters that we have assumed.
         For IGR~J11215$-$5952 is reported the terminal velocity in the polar wind region.}
\label{Summary HMXBs}
\begin{tabular}{lcccccccccc}
\hline
Source          &Type& Supergiant &Compact& $\dot{M}_{\rm tot}$      & $v_{\infty}$ & $\zeta$ &$\gamma$& $f$ &      $M_{\rm a}$       &    $M_{\rm b}$  \\
                &    &            &Object &(M$_{\odot}$~yr$^{-1}$)&km~s$^{-1}$&         &        &     &      (g)       &    (g) \\
\hline
Vela~X$-$1        &SGXB& B0.5~Ib    &  NS   & $2.1 \times 10^{-7}$   &$1600$& $1.1$  & $-1$     &$0.75$& $5\times 10^{18}$ & $5 \times 10^{21}$ \\
4U~1700$-$377     &SGXB&O6.5~Iaf$^+$&  NS?  & $1.3 \times 10^{-6}$ &$1700$& $1.2$  & $-6.5$  &$0.75$& $5\times 10^{16}$ & $2 \times 10^{19}$ \\
IGR~J11215$-$5952 &SFXT& B0.7~Ia    &  NS   & $1 \times 10^{-7}$   &$1500$& $1.1$  & $6$     &$0.75$& $5\times 10^{18}$ & $10^{20}$ \\
\hline
\end{tabular}
\end{center}
\end{table*}

In Table \ref{Summary HMXBs} we have summarized the wind
parameters obtained for the 3 sources.
This table shows the typical differences in the mass-loss rate and
terminal velocity expected from the CAK theory, similar values for
$\zeta$ and $f$, and different values for $\gamma$, $M_{\rm a}$,
$M_{\rm b}$, which seem to be proportional to the inverse of the
effective temperature of the supergiant. Moreover, we found that
the values of $\dot{M}_{\rm tot}$ that best reproduce the observed
light curves of the 3 HMXBs studied, are in agreement with the
hypothesis that the mass-loss rate derived by the
H$\alpha$ emission is overestimated by a factor 2--10, in
agreement with recent studies (see Section \ref{The effect of the
mass-loss rate}).
We can exclude that the observed light curves of the three HMXBs
studied can be reproduced with the same set of wind parameters.

In conclusion, the different values of $\gamma$, $M_{\rm a}$,
$M_{\rm b}$ obtained for the 3 HMXBs studied in this paper reveal
that, in the framework of our clumpy wind model, the properties of
the clumps of these 3 X--ray binary systems are slightly
different, independently of the orbital period. This discrepancy
could be due to the different spectral type of the 3 supergiants,
which could eject structurally inhomogeneus winds with slightly
different properties. We suggest as a possible cause of this
behaviour that the values of $\gamma$, $M_{\rm a}$, $M_{\rm b}$
could be related to the sonic radius 
where the clumps start \citep{Bouret-et-al.-2005}, 
which depends on the supergiant properties (Equation \ref{SonicPoint}).

\section*{Acknowledgments}

L.D. thanks Prof. A. Treves for very helpful discussions.
We thank the referee (L.B. Lucy) for useful comments 
which helped significantly improving our paper.
This work was supported in Italy by ASI contracts  I/023/05/0,
I/088/06/0 and I/008/07/0.

\bibliographystyle{mn2e}
%\bibliography{lducci_hmxbs.bib}

\bsp

\label{lastpage}

\end{document}